\documentclass[manuscript]{aastex}
\usepackage{amsmath,amssymb}
\usepackage{listings}
\usepackage{color,enumerate}
\usepackage{graphicx,lscape}

\newcommand{\etal}{et~al.~}
\newcommand{\hunit}{km~sec$^{\hbox{\scriptsize 
            -1}}$~Mpc$^{\hbox{\scriptsize -1}}$}

\newcommand{\kms}{\ifmmode\,{\rm km}\,{\rm s}^{-1}\else km$\,$s$^{-1}$\fi}

\newcommand{\magarc}{\ifmmode{{{{\rm mag}~{\rm arcsec}}^{-2}}}
             \else {{{mag}$~${arcsec}$^{-2}$}}\fi}
\def\Eq#1{Equation~(\ref{eq:#1})}
\def\eq#1{Equation~(\ref{eq:#1})}
\def\se#1{Section~\ref{sec:#1}}
\def\Fig#1{Figure~\ref{fig:#1}} 
\def\Table#1{Table~\ref{tab:#1}} 
\def\eg{{e.g.,~}} 
\def\ie{{i.e.~}}
\def\be{\begin{equation}} 
\def\ee{\end{equation}}
\def\ifm#1{\relax\ifmmode#1\else$\mathsurround=0pt #1$\fi}
\def\beqa{\begin{eqnarray}} 
\def\eeqa{\end{eqnarray}} 
\def \spose#1{\hbox to 0pt{#1\hss}}
\def \lta{\mathrel{\spose{\lower 3pt\hbox{$\sim$}}
     \raise 2.0pt\hbox{$<$}}}
\def  \gta{\mathrel{\spose{\lower 3pt\hbox{$\sim$}}
      \raise 2.0pt\hbox{$>$}}}
\def \ltsima{$\; \buildrel < \over \sim \;$} 
\def \lsim{\lower.5ex\hbox{\ltsima}} 
\def \gtsima{$\; \buildrel > \over \sim \;$} 
\def \gsim{\lower.5ex\hbox{\gtsima}}

\def \Msun {\ifmmode M_{\odot} \else $M_{\odot}$ \fi}

\def \ion#1#2{#1{\footnotesize{#2}}\relax} 
\def \ha {H$\alpha$} 
\def \hi{\ion{H}{I}}

\def \Rdo{R^c_{\rm d}}
\def \Rd{R_{\rm d}}
\def \Re{R_{\rm e}}
\def \Rp{R_{\rm p}}
\def \Rpe{R_{\rm p50}}
\def \Rpn{R_{\rm p90}}
\def \Mbar {M_{\rm bar}}
\def \Mgas {M_{\rm gas}}
\def \Mstar {M_*}
\def \farcm {\ifmmode \rlap.{' }\else $\rlap{.}' $\fi}            

\shorttitle{SDSS Scaling Relations of Spiral Galaxies}
\shortauthors{Hall et~al.}

\slugcomment{Submitted to MNRAS 2011, not for distribution} 
\begin{document}

\title{An Investigation of Sloan Digital Sky Survey Imaging Data
       and Multi-Band Scaling Relations of Spiral Galaxies (with
       Dynamical Information)}

\author{Melanie Hall and St\'{e}phane Courteau}
\affil{Department of Physics, Engineering Physics and Astronomy, 
Queen's University, Kingston, Ontario, Canada} 

\author{Aaron A. Dutton}
\affil{University of Victoria, 
Department of Physics and Astronomy, 
Victoria, British Columbia, Canada}

\author{Michael McDonald} 
\affil{Kavli Institute for Astrophysics and Space Research, 
Massachusetts Institute of Technology Cambridge, MA, USA} 

\author{Yucong Zhu} 
\affil{Department of Astronomy, Harvard University, Cambridge, MA, USA} 

\email{mhall@astro.queensu.ca, courteau@astro.queensu.ca, dutton@uvic.ca,
 mcdonald@space.mit.edu, yzhu@cfa.harvard.edu} 

       
\begin{abstract}
  We have compiled a sample of 3041 spiral galaxies with multi-band
  $gri$ imaging from the Sloan Digital Sky Survey (SDSS) Data Release
  7 and available galaxy rotational velocities, $V$, derived from \hi\
  line widths.  We compare the data products provided through
  the SDSS imaging pipeline with our own photometry of the SDSS
  images, and use the velocities, $V$, as an independent metric
  to determine ideal galaxy sizes ($R$) and luminosities ($L$).
  Our radial and luminosity parameters improve upon the SDSS DR7
  Petrosian radii and luminosities through the use of isophotal
  fits to the galaxy images.  This improvement is gauged via 
  $VL$ and $RV$ relations whose respective scatters are reduced
  by $\sim$8\% and $\sim$30\% compared to similar relations
  built with SDSS parameters.
  The tightest $VRL$ relations are obtained with the $i$-band radius,
  $R_{23.5,i}$, measured at 23.5 \magarc, and the luminosity
  $L_{23.5,i}$, measured within $R_{23.5,i}$.
  Our $VRL$ scaling relations compare well, both in scatter and slope, 
  with similar studies (such comparisons however depend sensitively
  on the nature and size of the compared samples). 
  The typical slopes, $b$, and observed scatters, $\sigma$, of the $i$-band
  $VL$, $RL$ and $RV$ relations are 
  $b_{VL}=0.27\pm0.01$, 
  $b_{RL}=0.41\pm0.01$,
  $b_{RV}=1.52\pm0.07$, and 
  $\sigma_{VL}=0.074$,
  $\sigma_{RL}=0.071$,
  $\sigma_{RV}=0.154$.
  Similar results for the SDSS $g$ and $r$ bands are also provided.
  Smaller scatters may be achieved for more pruned samples. 
  We also compute scaling relations in terms of the baryonic mass
  (stars + gas), $\Mbar$, ranging from $\Mbar \simeq 10^{8.7}$~\Msun
  to $10^{11.6}$\Msun.
  Our baryonic velocity-mass ($VM$) relation has slope $0.29\pm0.01$
  and a measured scatter $\sigma_{meas} = 0.076$ dex.   While
  the observed $VL$ and $VM$ relations have comparable scatter, the
  stellar and baryonic $VM$ relations may be intrinsically tighter,
  and thus potentially more fundamental, than other $VL$ relations
  of spiral galaxies.  
\end{abstract}



\section{Introduction}\label{sec:intro}

Fundamental scaling relations for spiral galaxies are known to emerge
from the combination of observed galaxy rotation velocity, $V$, total
luminosity $L$, and size, $R$.  For instance, the $VL$ relation, also
known as the Tully-Fisher relation (Tully \& Fisher 1997), probably
defines the fundamental plane of spiral galaxies.  That is, the
scatter of the $VL$ relation cannot be reduced by considering any
other third parameter (\eg Courteau \& Rix 1999, hereafter CR99;
Courteau \etal 2007, hereafter C07; Dutton \etal 2007, hereafter D07;
Pizagno \etal 2007).  The study of galaxy scaling relations also
enables a direct comparison with theoretical models of galaxy
formation (\eg Pizagno \etal 2005; D07; Avila-Reese \etal 2008,
hereafter AR08; Dutton \& van den Bosch 2009; Dutton \etal 2011,
hereafter D11).  These and other studies suggest that the simultaneous
matching of the $VL$ and $RL$ relations, whilst matching the observed
galaxy luminosity function and reproducing the shape of galaxy surface
brightness profiles is a challenging task. For example, for standard
disk galaxy models (\eg Mo, Mao \& White 1998; D07) with standard
stellar initial mass functions (i.e., Kroupa 2001/Chabrier 2003) to
match basic galaxy scaling relations, D07 and D11 showed that halo
expansion is required.  This may be realized through dynamical
friction on baryonic clumps and/or supernova driven mass outflows
  (e.g., Navarro, Eke, \& Frenk 1996; El-Zant, Shlosman, \& Hoffman
  2001; Mo \& Mao 2004; Governato \etal 2010; Cole, Dehnen \&
  Wilkinson 2011).

It has also been stated that the luminosity of a spiral galaxy is a poorer
tracer of its circular velocity than baryonic mass, $\Mbar$ (McGaugh
\etal 2000; McGaugh 2005).  The latter is defined as the sum of the
luminous mass, $\Mstar$, and the gas mass, $\Mgas$.  $\Mstar$ is
usually obtained by multiplying the total extrapolated luminosity,
measured in a specific wave band, by a suitable stellar mass-to-light
($M/L$) ratio.  $\Mgas$ is measured directly from the \hi\ flux, using
a correction factor of 1.4 to account for the mass fraction of helium. 
At low total galaxy mass, $\Mgas$ can be a significant fraction
of $\Mstar$, thus raising the question whether $L$, $\Mstar$ or
$\Mbar$ is a better match to $V$.  The latter, which is also known
as the ``baryonic Tully-Fisher'' or ``BTF'', may also have a different
slope than the standard $VL$ relation (Bell \& de Jong 2001;
Verheijen 2001; McGaugh 2005; Gurovich \etal 2010).  The samples
that have been used for BTF studies have typically included fewer
than $\sim 50$ galaxies based largely on multiple, heterogeneous samples. 
For their sample of 243 galaxies, McGaugh \etal (2000) found that
the BTF relation is more ``fundamental'' than the $VL$ relation;
deviations from the BTF relation may however exist (McGaugh \& Wolf 2010).
However, a BTF relation based on a large, statistical sample of galaxies
is still lacking.  The discrepancies between published BTF slopes and
scatters also motivate a new study with as large a galaxy sample with
accurate rotation velocities and \hi\ fluxes as possible.

Interest in galaxy scaling relations also stems from wanting accurate
distance estimators, which in turn is obtained via the suitable pairing
of a distance-dependent and distance-independent galaxy parameters
such as size, luminosity or colour with circular velocity.  The infrared
$VL$ relation has typical distance errors of $15-20\%$ (\eg Aaronson \etal 1979;
Pierce \& Tully 1988; Gavazzi \etal 1999; Masters \etal 2006; hereafter M06;  
Saintonge \& Spekkens 2011; hereafter SS11).  Very large data
samples and carefully measured galaxy parameters can reduce
sampling error in these studies.  The Sloan Digital Sky Survey
(Abazajian et al. 2009; hereafter SDSS) is currently the largest
data base of galaxy structural parameters\footnote{The NYU
Value-Added Galaxy Catalogue by Blanton \etal (2005) also offers
a cross-matched collection of galaxy catalogs, using the SDSS library
as a core.}.
It is thus relevant to ask if the SDSS library of galaxy
scaling parameters yields the tightest possible scaling relations. 

A main goal of this paper is indeed to investigate the quality
of galaxy scaling relations based on size, luminosity, and colour
derived from SDSS data products.  The intent is to compare SDSS
pipeline data products with similar measurements extracted from
SDSS images but using independent data reduction methods.

We develop our analysis of $VRL$ relations through two specific
channels: i) we first compare the multi-band data products from the
SDSS DR7 with our own galaxy structural parameters extracted from
SDSS FITS galaxy images, and ii) we determine which of the $VRL$
parameters yield the tightest luminous and baryonic scaling relations.
Limitations of the SDSS data products will be addressed along the
way (see also Masjedi \etal 2006; Lauer \etal 2007; Fathi \etal 2010). 

To set the foundations of the $VRL$ relation, we follow
Courteau \etal (2007, hereafter C07) whose sample comprised
1300 late-type galaxies.  The sample that we present here, 
having more than $\sim 3000$ galaxies, is a two-fold
increase over C07.  We follow most, though not all, of the
reduction methods and parameter corrections from C07.  For
instance, C07 used photometry and rotational velocities from
four separate sources (Mathewson \etal 1992; Tully \etal 1996;
Dale \etal 1999; Courteau \etal 2000).  A significant improvement
of this study over C07 is our use of strictly homogeneous data
for both the multi-band photometry (SDSS) and line widths (S05/S07).
C07 also used disk scale lengths as a measure of galaxy size in
order to facilitate comparisons with galaxy formation models (D07).
We consider disk scale lengths here as well, but will also extract
other radial metrics that yield tighter $VRL$ relations (see also SS11).

The available SDSS parameters that are of interest to us, namely
the galaxy size, $R$, and luminosity, $L$, are both distance-dependent.
Thus, in order to test which of our measurements or the SDSS
parameters yield the tightest scaling relations, we must compare
each value against an independent metric which we take here as
the distance-independent galaxy velocity.

Our study benefits from the availability of rotational velocities,
$V$, from spatially-unresolved neutral hydrogen (\hi) 21 cm spectral
line widths, for nearly 9000 spiral galaxies (Springob \etal 2005;
hereafter S05; Springob \etal 2007, hereafter S07).   Cross-correlation
of the S05 and S07 line width catalogs with the SDSS will define our
target sample.

Our paper is organised as follows: 
In \se{data}, we cross-correlate the S05 and S07 \hi\ line width
catalogs with the SDSS DR7; this yields a sample of 3041 disk galaxies
with accurate photometry and line widths.  We discuss the extraction
of radial and luminous parameters from SDSS images in \se{LightProfiles}. 
Final galaxy parameter corrections are applied in \se{Corrections}
and the data table of galaxy structural parameters is presented in
\se{FinalDataTables}.  In \se{ScalingMeasures}, we compare the SDSS
and our galaxy parameters against an independent foil, here chosen to
be the (distance-independent) rotational velocity, $V$, from S05/S07.
We determine the ``best'' scaling parameters with which to build
the tightest $VRL$ relation in \se{VRL} and present, in \se{BTF},
the largest baryonic Tully-Fisher sample to date.
A summary of our results is presented in \se{conclusion}. 


\section{Data Sample}\label{sec:data}

In order to simultaneously test the reliability of SDSS data products
and establish the most comprehensive $VRL$ scaling relations to date, we
seek not only the largest but the most homogeneous compilation of galaxy
structural parameters to date.  This can be done through the
cross-correlation of the large compilation of $\sim$9000 galaxies
within $cz < 28,000$ \kms with \hi\ line widths by S05 and S07 with
the multi-band $ugriz$ photometry provided by the SDSS.

The SDSS archive query with S05/S07 targets yielded FITS images for
4260 galaxies.  That sample was examined visually to eliminate galaxy
pairs, interacting galaxies and images with bright foreground stars,
and any galaxy whose light profile could not be extracted (\eg very
faint targets).  Due to the survey nature of the SDSS, target galaxies
may often appear close to the edge of the CCD image frame and
subsequently compromise the photometric analysis. A large fraction
($\sim$ 40\%) of the more luminous galaxies in our sample which suffer
from these ``edge effects'' had to be discarded.

We were left with a final sample of 3041 galaxies for which both
S05/S07 rotational velocities and acceptable SDSS FITS images are
available. We have subdivided this large sample, dubbed ``Sample A'',
into three subsets (Samples B, C, D) to investigate the effects
of distance uncertainties and inclination on galaxy scaling parameters.

The main sample and subsets include the following systems:

\begin{enumerate}[(A)]
\item \textit{Full Sample} - 3041 galaxies from S05/S07 found in
  SDSS DR7 with imaging suitable for surface photometry;
\item \textit{Best Inclinations} - 1725 galaxies from Sample A with
  moderate inclinations $40^{\circ}<i<75^{\circ}$.  This removes
  the nearly face-on galaxies that would suffer from large
  uncertainties in their deprojected rotational velocity, as
  well as the more inclined galaxies ($i>75^{\circ}$) whose 
  disk may be significantly obscured by line-of-sight dust
  and whose surface area for isophotal photometry is barely visible; 
\item \textit{Best Distances} - 1076 galaxies in Sample A with the
  best distance determinations from high quality spectroscopic data of
  S07 (referred to as ``SFI++'' by S07);
\item \textit{Best Inclinations and Distances} - 652 galaxies from the
  intersection of Samples B and C. Sample D is thus our ``Best
  Sample''.
\end{enumerate}


\section{Light Profile Extraction}\label{sec:LightProfiles}

\subsection{SDSS FITS Image Photometry}\label{sec:SBProfs}

We extract light profiles from the SDSS $gri$ images of all the
Sample A galaxies using the surface brightness profile extraction
methods presented in Courteau (1996) and adapted to SDSS images
by McDonald \etal (2009; 2011).  Given the lower signal-to-noise of the
SDSS $u$ and $z$ bands (Blanton \etal 2001), our analysis will
rely strictly on the $gri$ bands.

The image processing software \texttt{XVISTA}\footnote{see
\texttt{http://ganymede.nmsu.edu/holtz/xvista/}} is the
backbone of our surface brightness profile calculations. 
The \texttt{XVISTA} command \texttt{PROFILE} was used
to fit azimuthally-medianed elliptical isophotes through
the galaxy 2D light distribution. The position angle and
ellipticity of each isophote could vary while the centre was
kept fixed. The ellipticity which best defines the stable outer
disk was visually determined and the isophotal solution was extended
to faint light levels using that ellipticity 
to extract the deepest profile possible. The surface brightness
was calculated from the median sky-subtracted flux level over
an elliptical contour.  Intervening stars were masked and
isophotal twists due to bright spiral arms were also smoothed out by hand.  

The higher signal-to-noise of the $i$-band SDSS images, and the
lesser sensitivity of galactic dust at redder wavelengths, make the
$i$-band the filter of choice for galaxian surface photometry. The
profile templates (position angle and ellipticity) for each galaxy
is based on the $i$-band image and then applied to both $g$ and
$r$ images.  This ensures that galaxy colour profiles are
extracted over the same physical regions of a galaxy. The
surface brightness levels at each pixel were computed in each band
and scaled using the photometric zero-point, airmass and extinction
coefficients supplied in the SDSS-\texttt{drObj} photometric calibration
files.  The surface brightness profiles for the 3041 galaxies in Sample A
were all inspected individually for quality control. 

\subsection{SDSS Sky Measurements}\label{sec:SDSSsky}

The galaxy light profiles were all sky-subtracted; this operation
being clearly the largest source of error in surface brightness extraction.
The SDSS image headers already include an estimate of the sky background level. 
However, the sky backgrounds for bright extended galaxies in the SDSS
DR1 - DR7 are known to be systematically over-estimated resulting in
the under-estimation of galaxy fluxes and sizes by as much as 20-30\%
(Masjedi \etal 2006; Lauer \etal 2007; Abazajian \etal 2009; McDonald \etal 2011).
Conversely, for fainter systems such as LSB galaxies, West \etal (2010)
report an over-subtraction of the sky level in the SDSS photometric pipeline.
We now present our investigation of sky subtraction uncertainties
based on DR7 SDSS $i$-band galaxy images\footnote{Note that background
subtraction improvements in DR8 have resulted in more reliable
photometry of large galaxies (Blanton \etal 2011).  The treatment
about sky uncertainties presented in our paper is still relevant
for smaller galaxies, whether in DR7 or DR8.}.

\begin{figure*}[htb]
\centering
\includegraphics[width=\textwidth]{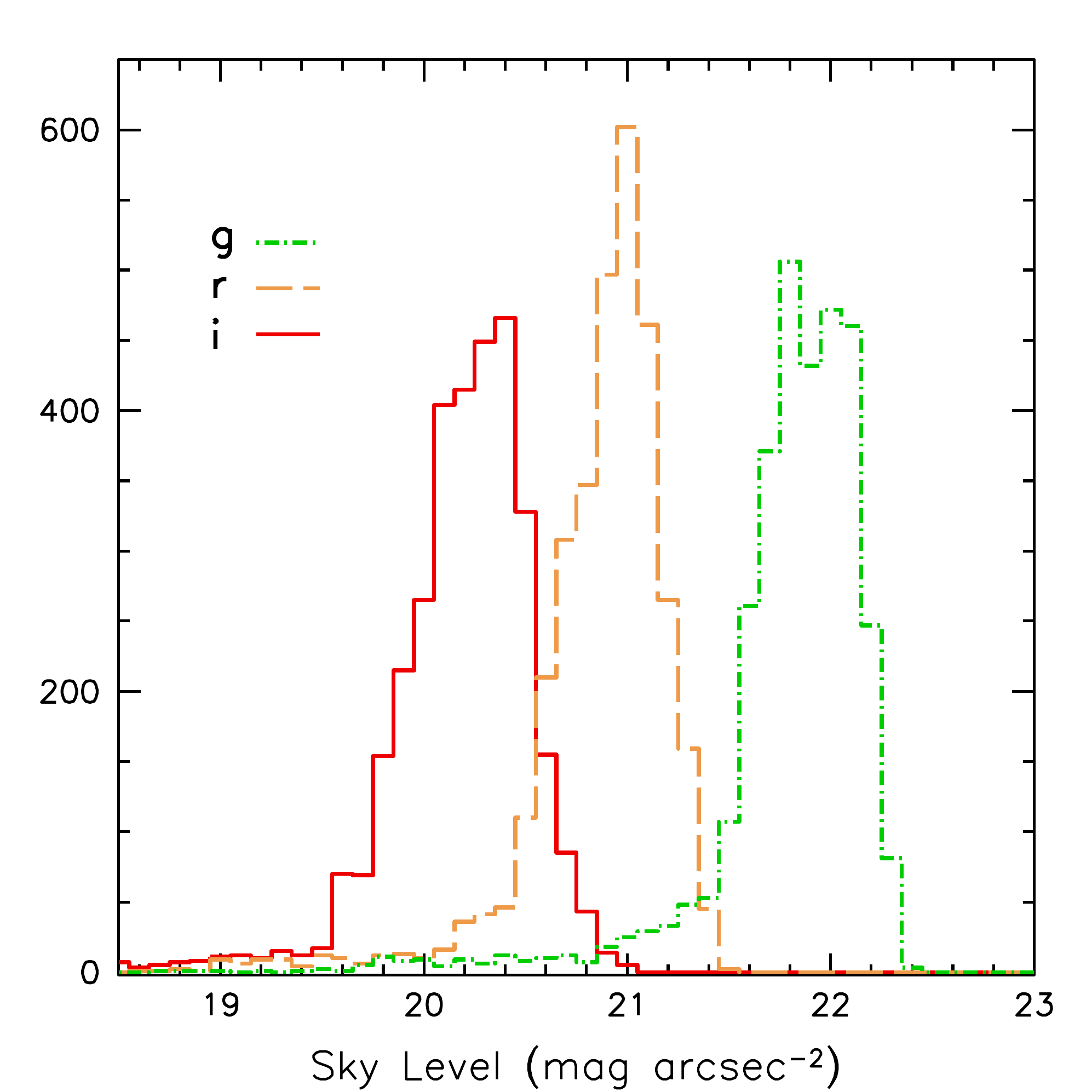}
\caption{Distribution of the $g$- (\textit{green dot-dashed
    line}), $r$- (\textit{orange long-dashed line}) and $i$-band
    (\textit{red solid line}) SDSS sky levels for the 3041 galaxies
    in Sample A.}
\label{fig:skyHist}
\end{figure*}

The SDSS sky background level, $\rm{sky_{SDSS}}$, is measured from
an initial estimate of the median pixel value of every fourth pixel
in the image clipped at $\sim$2.3 $\sigma$, where $\sigma$ is the
rms noise of the image. The brightest stars, but not necessarily
their wings, are thus removed from the image and the sky level is
recomputed to provide a final background estimate.  The different
sky levels as provided by SDSS for Sample A galaxies are shown
in \Fig{skyHist} for the three SDSS $gri$ bands.

To investigate the deviations in sky measurements, we compute two
other sky measurements from the SDSS $i$-band FITS images: i) 
${\rm sky_{Full}}$, is the median pixel value over the full 
image frame and, ii)  $\rm sky_{True}$ is the
median sky background level from five boxes placed interactively
away from the galaxy image.  The latter was computed for a sample
of 30 galaxies with the largest and smallest angular sizes. 
In principle, $\rm sky_{True}$ provides the most realistic 
estimate of the sky background. 

We measure the difference $\Delta$sky in $i$-band sky levels
with respect to $\rm sky_{True}$ as, 

\be
  \Delta \rm sky_i = \frac{\rm sky_i - \rm sky_{True}}{\rm sky_{True}}.
\label{eq:percentSky}
\ee

\begin{figure*}[htb]
\centering
\includegraphics[width=\textwidth, trim=0cm 1cm 5cm 4cm ]{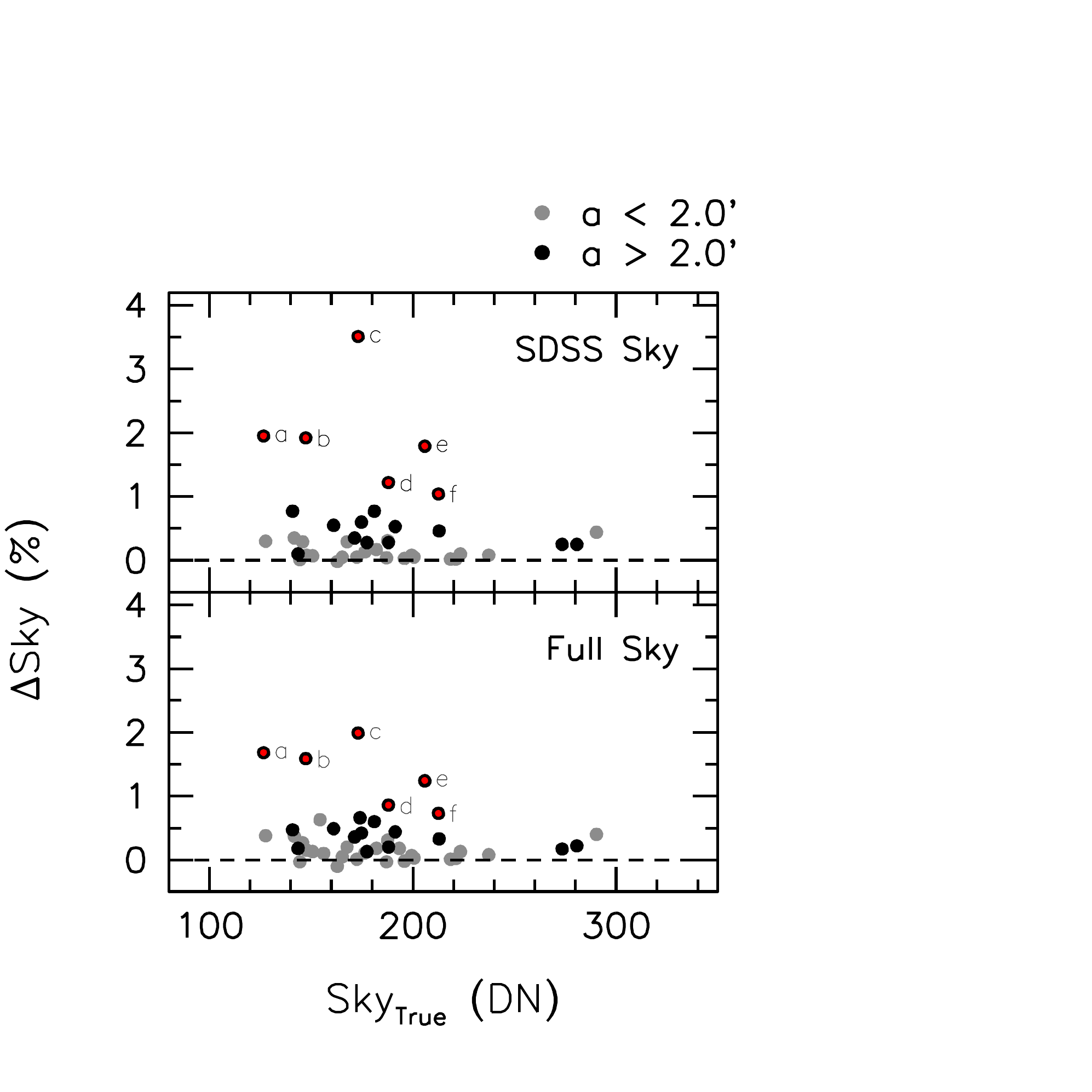}
\caption{Comparison of the SDSS and Full sky levels
  against the manually-computed sky measurements for thirty
  of the largest (black points) and smallest (blue points)
  galaxies in the sample. The labelled points correspond
  to (a) UGC 7524, (b) UGC 2173, (c) UGC 8334, (d) UGC 5882, 
  (e) UGC 7989 and (f) UGC 6346, which show
  the largest deviations from the manually-extracted sky
  levels in the selected sample.  The six galaxies all
  exceed 7.5\arcmin\ in diameter.
}
\label{fig:skyComp_man}
\end{figure*}

\Fig{skyComp_man} shows the results for $\Delta {\rm sky_{SDSS}}$ and
$\Delta {\rm sky_{Full}}$ for our subsample of 30 selected galaxies.
First, we note that the largest deviations in both panels
are for galaxies whose angular diameter exceeds 7\farcm5.
Fortunately, most of our galaxies have much smaller sizes.

The top panel of \Fig{skyComp_man} confirms that the SDSS
sky levels are overestimated with respect to our ``best''
estimates, $\rm sky_{True}$.  The percent difference for
most small galaxies is however less than $\sim$0.2\%.
The offset between the ``Full Sky'' estimate and our best
sky measurement (bottom panel) is surprisingly comparable
to that with the SDSS sky level (top panel).  In most cases,
the deviations are less than $0.4$\%.

\begin{figure*}[htb]
\centering
\includegraphics[width=\textwidth]{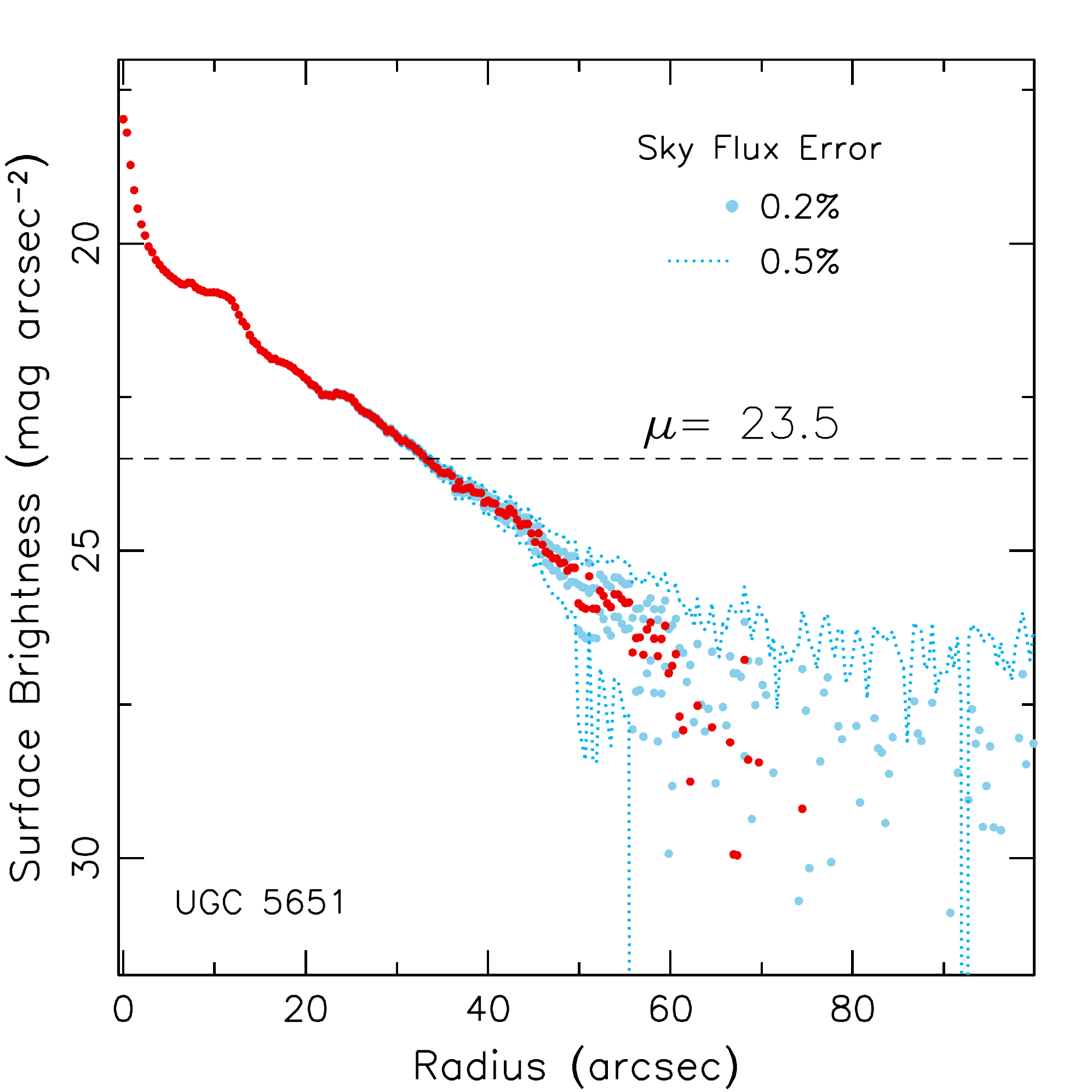}
\caption{$i$-band surface brightness profiles for the galaxy UGC
  5651 showing the 0.2\%, 0.5\% and 1.0\% sky flux error envelopes as
  solid and dotted lines respectively. The $\mu=23.5$ \magarc\ surface
  brightness level is shown as a horizontal dashed line.}
\label{fig:skySub}
\end{figure*}

The effect of over- and under-subtracted skies on
a typical light profile is shown in \Fig{skySub} for
sky background errors at the 0.2\% (light blue points),
0.5\% (dashed line) and 1.0\% (dark solid line) levels
on the $i$-band surface brightness profile of UGC 5651
(an intermediate-size Sc galaxy centered on the image frame). 
The dashed black line marks the $\mu=23.5$ \magarc\ level
and it can be seen that most isophotal measurements above
the $\mu=23.5$ \magarc\ level are largely free of normal
sky subtraction errors.  We will compute in \se{radii}
the effect of sky subtraction errors on measured sizes
and fluxes. 

Based on these tests, we have adopted the SDSS sky levels, 
$\rm sky_{SDSS}$, for the sky background subtraction of
our surface brightness profiles.  Caution must still be
taken when utilizing SDSS sky background estimates,
especially for large galaxies.  

\begin{figure*}[htb]
\centering
\includegraphics[width=\textwidth, trim=0cm 0cm 0cm 10cm]{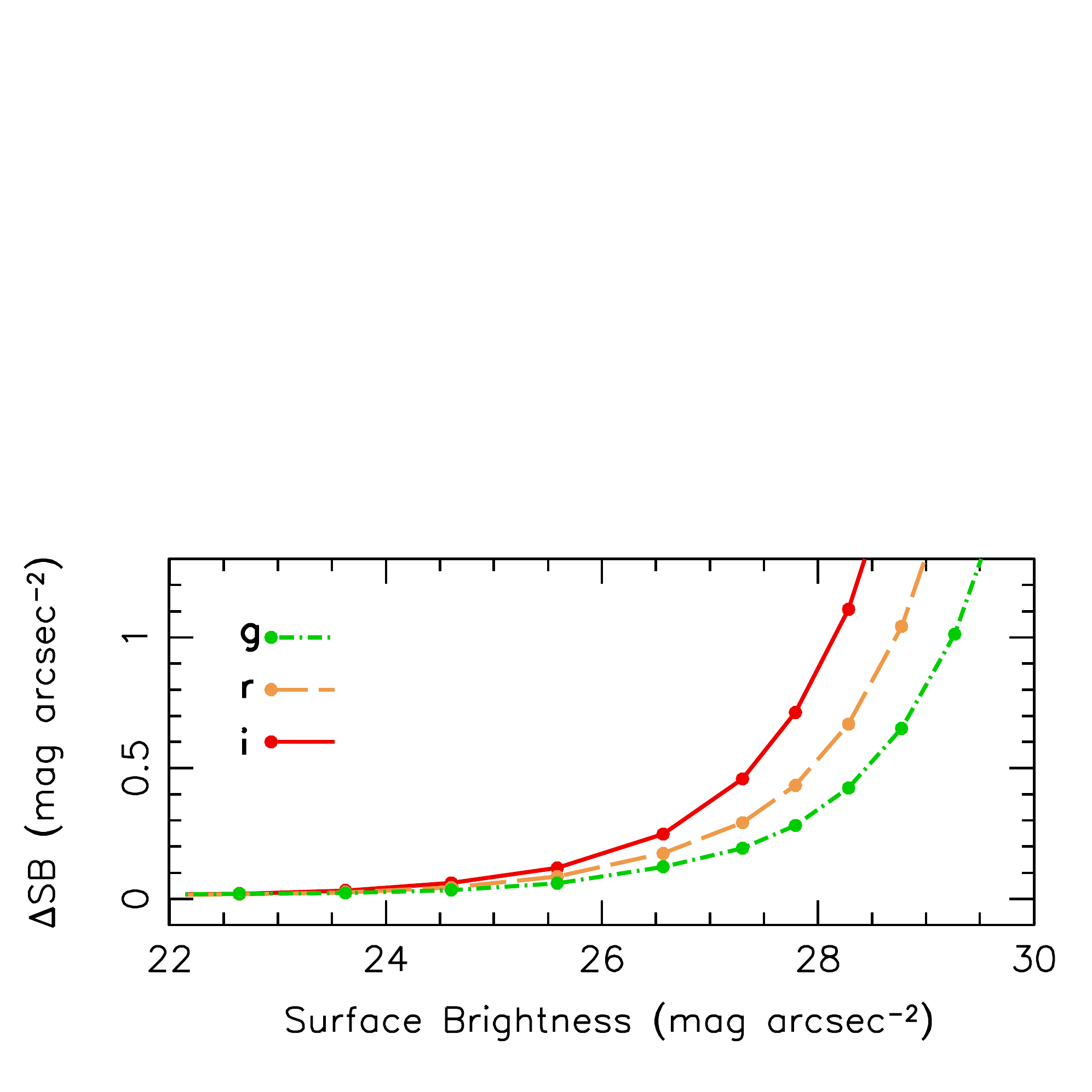}
\caption{Median surface brightness error as a function of surface
  brightness for all Sample A galaxies. The $g$-, $r$- and $i$-band
  data are shown as green dot-dashed, orange long-dashed
  and solid red lines respectively.}
\label{fig:profMedians}
\end{figure*}

\Fig{profMedians} shows the median surface brightness errors
versus median surface brightness in the $gri$ bands for all
the galaxies in Sample A.  Our sky-subtracted galaxy light
profiles are quite stable and reliable down to surface brightnesses
of $\mu \simeq 27$~\magarc, where surface brightness errors
typically do not exceed 0.4~\magarc.

\subsection{Structural Parameter Extraction}\label{sec:ParamExtract}

We now present the measurements of structural parameters in the
$gri$ bands for the 3041 galaxies in Sample A.  We seek isophotal, 
effective and physical size and luminosity parameters for the construction
of $VRL$ relations.

Isophotal parameters are measured at a specific surface brightness
level, effective parameters are measured at a radius which encloses
half of the total light, and physical sizes can refer to a disk scale
length or an exact radius in kpc. 

The total luminosity of a galaxy can be estimated by extrapolating
its light profile to infinity assuming an extended exponential disk.
Given that the SDSS photometry is already fairly deep (\Fig{profMedians}),
the profile extrapolation adds only a few percent of the light to the total.
The exponential disk model is given by 
\be
I(R)=I_0 \exp(-R/R_{\rm d}),
\label{eq:expDcounts}
\ee
where $I_0$ is the extrapolated central surface brightness
in counts and $R_{\rm d}$ is the scale length of the exponential
disk. In magnitude units, \Eq{expDcounts} becomes
\be
\mu(R)=\mu_0+1.086(R/R_{\rm d}),
\ee 
where the units of $\mu$ are \magarc. The disk light, $L_{\rm d}$,
in counts, is the cumulative surface brightness under the
exponential fit from $r=0$ to infinity,
\be
L_{\rm d}=\int^{\infty}_{0}I(r) dr.
\ee

Disk extrapolation fits are highly susceptible to the fit baseline
and the presence of spiral arms.  Indeed, disk scale lengths ($R_{\rm d}$)
and central surface brightnesses ($\mu_0$) may vary by as much as 20\%
from author to author (Knapen \& van der Kruit 1991; Courteau 1996;
Courteau \etal 2011). Here we adopt the interactive ``marking the disk''
technique to fit an exponential function to the disk surface brightness profile
(Giovanelli \etal 1999; C07).  The inner and outer radii for the
fit baseline encompass the region of the SB profile that is dominated
by the disk.  Full 2D bulge-disk model decompositions of the SDSS
images will be reported elsewhere; for our current purposes, the
``marking the disk'' technique is fully adequate.

Besides the measurement of a disk scale length and disk central
surface brightness, we can also compute from these fits the 
isophotal radius, $R_{23.5}$, and apparent magnitude, m$_{23.5}$,
within the 23.5 \magarc\ isophote. This straightforward measurement
does not depend on any disk fitting or parameterisation.  Other
isophotal radii can be measured but $R_{23.5}$ can be shown to yield
smallest scatter of the RL relation (\eg Courteau 1996; \se{radii}). 

We also compute the half-light (or effective) radius $R_e$
containing 50\% of the total extrapolated light. The effective surface
brightness is defined as $\mu_{\rm e}=\mu(R_{\rm e})$.
We also define other parameters: m$_{\rm ext}$ is the total magnitude
of the extrapolated light profile and m$_{2.2}$ the magnitude
integrated within the radius $R=2.2R_{\rm d}$. The latter corresponds
to the peak of the rotation curve of an idealized pure exponential
disk (\eg Freeman 1970; Binney \& Tremaine 1987).

We now have various sets of radii and apparent magnitudes based
upon our extrapolated light profiles for 3041 galaxies in $g$, $r$
and $i$-bands.  Colours can also be computed from various magnitude
combinations.  In \se{ScalingMeasures} we quantify the robustness
of these structural parameters in the context of the $VRL$ scaling relations.

\begin{figure*}[htb]
\centering
\includegraphics[width=\textwidth]{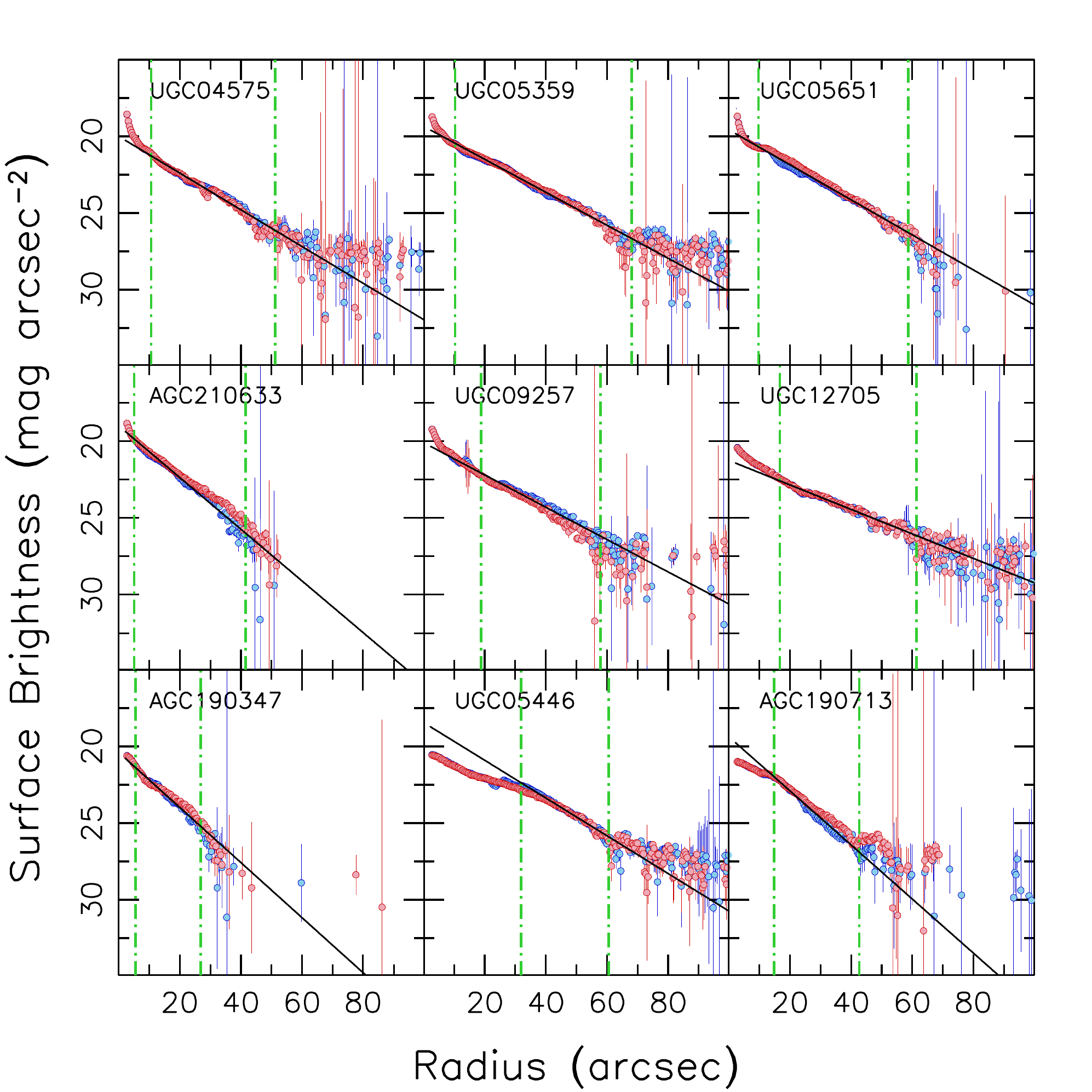}
\caption{Sample of nine surface brightness profiles measured by
  authors MH in blue and YZ in red. The solid line shows the
  exponential disk extrapolation and the disk fit baseline is
  delineated by the green vertical dot-dashed lines.
}
\label{fig:S05inS07}
\end{figure*}

Our analysis benefits from the independent reduction of
211 galaxy light profiles by two of us (MH and YZ).  The light
profiles for nine of these galaxies are presented in \Fig{S05inS07}
as blue (reduced by MH) and red (reduced by YZ) profiles.  Comparing
these profiles shows that differences between MH and YZ, if any, are
small and purely random, despite slightly different treatments in
the few interactive aspects of the SB profile extraction.
These two treatments differ only in the smoothing of isophotal
twists and luminosity extrapolation of the outer disk (see \se{SDSSsky}).
MH extracted the light profiles for all 3041 galaxies in Sample A
and we adopt her final catalog.  The extensive comparisons with YZ
reinforce the reliability of our sample. 


\section{Final Data Corrections}\label{sec:Corrections}

We now correct the extracted structural parameters
(\eg sizes, luminosities, surface brightnesses) following
the prescription of C07 but now accounting for the SDSS filter system.
The velocities derived from the \hi\ line profiles are corrected
according to the prescriptions of S05 and S07.  The inclination
of each galaxy is measured from the ellipticity of the
outer disk isophotes of the SDSS $i$-band image and corrected for
projection effects using Holmberg's oblate spheroid description,
\be
\cos i=\sqrt{ {(1-\epsilon)^2 - q_0^2}\over{1-q_0^2}}),
\label{eq:incl}
\ee
where the ellipticity, $\epsilon=1- b/a$, 
$q_0=c/a$ is the axial ratio of a galaxy viewed edge-on,
$a$ and $b$ are the disk major and minor axes, and $c$
is the polar axis.

In order to determine the variation of projected ellipticity with
colour or morphological type, we have used the Galaxy Zoo classification
(Lintott \etal 2008).  We selected all of our galaxies classified
as edge-on by $>80$\% of all the Galaxy Zoo survey participants
(and eliminated all those classified as ellipticals or merger
by $>90$\% of them) yielding a sample of 1377 candidate edge-on
galaxies.  We then visually inspected the SDSS images for all
these galaxies and eliminated a further 506 systems deemed not
perfectly edge-on.  For each of the remaining 871 edge-on galaxies,
an ellipticity was computed using the adaptive second moments,
$e_+$ and $e_x$, provided by the SDSS pipeline.  From these
moments, the adaptive ellipticity was computed following
Kautsch (2009). Finally, this ellipticity was converted
to an axial ratio ($q_0=1-e$).

\begin{figure*}[htb]
\centering
\includegraphics[width=\textwidth]{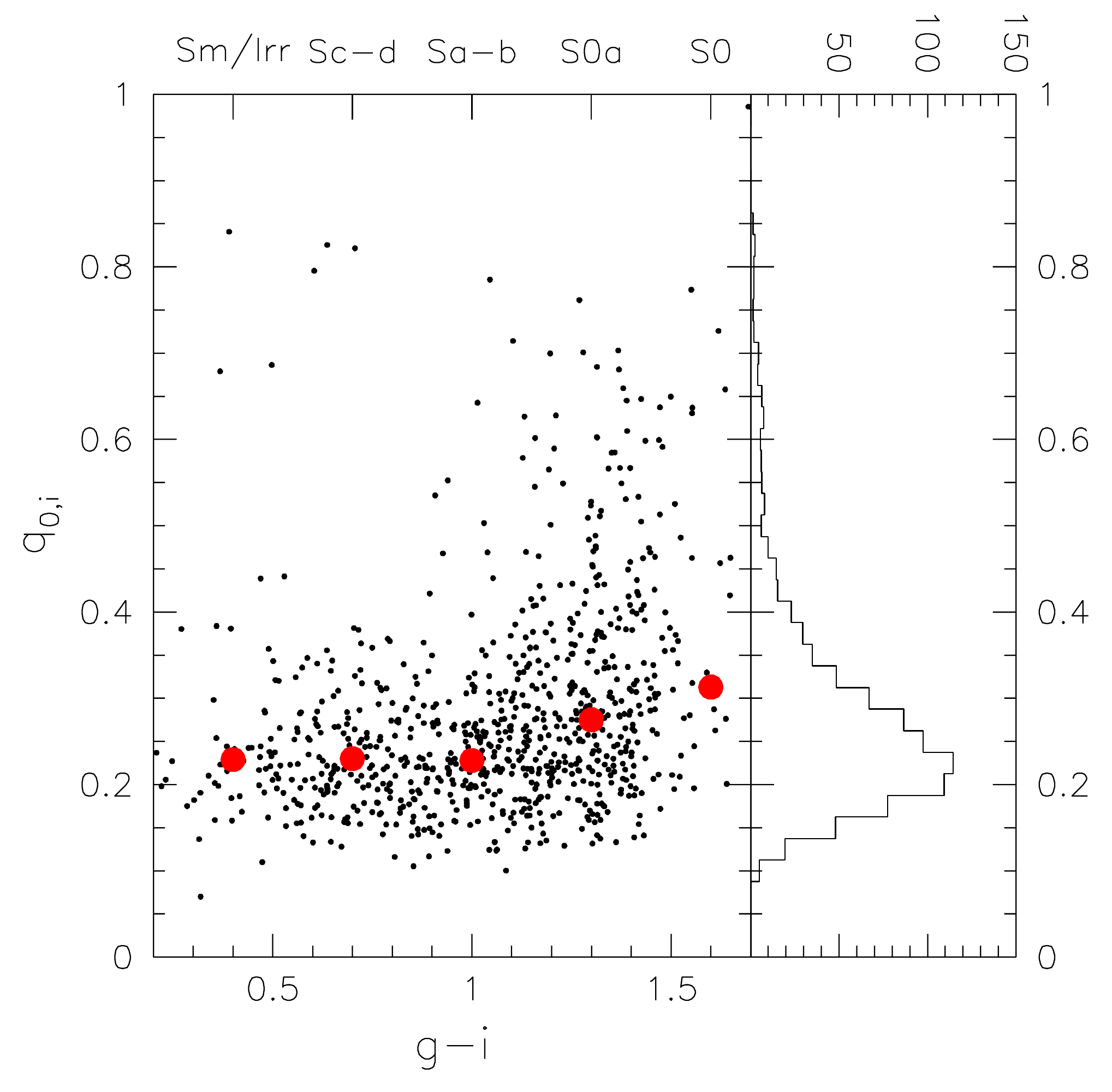}
\caption{Distribution of SDSS-derived axial ratios versus mean
$g-i$ galaxy colour (computed at the effective radius) for
871 edge-on galaxies as classified by the Galaxy Zoo and ourselves.
The red dots are median values of the disk flattening, $q_0=c/a$,
in bins of $g-i$ colours and morphological types (upper axis).
The histogram on the right includes all 871 galaxies and peaks at $q_0 = 0.23$.
Compare with the similar histograms by Lambas \etal (1992; Fig. 6c)
and Giovanelli \etal (1994; Fig. 7).
}
\label{fig:edgeon}
\end{figure*}

\Fig{edgeon} shows the variation of SDSS-derived axial ratios
in the $i$-band versus mean galaxy colour, computed at the effective
radius, for 871 edge-on galaxies each depicted as a black dot.
The morphological correspondance (top axis) was derived from
SDSS colours tabulated in McDonald \etal (2011). 
The red dots are median averages per given morphological bin.
The trend of $q_0$ naturally increases with redder, progressively
bulge-dominated galaxies.  
The $q_0$ distribution is relatively flat for blue (late-type)
galaxies, with an upturn at $g-i \sim 1.3$, corresponding to Sa/S0a galaxies. 
Several S0 galaxies have $q_0 \gta 0.8$, reminiscent of the Sombrero galaxy. 
The few late-type systems with $q_0 > 0.5$ are likely a failure in the
adaptive moment pipeline.  Note that similar $q_0$ distributions are
obtained for the $g$ and $r$ bands. 

Based on \Fig{edgeon}, we adopt $q_0 \simeq 0.13$ as the minimum axial
ratio, and thus intrinsic thickness, of spiral galaxies. 
This value agrees with a similar study by Giovanelli \etal (1994), 
though a value as high as 0.2, as reported by Lambas \etal (1992),
is also realistic considering the intrinsic errors on galaxy radii.
For their study of the baryonic Tully-Fisher relation (see \se{BTF}),
Stark \etal (2009) adopted $q_0=0.15$.  The effect of using $q_0=0.13$
or 0.2 is inconsequential for our study.

The apparent ($gri$-band) magnitudes have been homogeneously corrected
for internal extinction $A_i$, external galactic extinction $A_g$ and
$k$-correction $A_k$ according to
\be m=m_{\rm obs}-A_i-A_{g}-A_k.
\label{eq:mobs}
\ee
The inclination-dependent internal extinction correction in magnitudes
is given by 
\be
A_i=\gamma \log (b/a) 
\ee
where $\gamma $ is interpolated in the $g$, $r$ and $i$ bands from the 
corrections of Tully \etal (1998) given the corrected line widths $\log W$
from S05,
\beqa
\gamma_{g}=1.51+2.46(\log W-2.5), \\
\gamma_{r}=1.25+2.04(\log W-2.5), \\
\gamma_{i}=1.00+1.71(\log W-2.5).
\eeqa 

The Galactic extinction, $A_{g}$, from Schlegel \etal (1998)
is provided by the SDSS.  The cosmological $k$-correction, $A_k$,
is calculated from the template of Blanton \& Roweis (2007)
for each SDSS filter.

The absolute $gri$ magnitudes, $M$, are calculated in their
respective band as 
\be
M=m-5\log D_{\rm L}+25; \; \; \; \;\;\;\; D_{\rm L} =
V_{\rm CMB}H_0^{-1}
\label{eq:distance}
\ee
where the Hubble constant $H_0=71$ \hunit\ (Komatsu \etal 2011) 
and $D_{\rm L}$ is the distance in kpc derived from
the velocities in the rest frame of the cosmic microwave
background, $V_{\rm CMB}$, as compiled by S05 and S07.
For calibration to solar units, we use $M_{\odot,g}$=5.12,
$M_{\odot,r}$=4.68 and $M_{\odot,i}$=4.57 (York \etal 2000); 
we use $D_{\rm L}$ in the conversion from apparent to absolute
quantities. 

We correct both effective ($\mu_{e}$) and central ($\mu_{0}$) surface
brightnesses for internal and external extinction following C07:
\beqa
\mu^c_{e}=\mu_{e}+0.5\log(a/b)-A_{g}-2.5\log(1+z)^3,\\
\mu^c_{0}=\mu_{0}+0.5\log(a/b)-A_{g}-2.5\log(1+z)^3.
\eeqa
These are reported in a data table (\Table{datatable} as presented
\se{FinalDataTables}) but not used for further analysis in this
paper, except in \Fig{VRLmuE}. 
Radii measurements may be corrected following
Giovanelli \etal (1995), as also reported in C07.
However, we shall find in \se{R-L} that such corrections
do not yield a reduction of the $RL$ and $RV$ relations scatter.
In view of the uncertain nature of most radial scale corrections,
our analysis relies on uncorrected radii. 

\subsection{Parameter Uncertainties}\label{sec:errors}

We follow M06, C07, AR08, SS11 and many other studies of
galaxy scaling relations in estimating the uncertainties
for each scaling relation parameters $\log V$, $\log R$ and $\log L$.
Uncertainties in $\log L$ assume 10-15\% errors for each of the parameters
$b/a$, $\gamma$, $A_{\rm gal}$, $A_{\rm k}$ and $D$ in \Eq{mobs}.
That is,
\be
 \sigma_{\rm logL}^2 \simeq \sigma_{\rm mag}^2 + \sigma_{\gamma}^2 + 
 + \sigma_{\rm \log(b/a)}^2 + \sigma_{\rm A_{gal}}^2 + \sigma_{A_k}^2
 + \sigma_{D}^2,
 \label{eq:elogL}
 \ee
where $\sigma_{\rm mag}$ is the statistical error in the raw
magnitude measurement.  A more complete assessment of $VRL$
parameter errors is found in SS11.  Our results are however
practically the same.

Typical measurement errors on $L$ amount to 20\%, or dlog${L} = 0.09$.
This value is also obtained by M06 and SS11 for their measurements
of the absolute luminosity, $M_I$, in the $I$-band.
The uncertainties on the rotational velocities as compiled by S05/S07
amount to $\sim$5\%, or dlog$V=0.02$ (SS11 use dlog$V=0.03$).  We
also assume a 15\% uncertainty, or dlog$R_d=0.07$ on disk scale lengths
(e.g., MacArthur \etal 2003; Courteau \etal 2011) and a 7\% error,
or dlog$R$=0.03, on isophotal radii (same as SS11).

In the $VRL$ fits that follow (\se{VRL}), the quantities dlog${L}$
and dlog$V$ are computed independently for each galaxy while
dlog$R$ for various radii measurements ($R_{23.5}$, $R_d$, $R_e$)
is set to the mean values above for each galaxy.  Using this method
or assigning a constant error to each parameter plays no role in
the final scaling relations. 

\section{Data Table}\label{sec:FinalDataTables}

\begin{figure*}
\centering
\includegraphics[width=\textwidth, trim=0cm 0cm 3cm 0cm]{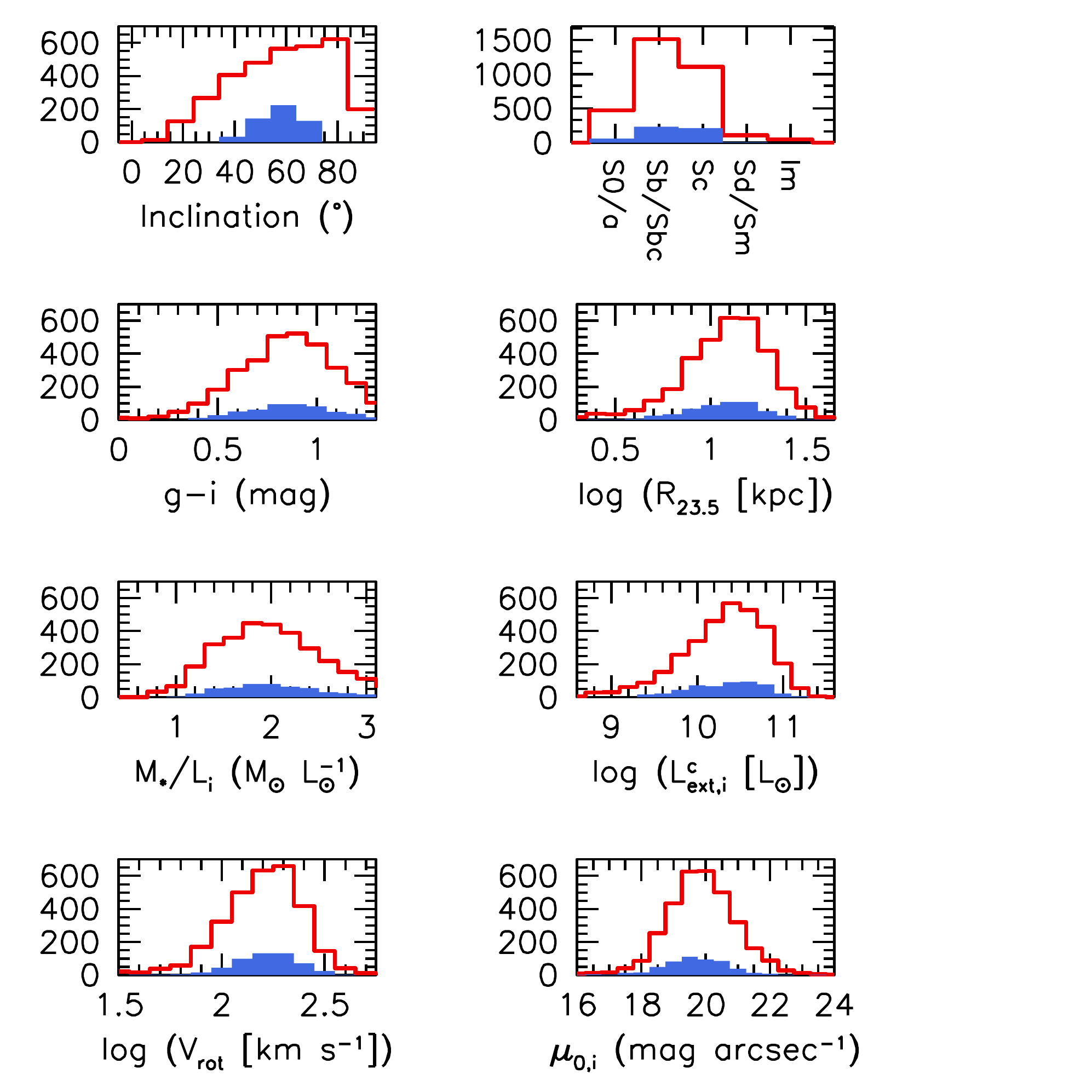}
\caption{Distribution of physical parameters for $3041$ galaxies
  in Sample A (red) and $652$ galaxies in Sample D (blue).}
\label{fig:bigHist}
\end{figure*}

We now present our extensive compilation of galaxy structural
parameters. The recessional and rotational velocities, \hi\ flux,
and morphological classification are from S05/S07.  All other
structural parameters have been derived by us from SDSS $gri$
images as described in \se{ParamExtract}.
The distribution of several of these parameters is shown in 
\Fig{bigHist} for the full Sample A in red and for our
Sample D with best constrained distances and inclinations
in blue. \Fig{bigHist} shows a broad parameter coverage, 
though none of which can be deemed complete. The parameter distributions
are roughly (log-)normal with an emphasis on Sb/Sc morphological types.

\Table{datatable} gives a list of structural parameters for all the Sample A
galaxies.  The first few entries of the catalog are shown here;
please see the electronic version for the full list. The entries
are arranged as follows:

\begin{description}
\item[{\sl Col. (1):}] UGC or AGC galaxy name;

\item[{\sl Col. (2):}] $i$, the corrected galaxy inclination in degrees using $q_0=0.13$
   in \eq{incl};

\item[{\sl Col. (3):}] $V_{\rm CMB}$, the recession velocity of the galaxy relative to the
  cosmic microwave background frame in \kms\ as calculated by S05 and S07.  The S07
  estimates of $V_{\rm CMB}$ supersede S05 in the case of multiple measurements;

\item[{\sl Col. (4):}] $V_{\rm rot}$, the corrected rotational velocity in \kms\
  from the \hi\ line width profiles measured at 50\% of total flux by S05 \& S07
  ($W_{\rm F50}$). S05/S07 line widths are corrected for instrumental effects
  and redshift broadening.  $V_{\rm rot}$ is corrected for the inclination
  of the disk and distance but not for turbulent motion in the \hi\ gas.
  Sample D uses only rotational velocities from S07;

\item[{\sl Col. (5):}] T, the RC3 morphological type as listed by S05 \& S07;

\item[{\sl Col. (6):}] $\Re$, the half-light radius based on the total
   extrapolated $i$-band luminosity of the galaxy in arc seconds;

\item[{\sl Col. (7):}] $R_{23.5}$, the radius at the $i$-band 23.5
  \magarc\ isophote in arc seconds;

\item[{\sl Col. (8):}] $\Rd$, the disk scale length of the galaxy in
  arc seconds;

\item[{\sl Col. (9):}] m$^c_{\rm ext}$, the total apparent magnitude of the galaxy
  including the extra light from the disk extrapolation to infinity;

\item[{\sl Col. (10):}] m$^c_{\rm 23.5}$, the apparent magnitude within the $i$-band
  23.5 \magarc\ isophote;

\item[{\sl Col. (11):}] m$^c_{\rm 2.2}$, the apparent magnitude within 2.2$R_{\rm d}$ of
  the surface brightness profile. Non-entries in this column are due to a failed disk
  extrapolation, described in \se{magnitudes};

\item[{\sl Col. (12):}] $\mu_0^c$, the corrected central surface brightness in \magarc;

\item[{\sl Col. (13):}] $\mu_{\rm e}^c$, the corrected effective surface brightness
  measured in \magarc;

\item[{\sl Col. (14):}] $g-r$, the colour of the galaxy calculated from the difference
  in $g$ and $r$ extrapolated magnitudes $m_{\rm ext}$;

\item[{\sl Col. (15):}] $g-i$, the colour of the galaxy from the $g$ and $i$ bands as
  above;

\item[{\sl Col. (16):}] $C_{28,i}$, the galaxy light concentration, $C_{28}=5 \log
  (R_{80}/R_{20})$, where $R_{20}$ and $R_{80}$ enclose 20\% and 80\%
  of the extrapolated $i$-band light, respectively;

\item[{\sl Col. (17):}] $\log M_{\hi}$, the neutral hydrogen gas mass in \Msun\,
  derived by S05 from the corrected \hi\ line flux measurements;

\item[{\sl Col. (18):}] $\log M_*$, the stellar mass of the galaxy in units
  of \Msun from the extrapolated $i$-band luminosity $L^c_{\rm ext,i}$ times
  the stellar mass-to-light ratio $M_*/L_i$ prescription of
  Bell \etal (2003) with a Chabrier (2003) IMF; 

\item[{\sl Col. (19):}] $\log M_{\rm bar}$, the baryonic mass of the galaxy
  in units of \Msun. \se{BTF} discusses the sum of the stellar, \hi\ and
  gas masses to make up the total baryonic mass, 
  $M_{\rm bar}=M_*+1.4M_{\hi}$; 
\item[{\sl Col. (20):}] $\log M_{\rm T}$, the total mass of the galaxy
  computed at $R=R_{23.5}$ according to
  $M_{\rm T}(R)=  2.33 \times 10^5 R V_{\rm rot}^2 / \sin^2(i)$ \Msun
  where $R$ is the radius in kpc and $V_{\rm rot}$ is the observed rotation
  velocity in \kms. 
\end{description}


\section{SDSS Comparisons}\label{sec:ScalingMeasures}

Below, we compare our detailed structural parameters extracted
from SDSS galaxy images with SDSS pipeline data products for
the same galaxies. 

\subsection{ The SDSS Data Products}\label{sec:sdssProd}

The SDSS measurements of interest to us are the $g$, $r$, and
$i$ band {\it Petrosian} parameters. The Petrosian radius, $\Rp$,
is the radius at which ratio of the local surface brightness averaged
over the annulus $\Rp$ is equal to 0.2 times the mean surface brightness
within $\Rp$ as measured by the automated SDSS pipeline (Blanton \etal
2001; Strauss \etal 2002; Yasuda \etal 2001). The Petrosian magnitude,
m$_{\rm p}$, is measured within the $circular$ aperture of {\it radius} 2$\Rp$. 
The Petrosian radii $\Rpe$ and $R_{\rm p90}$ encompass 50\% and 90\%
of the total light measured within $\Rp$\footnote{The online 
SDSS archive tags are \texttt{petroRad, petroR50, petroRpn}
and \texttt{petroMag}.}.

We first find that the SDSS data products are burdened by the
mis-identification of galaxy objects in the SDSS SQL query
method, leading to erroneous measurements of radius and light
for numerous objects.  Some of the faulty measurements are due to
the ``shredding'' caused by the deblending algorithm in the SDSS
reduction pipeline. Overlapping objects are deblended to separate
underlying components and extract proper measurements. Occasionally
large galaxies may be interpreted as multiple systems and are thus
``shredded'' by the algorithm.  The 1\% ``shredding'' occurrence
in the SDSS DR1 reported by Blanton \etal (2001) is also found in DR7.
We confirmed this ``shredding'' effect in the SDSS DR7; we found
object-type mis-identifications for objects classified as galaxies
that are either star-forming regions or foreground stars, both
of which are more concentrated than a typical galaxy.

\begin{figure*}[htb]
\centering
\includegraphics[width=\textwidth]{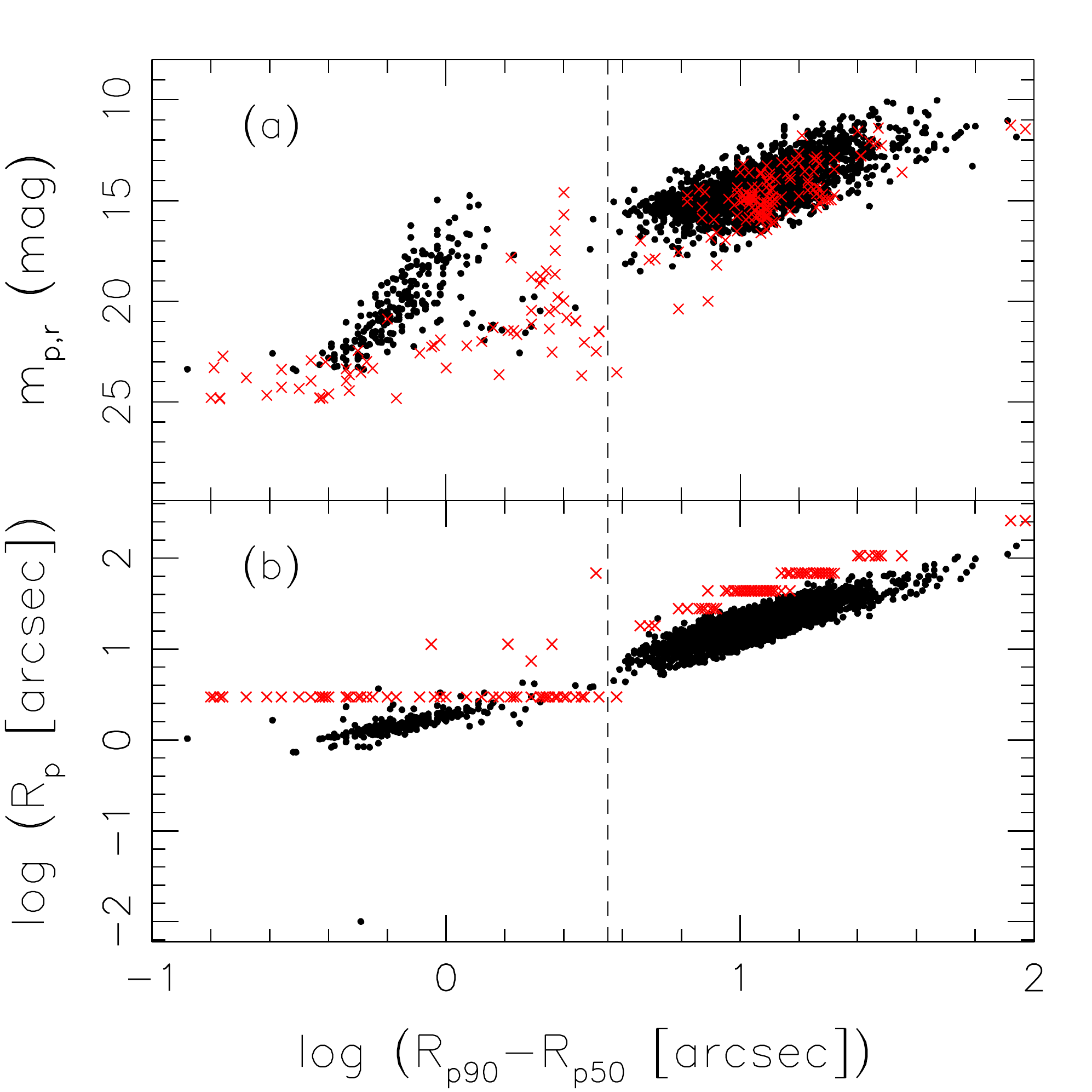}
\caption{SDSS Petrosian radius $\Rp$ and apparent
  magnitude m$_{\rm p,r}$ against the ``concentration''
  $\log (\Rpn-\Rpe)$ in the $r$-band .
  Red crosses represent galaxies with failed measurements
  of $\Rp$ in the SDSS pipeline.  
  The mis-identified targets, which are mostly compact
  objects, lie to the left of the vertical dashed line
  at $\log R_{\rm p90}-R_{\rm p50} = 0.55$.
  These and red cross objects were discarded from
  our ``clean'' SDSS sample.}
\label{fig:purgeSDSS}
\end{figure*}

To identify the deviant SDSS pipeline data, we plot in \Fig{purgeSDSS}
the logarithmic difference between the 90\% and 50\% light radii,
$R_{\rm p90}-R_{\rm p50}$, in the $r$-band against both the magnitude
m$_{\rm p}$ in the upper panel (a) and the logarithmic Petrosian
radius, $\Rp$, in the lower panel (b). The clear separation at
$\log R_{\rm p90}-R_{\rm p50} = 0.55$, shown as a dashed vertical
line, is the clear distinction between correctly identified galaxies
and the mis-identified compact objects.  Objects located left of
this line were discarded. 

We also examined the $r$-band Petrosian radii and their associated
errors.  Faulty computations of the Petrosian radius or magnitude
are given a default -1000 error; those are shown as red crosses
in \Fig{purgeSDSS}.  Many failed calculation of Petrosian radius
are given a SDSS default value of 3 arc seconds; hence the
horizontal array of crosses near $\log R_{\rm p}=0.48$ in panel (b).
These faulty measurements apply mostly to faint galaxies whose
flux is below a nominal threshold within the Petrosian aperture.

Thus, we weed out SDSS data based on the following criteria: i)
mis-identified non-galaxian objects with $\log R_{\rm p90}-R_{\rm
p50} < 0.55$, and ii) failed computation of the Petrosian radius
with errors of $R_{\rm p}<0$ arc seconds in the $r$-band.  This
eliminates 14\% of the SDSS pipeline data products; we are left
with a ``clean'' SDSS sample of 2605 galaxies.

The computation of SDSS data products within a {\it circular} aperture
compared to our structural parameters extracted from isophotal 
ellipse fitting is an additional source of confusion in the
comparison of our data.
Circular and isophotal apertures clearly measure different
portions of the galaxy.  Projection effects for a galaxy bulge
and disk within a circular or elliptical aperture yield very
different luminosity profiles (\eg Bailin \& Harris 2009). 

\begin{figure*}[htb]
\centering
\includegraphics[width=\textwidth]{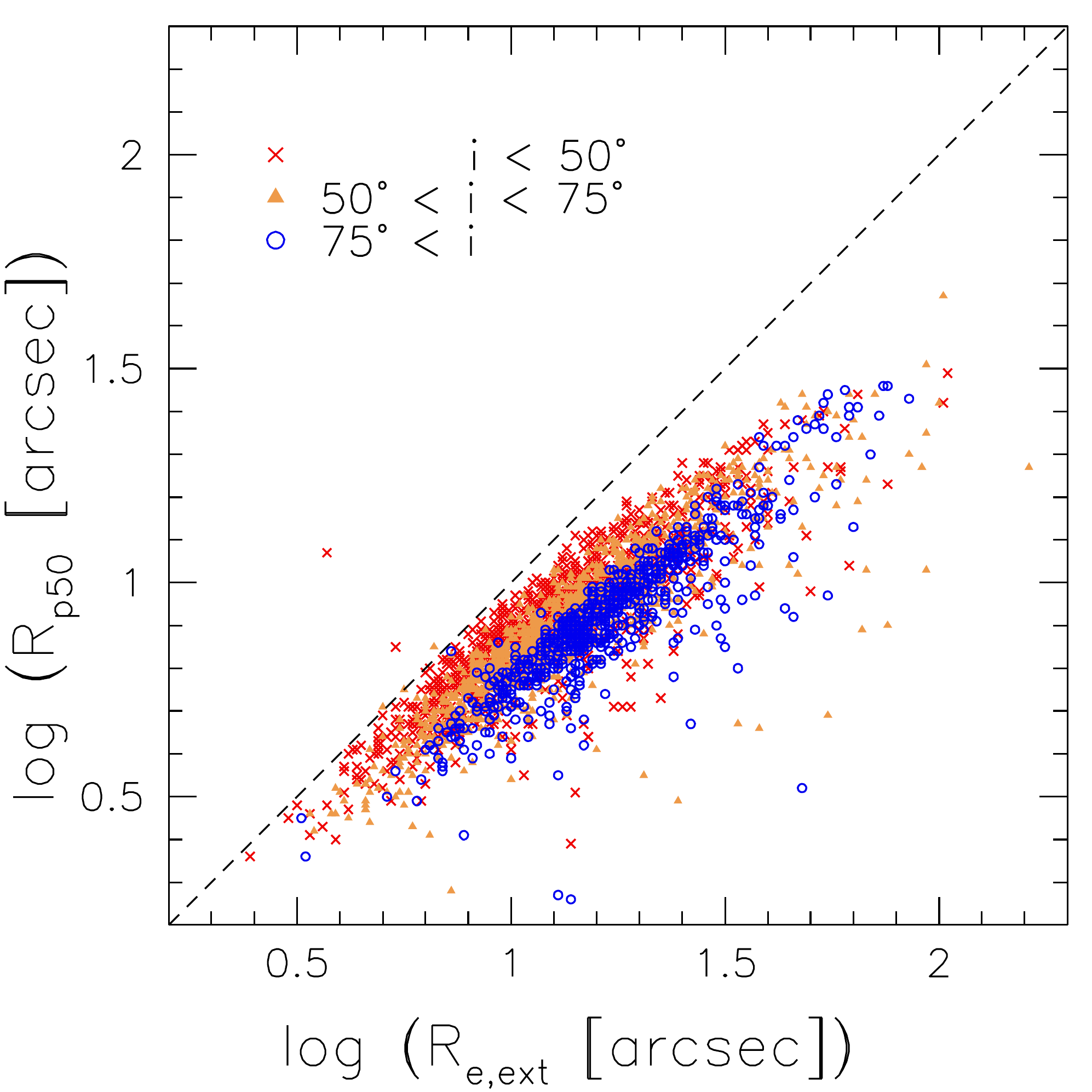}
\caption{Comparison of the $i$-band effective radii $\Re$
  and the SDSS effective Petrosian radius $R_{\rm p50}$ in three
  inclination bins: $i<50^{\circ}$ (\textit{red crosses}), 
  $50^{\circ}<i<75^{\circ}$ (\textit{orange triangles}) and
  $i>75^{\circ}$ (\textit{blue circles}).  Inclination is the 
  main source of scatter between the Petrosian (circular 
  aperture) and isophotal (elliptical aperture) radii.}
\label{fig:ReVSRp50}
\end{figure*}

Such inclination effects between the SDSS Petrosian half-light radius, $\Rpe$,
and the half-light radius, $\Re$, measured from isophotal fitting are explored
in \Fig{ReVSRp50}.  The Petrosian radius is systematically smaller than $\Re$.
More importantly, this offset increases with inclination, from red crosses
($i<50^{\circ})$ to blue circles ($i>75^{\circ}$) in \Fig{ReVSRp50}, as
the Petrosian aperture progressively samples less light (compared to the
isophotal aperture) at higher tilt.  Slight curvature may also be detected
in each inclination distribution but overall, there exists a direct mapping
between $\Rpe$ and $\Re$ as a function of inclination.  The scatter between
a Petrosian radius and an isophotal radius is entirely dominated by inclination. 

\subsection{Comparison of Radial Measurements}\label{sec:radii}

\begin{figure*}[htb]
\centering
\includegraphics[width=\textwidth]{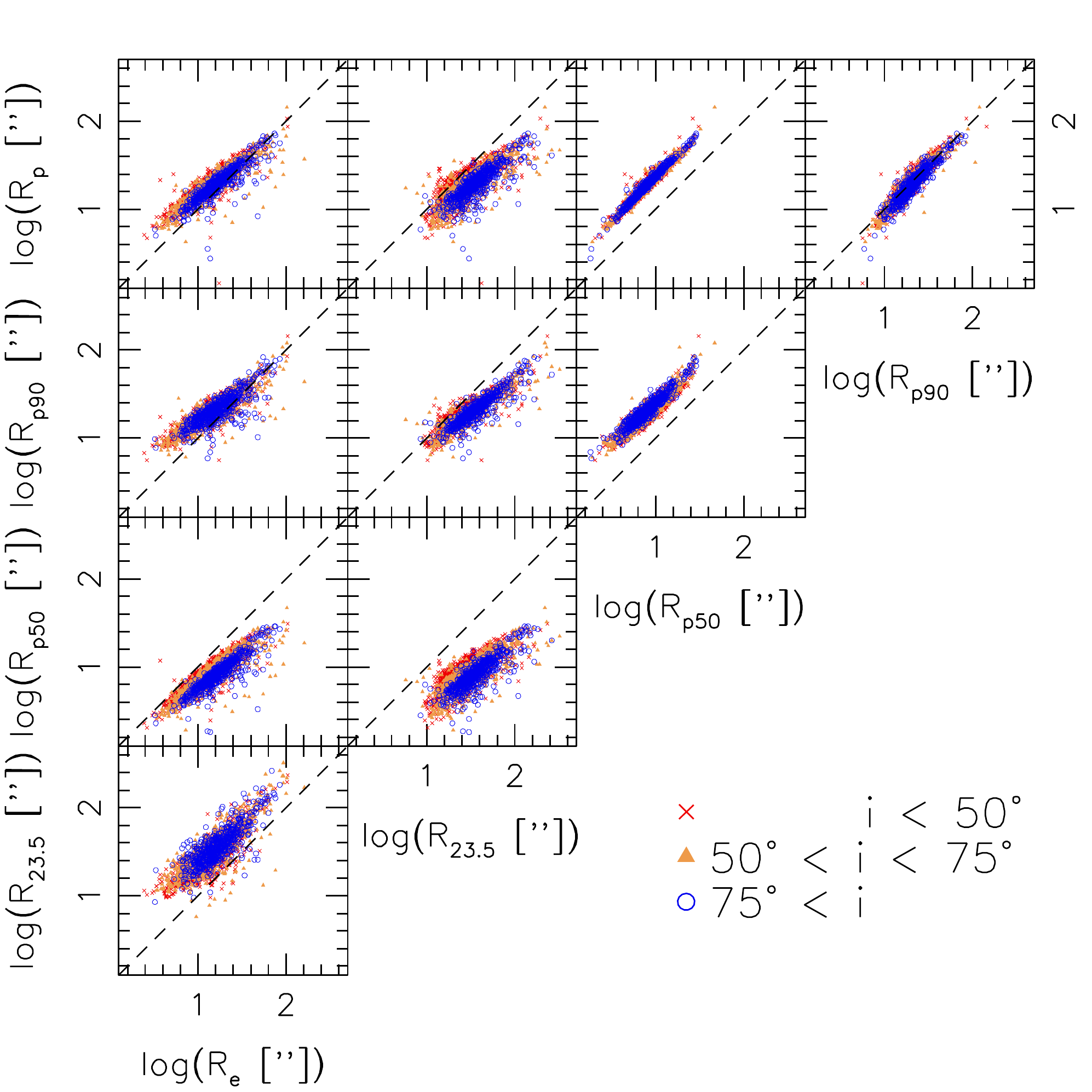}
\caption{Comparison of $i$-band radial parameters
  $R_{\rm p}$, $R_{\rm p50}$ and $R_{\rm p90}$
  from our ``clean'' SDSS data sample with our 
  measurements of $R_{\rm 23.5}$, $\Re$.
  No corrections have been applied to the data.
  The black dashed line has slope unity.
}
\label{fig:RadvsRad}
\end{figure*}

We now compare the ``clean'' $i$-band SDSS Petrosian radii
for 2605 galaxies with our suite of radial parameters, namely 
$\Re$, $R_{23.5}$ and $\Rd$ computed for Sample A (\se{LightProfiles}). 
\Fig{RadvsRad} shows the raw Petrosian radial measures
$\Rp$, $\Rpe$ and $\Rpn$ from the SDSS pipeline against
two of our own radial measurements $R_{23.5}$ and $\Re$,
extracted from our $i$-band light profiles. The red dashed
lines have slope unity. 

We first observe that inclination drives the scatter in
all the correlations between Petrosian (\ie circular) 
and isophotal (\ie elliptical) radii although that
effect is strongest for $\Rpe$. 
Of course, no such dependence is seen for $\Re$ versus
$R_{23.5}$ as both measures are derived from isophotal fits.
There is also no one-to-one correlation between our radial
measures and the SDSS Petrosian radii.  Therefore, any
scaling relation based on SDSS or isophotal sizes ought
to yield different parameters (slope, zero-point, scatter).

The Petrosian radius $\Rp$ correlates well with the two other
Petrosian radii $\Rpe$ and $\Rpn$, as one might expect since 
the latter two derive from $\Rp$.  The bright end of
the diagrams involving Petrosian radii could be slightly
skewed to smaller values due to the reported over-sky subtraction
errors for large objects (see \se{SDSSsky}).  The effect of an
over-subtracted sky on the measured isophotal radii, and fluxes,
was demonstrated in \Fig{skySub} for the sample galaxy UGC 5651.

\Table{skySubErr} presents the propagated error
in radius $\Delta\Re$ and $\Delta{R_{23.5}}$ as well as the
magnitude differences $\Delta${m$_{23.5}$} and
$\Delta${m$_{\rm ext}$} due to sky uncertainty.  A 0.2\% sky
error typically yields 1-2\% radial scale variations or magnitude
differences less than 0.03 mag.  These small values are representative
of our data; 1\% sky errors would be dramatic but they are,
fortunately, unrealistic.  As observed earlier (\se{SDSSsky}),
the isophotal parameters, $R_{23.5}$ and m$_{23.5}$
are least affected by typical sky background errors.

\begin{figure*}[htb]
\centering
\includegraphics[width=0.9\textwidth, trim=0cm 0cm 2cm 0cm]{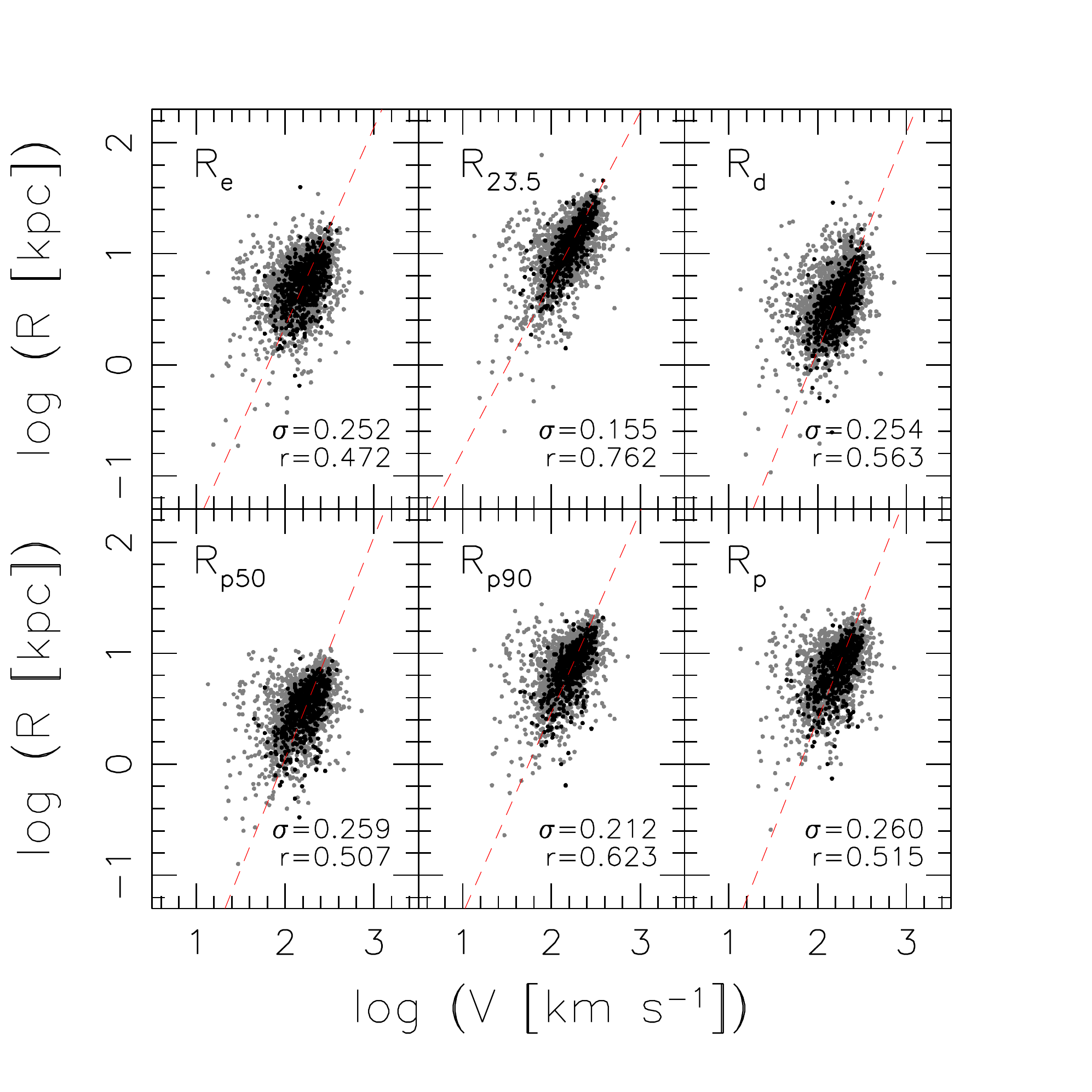}
\caption{The comparison of uncorrected $i$-band radial measurements
  against the deprojected rotational velocity $V_{\rm rot}$. Black
  points show our best Sample D overlaid onto Sample A shown in
  gray. The red dashed line represents the orthogonal linear fit
  to Sample D. The standard deviation $\sigma_{RV}$ and Pearson $r$
  correlation displayed in the bottom right corner of each panel.
  The top and bottom panels show $\Re$, $R_{23.5}$ and $\Rd$, 
  and the three Petrosian radii for the ``clean'' SDSS sample, 
  respectively.}
\label{fig:radvel_best}
\end{figure*}

\Fig{RadvsRad} highlights the comparison and differences
between radii derived from circular and isophotal apertures.
However, to decide which of these measurements yields the
tightest galaxy relations requires an objective comparison
against an uncorrelated variable, here chosen to be the
deprojected rotational velocity, $V$.
\Fig{radvel_best} shows the distribution of $RV$ data
for each $i$-band radial parameter; Sample A (all the available
data) is shown with gray points overplotted whereas Sample D
(best inclination and distance data) is in black.  The top panels
show $RV$ scaling relations with isophotal radii and the disk scale
length; similarly for the bottom panels with Petrosian radii for
the clean SDSS sample.  The red dashed line in each panel is an
orthogonal linear fit, with bootstrap re-sampling (see C07),
of the $RV$ relations for Sample D.

We evaluate the tightness of each fit with both the one-$\sigma_{RV}$
standard deviation about the best orthogonal fit and the Pearson $r$
coefficient. 
The listed values of $r$ and $\sigma_{RV}$ are those for Sample D. 

The tightest $RV$ relation is clearly that which involves $R_{23.5}$, 
in agreement with Saintonge \etal (2008) and SS11.  We discuss their
results in \se{VRL}.  Of the Petrosian radii, $R_{\rm p90}$, yields
the tightest $RV$ relations. 

\begin{figure*}[htb]
\centering
\includegraphics[width=0.9\textwidth, trim=1cm 1cm 7cm 5cm]{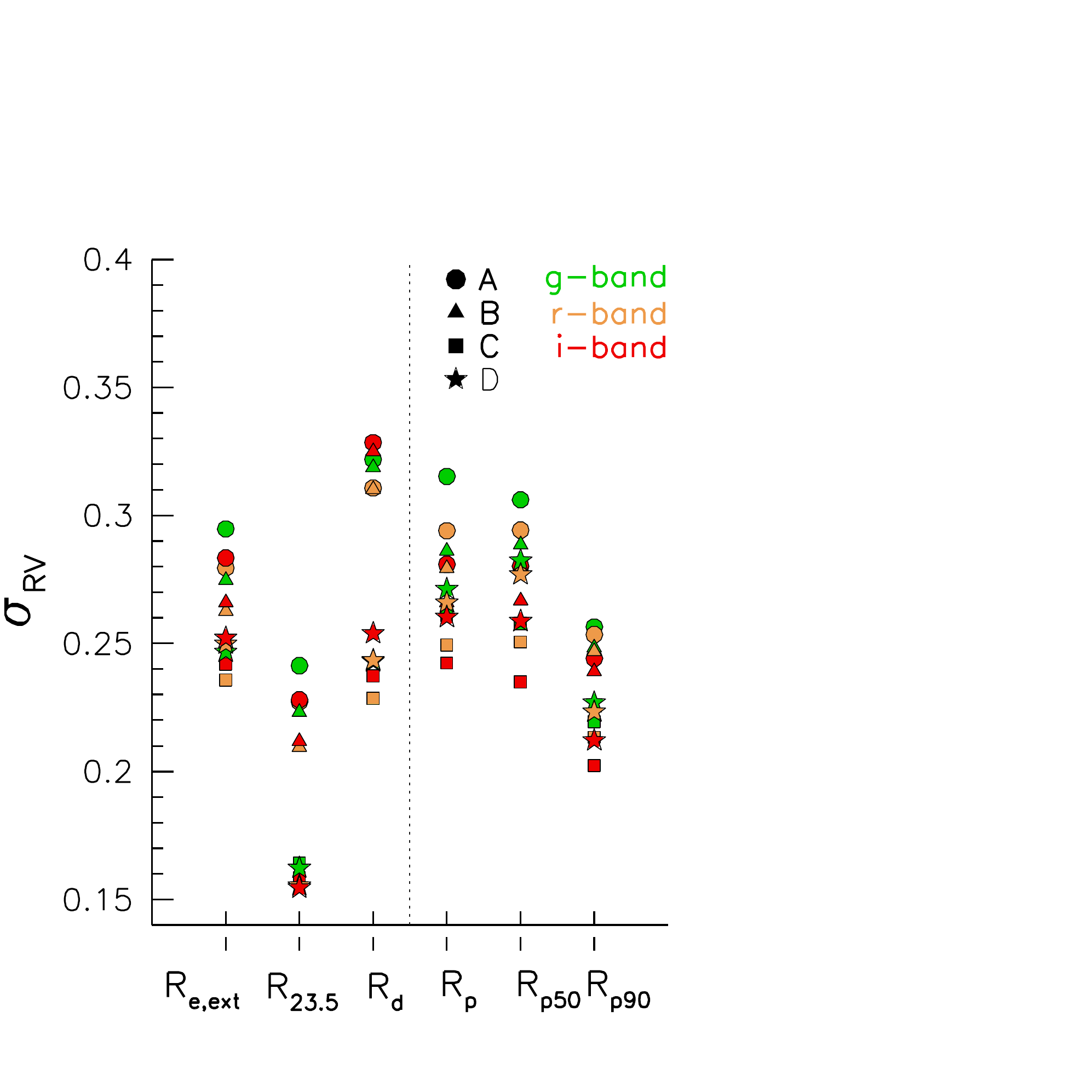}
\caption{The 1$\sigma$ standard deviations of all variants of the
  $RV$ relation plotted against rotational velocity in
  \Fig{radvel_best}. Measurements are plotted for each parameter in
  the $g$, $r$ and $i$-bands in green, orange and red
  respectively. Sample A is plotted in circles, B as triangles, C as
  squares and D as five-point stars.}
\label{fig:radStats}
\end{figure*}

\Fig{radStats} shows the variation of $\sigma_{RV}$ for the 
three $gri$ radii shown in green, yellow and red colours respectively.
The four sub-samples are plotted as A (circles), B (triangles),
C (squares) and D (stars). The vertical dashed line separates
our radial measurements from the SDSS Petrosian radii.

Overall, Sample C and D yield comparably tight $RV$ relations
indicating that the inclination cuts in Sample D do not improve
the trends for the $RV$ relation significantly. 
Other $RV$ relations with $\Re$, $\Rd$ and the Petrosian
radii are however improved when high and low inclinations are
weeded out.  Still, the biggest effect in obtaining a tighter
$RV$ relation results from eliminating galaxies with uncertain distances. 

Parameters derived from the extrapolated light profiles, such as $\Re$
and $\Rd$, are especially sensitive to the vagaries of the
``disk-like'' exponential region in such profiles.  Indeed both $\Re$
and $\Rd$ show a poorer scatter in \Fig{radStats}.  $\Rd$ is also
affected by projection effects, internal extinction by dust and
stellar populations.  For instance, there is a real dust extinction
gradient from center (opaque) to edge (transparent) in galaxy
disks. The measured scale lengths are larger, and central surface
brightnesses are smaller than their intrinsic dust-free values.
Empirical corrections (Giovanelli \etal 1995; Masters \etal 2003;
Graham \& Worley 2008) or corrections based on radiative transfer
models (Popescu \etal 2005; Driver \etal 2007; Gadotti \etal 2010) for
extinction effects on $\Rd$ as a function of inclination have been
proposed, but these remain grossly uncertain.

The isophotal radius $R_{23.5}$ is the clear winner with this test,
for all bands and all samples.  The SDSS Petrosian radius, $\Rpn$,
also fares well though the unreliable object identification and
deprojection effects for tilted galaxies (\se{sdssProd}) make
the interpretation of this parameter less simple than $R_{23.5}$.

\begin{figure*}[htb]
\centering
\includegraphics[width=\textwidth]{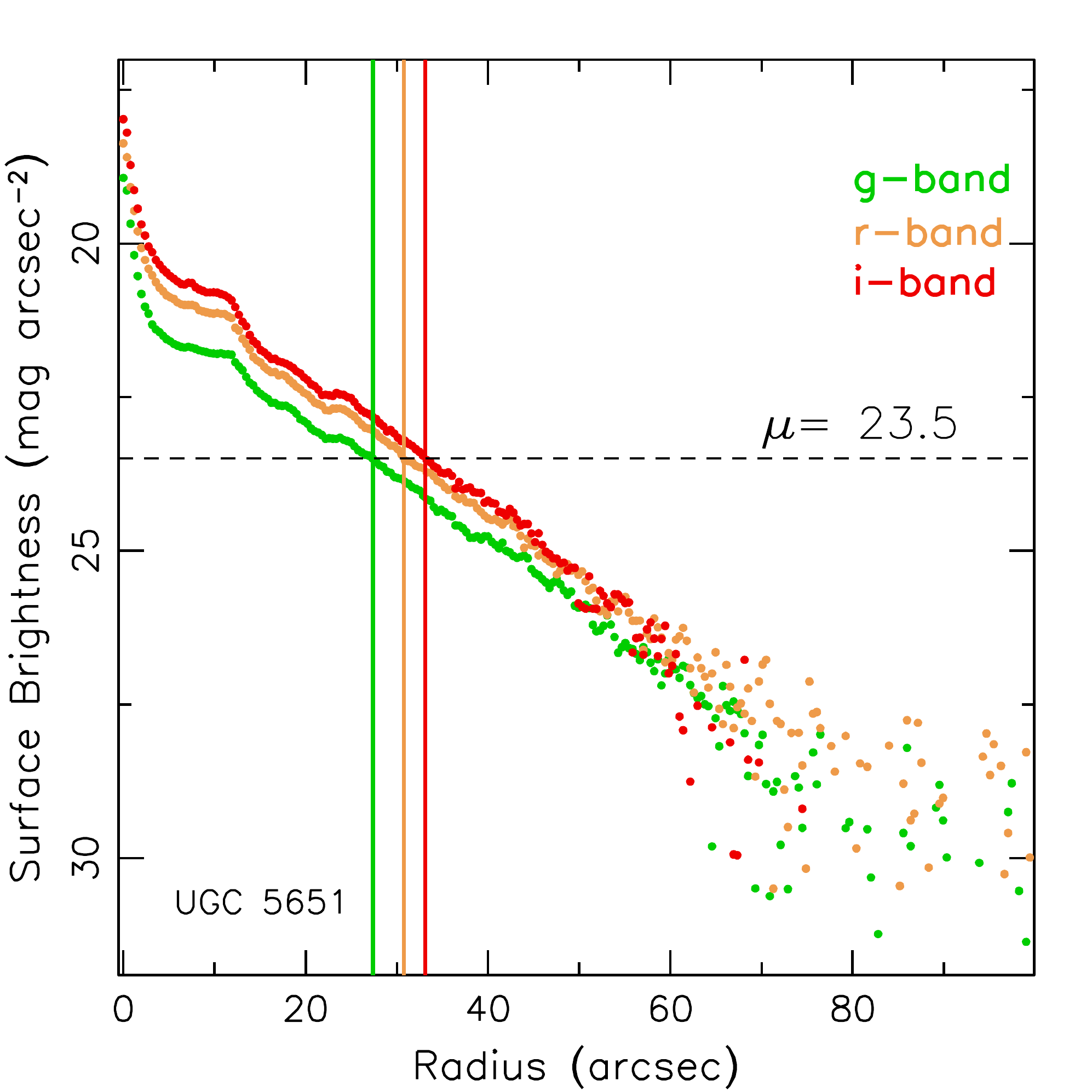}
\caption{Surface brightness profiles for the galaxy UGC 5651. The
  isophotal radius, $R_{23.5}$, corresponds to the location where
  the surface brightness profile reaches 23.5 \magarc\ in each band.
  These are depicted by the coloured vertical lines.}
\label{fig:R235SBcuts}
\end{figure*}

In seeking a direct match with galaxy formation models
(\eg Mo, Mao, \& White 1998; D07; D11), numerous scaling
relation studies have relied on $\Rd$ as the fiducial galaxy size
(\eg C07 and references therein).  This is because the scale length indicates
the change of surface density with radius for any galaxy;
as a \textit{global} tracer of galaxy structure, $\Rd$ is a
straightforward prediction of galaxy formation models. 
Different stellar populations however have different scale
lengths.  The space density of a given stellar population
in a disk is mostly controlled by the angular momentum evolution
of that disk (Foyle \etal 2008; Ro{v s}kar \etal 2008; Dutton 2009).
$R_{23.5}$, which measures the \textit{local} surface
density of a disk, is also sensitive to the local stellar
populations (see \Fig{R235SBcuts}).  The local space density
is more sensitive to local star formation conditions than
those averaged at $\Rd$.

Whether $R_{23.5}$ or $\Rd$ is preferred for either the
tightest $RV$ relation that it yields or for a more direct
connection with theory, the interpretation of the $RV$
relation relies on a full understanding of the biases
that each method entails.

\subsection{Comparison of Luminosity Measurements}\label{sec:magnitudes}

\begin{figure*}[htb]
\centering
\includegraphics[width=\textwidth]{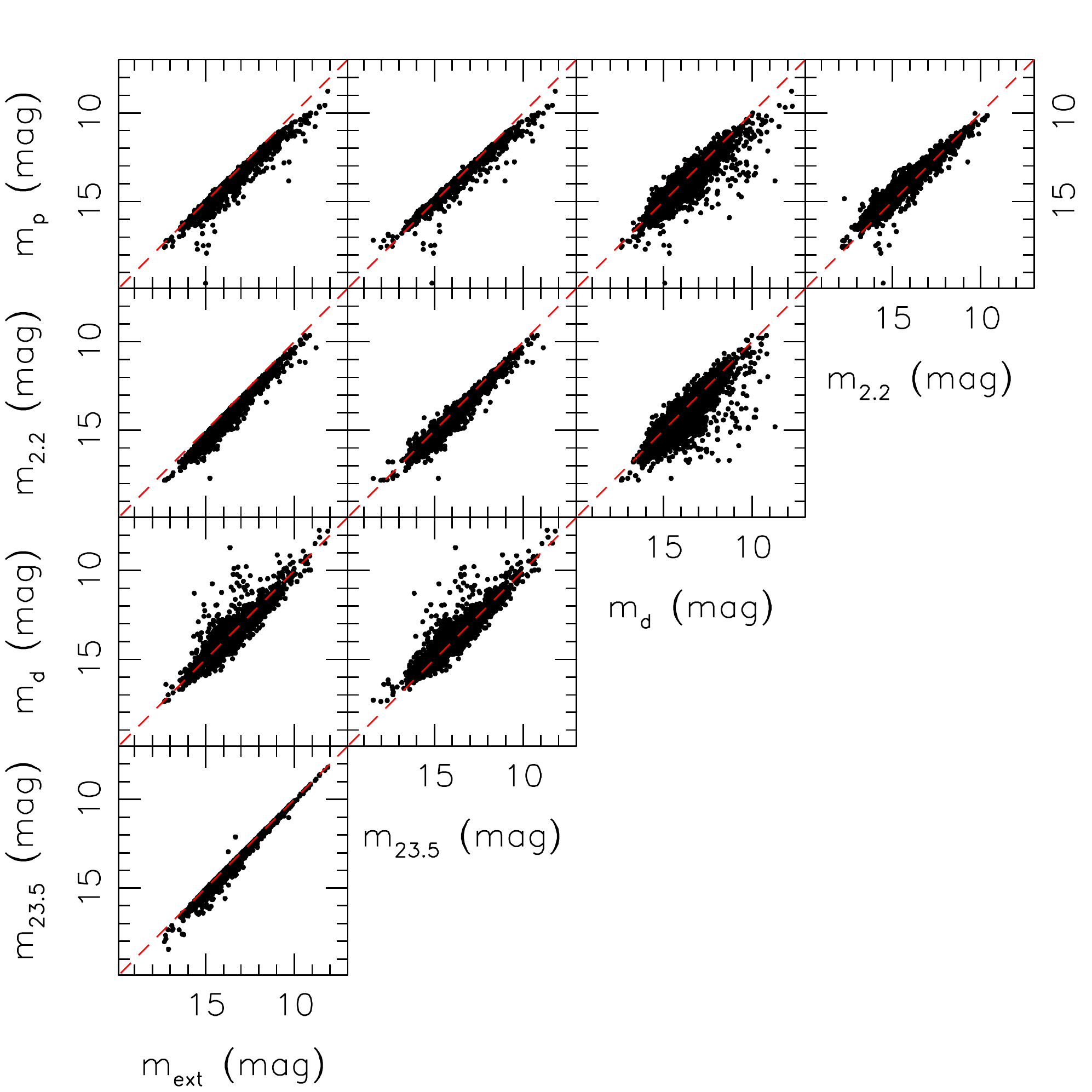}
\caption{Comparison of the apparent Petrosian magnitude m$_{\rm p}$
  from the ``clean'' sample of SDSS data products and our measurements
  of m$_{\rm ext}$, m$_{\rm 23.5}$, m$_{\rm 2.2}$ and m$_{\rm
    cut}$. The red dashed line represents a one-to-one correlation.}
\label{fig:MagVSMag}
\end{figure*}

In order to determine the best luminosity measure from
circular (Petrosian) or elliptical apertures, we compare
the \textit{clean} SDSS Petrosian apparent magnitudes,
m$_{\rm p}$, against our suite of derived apparent magnitudes
m$_{23.5}$, m$_{\rm ext}$, m$_{\rm d}$ and m$_{2.2}$. \Fig{MagVSMag}
shows these five $i$-band measurements against each other; the red
dashed line has slope unity.

The top row in \Fig{MagVSMag} shows that the best match with
the Petrosian magnitude, m$_{\rm p}$, is obtained with $m_{23.5}$. 
The top panels all show a departure from the one-to-one
line at the bright end. This is still due to the 
``shredding'' effect (see \se{sdssProd}) for large galaxies.
For non-shredded systems, the Petrosian flux accounts for
$\sim$ 98\% of the total flux of a galaxy (see Shen \etal 2003). 

The disk magnitude m$_{\rm d}$ shows the largest scatter with
other luminosity measurements, as it is tied to the uncertain disk
scale length $\Rd$ via \Eq{expDcounts} and does not account for
the bulge luminosity. 
A superb correlation between $m_{23.5}$ and $m_{\rm ext}$
exists; $m_{2.2}$ however yields noisier correlations.

\begin{figure*}[htb]
\centering
\includegraphics[width=\textwidth]{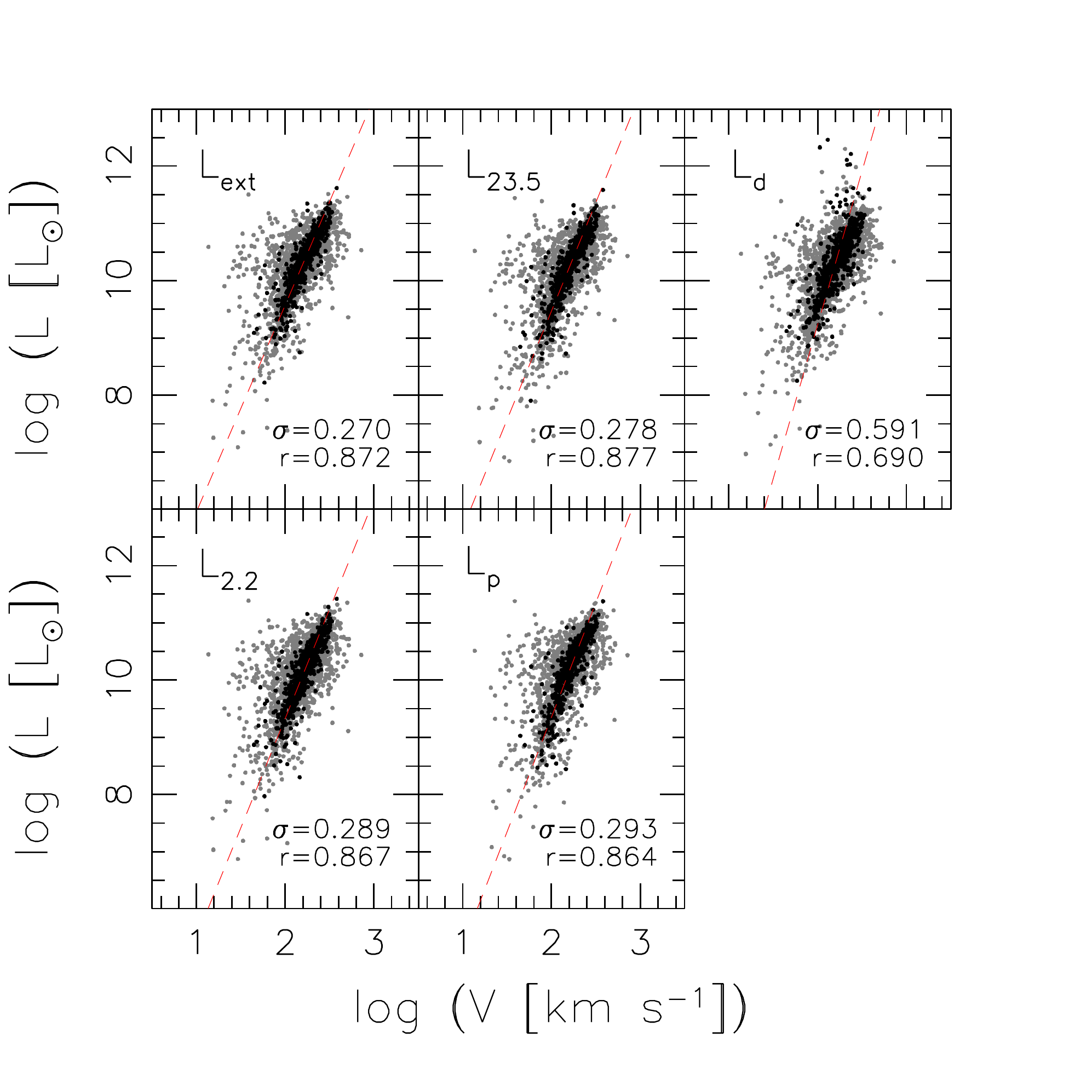}
\caption{Comparison of the corrected $i$-band luminosity
  measurements against the rotational velocity, $V_{\rm rot}$. Sample
  D is shown as black points overlaid on Sample A in gray points. The
  red dashed line represents the orthogonal fit to Sample D.  The 
  standard deviation $\sigma_{LV}$ and Pearson $r$ correlation coefficient
  for the fit to sample D are shown at the bottom of each panel. 
 } 
\label{fig:lumvel_best}
\end{figure*}

In \Fig{lumvel_best} we examine which $i$-band luminosity
parameter yields the tightest $LV$ scaling relation.  The dashed
red line is the orthogonal bootstrap fit through Sample D, shown
with black points. The full Sample A is shown in gray.
The extrapolated luminosity, $L_{\rm ext}$, and the isophotal
luminosity, $L_{23.5}$, yield mathematically tightest $LV$ relations;
however, the Petrosian radius, $L_p$, and $L_{2.2}$ are practically as good. 

\begin{figure*}[htb]
\centering
\includegraphics[width=0.9\textwidth, trim=1cm 1cm 7cm 5cm]{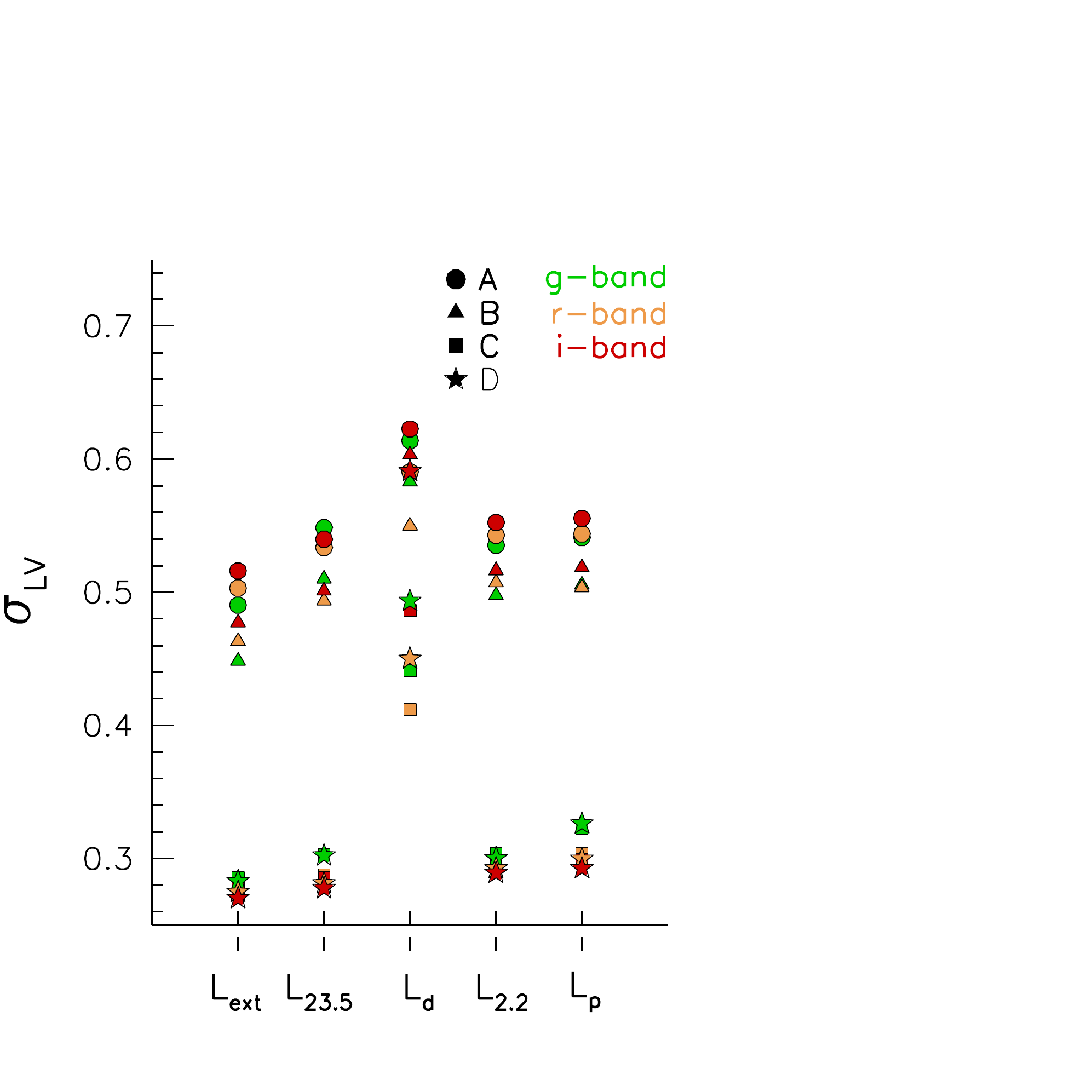}
\caption{The 1$\sigma$ standard deviations of all luminosity measures
  compared against rotational velocity in \Fig{lumvel_best}. 
  Symbols are as in \Fig{radStats}. 
}. 
\label{fig:lumStats}
\end{figure*}

\Fig{lumStats} shows the 1$\sigma$ standard deviation for the 
$LV$ fits in each $gri$ band and for each galaxy sub-sample A-D.
Sample D at $i$-band yields tightest relations, as does the use of $L_{ext}$.

Overall, all the luminosity metrics, with the exception of the
disk luminosity, $L_{\rm d}$, yield comparably tight $LV$ relations. 
Technically the total extrapolated luminosity of the galaxy
$L_{\rm ext}$ and $L_{23.5}$ yield slightly tighter $LV$
relations than our other tested luminosities. 
Luminosity measurements are rather stable due to their $global$ 
(cumulative), rather than local, nature. 

We have seen in \Table{skySubErr} the effect of
a $\pm0.2$\% and $\pm1$\% sky uncertainty on the radial
and apparent magnitude measurements for the galaxy UGC 5651.
The luminosity within 2.2 disk scale lengths, m$_{2.2}$, 
is mostly affected by uncertain disk scale lengths. 
Extrapolated magnitudes are less sensitive with $\sim0.03$ mag
uncertainty for a typical 0.2\% sky error.  Despite small
luminosity errors, we stress the importance of well-measured
sky levels, and caution against SDSS luminosity estimates
for the brightest and biggest galaxies in our sample.


\section{VRL Relation}\label{sec:VRL}

We now use the best galaxy structural parameters, $R_{23.5}$,
$\Rd$, $L_{\rm ext}$ and $L_{23.5}$ as determined in \se{ScalingMeasures},
and the corrected galaxy rotational velocity $V_{\rm rot}$ (hereafter $V$) 
for a detailed analysis of $VRL$ scaling relations.

We construct various scaling relations $VL$, $RL$ and $VR$
which we refer to collectively as the $VRL$ relation.
We also adopt orthogonal linear fits to model our scaling
relations.  Fit differences result from different modeling
techniques; e.g., bisector fits yield a steeper
slope than orthogonal fits.  Given that the $VRL$ scaling parameters
are not fully independent (e.g., correlated via inclination and distance),
neither a forward or inverse fit would work.  Bisector fits,
which average the forward and inverse fits, are therefore
also incorrect by construction.  Indeed, the bisector fit of a perfectly
uncorrelated distribution of two variables has an absolute slope of
one (C07; Hogg \etal 2010).  Lacking an accurate covariance matrix,
we adopt the orthogonal fit, as did C07 and SS11, as the least
biased of the suite of numerical fitting methods.  The $VRL$ 
parameter errors discussed in \se{errors} are used for, but play
little role in, our orthogonal fits. 

\Fig{VRLBest} shows the logarithmic form of the $VRL$ relation,
using the scaling parameters $L_{\rm ext}$, $R_{23.5}$ and $V$, for our
moderate inclination, best determined distance Sample D in black. The
full Sample A is shown in gray.  The red line is the orthogonal fit
to Sample D's parameter distribution and the 2-$\sigma$ bootstrap
errors are shown as dotted lines.  See C07 for details about the
fitting procedure. 
The direction and magnitude of a 20\% distance error are shown
with an arrow in each panel's corner (see \se{distances}).

Tables \ref{tab:VL} - \ref{tab:RVgri} present the fit slopes and
zero points, as well as the forward $Y|X$ and inverse $X|Y$
1-$\sigma$ deviations and overall Pearson $r$ coefficients
of the $gri$-band $VRL$ relations. 
The $\sigma_{X|Y}$ and $\sigma_{Y|X}$ standard deviations are
presented to facilitate comparisons with other authors and/or
theoretical models (D07; D11).  Any observational limit (\eg magnitude
cut) will bias the estimate of $\sigma_{X|Y}$ against $\sigma_{Y|X}$.
The $X|Y$ and $Y|X$ sigmas also differ in proportion to the
slope of each scaling relation.

To compare our standard deviations with those of other authors,
let us write the slope and scatter transformations for systems
relying on magnitude or luminosity.  Ignoring zero-points, the $VL$
relation ($\log V=b_{VL} \log L$) is directly related to the $VMag$
relation ($\log V=b_{VMag} Mag$) via $Mag=-2.5 \log L$ or
$b_{VL}=-2.5\times b_{VMag}$.  Similarly, the scatter
$\sigma_{VL}=\sigma_{MagV}/b_{MagV}$.

Given that previous scaling relation studies (\eg C07; AR08; D07; SS11)
have used both $\Rd$ and $R_{23.5}$ as radial metrics, the various
scaling rleations that we present in Tables \ref{tab:VL} - \ref{tab:RVgri}
will showcase those two radii where relevant. 

\subsection{The V-L (Tully-Fisher) Relation}\label{sec:V-L}

The $VL$ relations from \se{magnitudes} are modeled as $\log V=a+b \log L$.
We tabulate in \Table{VL} our best fit values for $a$ and $b$ in the $i$-band
as a function of luminosity (whether $L_{\rm ext}$ or $L_{23.5}$).
For our best $i$-band Sample D with 652 galaxies, we find
\be
 V \propto L_{\rm ext}^{0.29\pm0.01} \;\;\;\;\;\;\;\;\;\;\;\; 
 V \propto L_{\rm 23.5}^{0.27\pm0.01}.
\label{eq:VLrel}
\ee

The slopes and scatters for the multi-band $gri$ $VL$ relations
are tabulated in \Table{VLgri}.
The tightest relation is indeed obtained in the $i$-band, and 
slopes are slightly shallower from $g$ to $i$ band. 

These $i$-band slopes are identical to those reported by M06, C07 and SS11
(among others).  The consistency of the $VL$ slopes from these
and other studies based on rather different samples attests to
the robustness of the $VL$ relation and its independence to
numerous ``third parameters'' (e.g., Courteau \& Rix 1999; D07).
The $VL$ scatters reported in the literature may differ
more significantly as a result of various selection and/or 
``modeler'' biases.  Since we compare our scatters with those
of M06, C07 and SS11 below, a brief reminder about each
sample is warranted here:

C07 studied the $VRL$ scaling relation of 1300 late-type galaxies
with a mixture $L_{23.5}$ and $L_{\rm ext}$ luminosities and \ha\ and
\hi\ rotation curve velocities.  M06 used a subset of 807 so-called
``template'' cluster galaxies from the S07 catalog.  They extracted
their own $I$-band total magnitudes truncated at 8$R_{\rm d}$ and
corrected for Galactic and internal extinction and k-cosmological term.
These magnitudes were also corrected for morphology\footnote{While
appropriate for distance scale studies, a morphological correction 
to galaxy magnitudes removes all astrophysical signatures
of morphology on scaling relations.} as well as incompleteness bias.
SS11 used the entire S07 catalog (template and non-template galaxies). 
Like C07, they did not correct for morphological type and
incompleteness bias, but unlike C07, their final fits invoked outlier
clipping. 
The \hi\ line widths for M06, SS11
and the current study all come from the same source (S07). 

In order to compare scaling relation scatters obtained from orthogonal fits,
we were fortunate to use the original data files
from M06 (publicly available; 807 galaxies) and
SS11 (A. Saintonge, private comm.; restricted set of 665 template galaxies).  

For our sample, we find $\sigma^{\textrm{ortho}}_{VL}=0.074$ 
(\Table{VL}).  Meanwhile, our orthogonal $VL$ fits to the M06
and SS11 linewidths and $I$-band magnitudes of their template
galaxies yield $\sigma^{\textrm{ortho}}_{VL}=0.065$ in both
cases.  Considering the similar samples, we attribute most of
the scatter difference with M06 to their morphological correction
and with SS11 to their outlier clipping procedure.  Our tests have
also revealed scatter differences of 0.02 dex between our respective
orthogonal fitting engines (i.e., same data, different software).

C07 found an even smaller observed scatter of $\sigma_{VL}=0.057$.
Note that C07 did not correct for morphological differences
or for incompleteness bias, or sigma-clip their scaling relations.
We speculate that the scatter difference is here due to the
morphological make-up between the S05/S07 samples and the ones 
collected in C07.  The former is morphologically broader than C07. 
For instance, restricting our sample to only Sc galaxies reduces
the $VL$ scatter by 0.005 dex.  Tailoring one's data
set according to specific criteria will obviously affect the final
scaling relations and complicate comparisons amongst even similar studies.

Other related results are also found in the literature.  Kannappan
\etal (2002) obtained a $B$-band $VL$ slope of $0.29$ and scatter
of $\sigma_{VL}=0.08$ for 68 spiral galaxies. AR08 found a
$B$-band $VL$ slope of $0.314\pm0.015$ and an intrinsic scatter of
$\sigma_{VL}=0.063$ for 76 disk galaxies.  The TF study of 162 disk
galaxies with \ha\ rotation curves and SDSS imaging by Pizagno \etal
(2007; hereafter P07) allows a more direct comparison with our SDSS
based results.  Their derived bivariate fit $gri$ $VL$ slopes are
$b_{VL,g}=0.340$, $b_{VL,r}=0.338$ and $b_{VL,i}=0.325$, all steeper
than ours (\Table{VLgri}).  The scatter in each relation $\sigma_{VL,g}=0.073$,
$\sigma_{VL,r}=0.063$ and $\sigma_{VL,i}=0.061$, is also smaller
than ours, partly due to their steeper slopes, different fitting
methods (bisector vs orthogonal), and their more careful sample selection. 

While many of the studies above either seem to agree or differ
slightly, we stress that the comparison of scaling relation parameters
depends highly on the selection and manipulation of each data sample.

\subsection{The R-L (Size-Luminosity) Relation}\label{sec:R-L}

\begin{figure*}[htb]
\centering
\includegraphics[width=\textwidth]{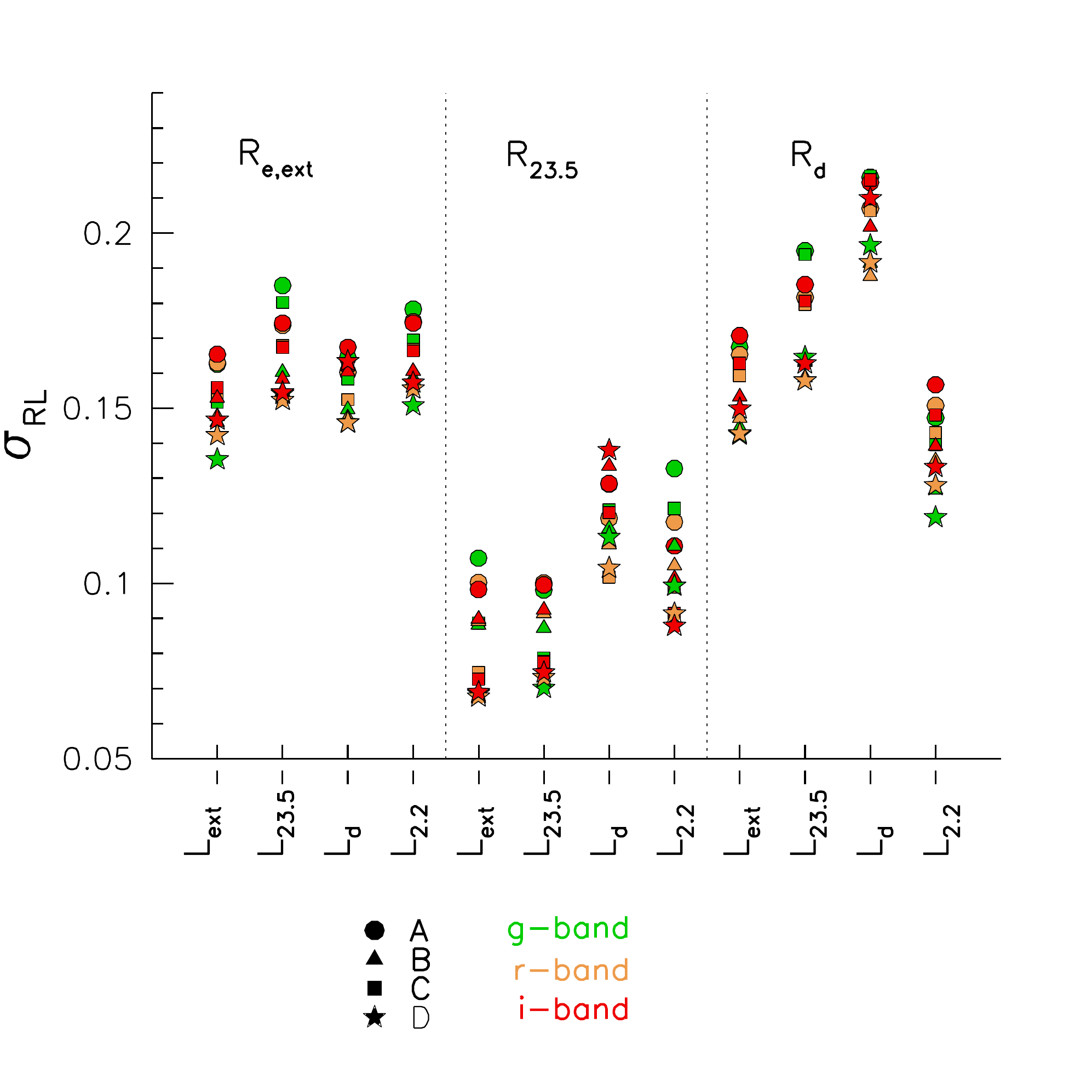}
\caption{The 1-$\sigma$ standard deviations of all variations of the
  $RL$ relation labelled across the x-axis with their $RL$
  scatters. Measurements in the $g$, $r$ and $i$ bands are shown in
  green, orange and red respectively with Sample A plotted in
  triangles, B in circles, C in squares and D in five-point stars.}
\label{fig:RLStats}
\end{figure*}

Much like the $VL$ relation, the size-luminosity $RL$ relation
is a fundamental constraint to galaxy formation models (P07, AR08, D07, D11). 
We compute the $RL$ relation for the combinations of our radial
parameters $\Re$, $R_{23.5}$ and $\Rd$ with the luminosity
parameters $L_{\rm ext}$, $L_{23.5}$, $L_{\rm d}$ and $L_{2.2}$.
The 1-$\sigma$ scatter of each $RL$ relation is shown in \Fig{RLStats},
where each subsample A-D is separated by point type and each band
by colour (see caption).  Of our three radial measurements,
the isophotal radius $R_{23.5}$ yields the tightest $RL$ relation.  

We present the four combinations of the $RL$ relation with $R_{23.5}$,
$\Rd$, $L_{23.5}$ and $L_{\rm ext}$ in \Table{RL}.  We have
also computed fits using the corrected radii $R^c_{23.5}$ and $\Rdo$,
as decribed in \se{Corrections}; however, these corrections provide
no gain in scatter reduction.  In view of their uncertain nature, 
we use uncorrected radii in what follows.  We express
our results as $R\propto L^{b}$ for our best $i$-band Sample
D of galaxies with uncorrected radii,
\beqa
R_{\rm d} \propto L_{\rm ext}^{0.41 \pm 0.02} &\;\;\;\;\;\;\;\; &  
R_{\rm d}\propto L_{\rm 23.5}^{0.36 \pm 0.20} \\
R_{\rm 23.5}\propto L_{\rm ext}^{0.44 \pm0.07} &\;\;\;\;\;\;\;\;&
R_{\rm 23.5}\propto L_{\rm 23.5}^{0.41\pm0.07}
\label{eq:RLrel}
\eeqa
As both radii and luminosities depend on the distance estimate
(\se{Corrections}), distance errors contribute weakly to the 
scatter in the $RL$ relation over all subsamples A-D. We show in
\Fig{VRLBest} and in \se{distances} that distance errors scatter
along the direction of the $RL$ relation, thus introducing minimal
spread to the relation.  
The larger Sample B (moderate inclination galaxies only)
is thus often just as tight as the smaller, distance-pruned, 
Sample D of galaxies. 

The measured slopes in \Eq{RLrel} are in broad agreement with
previously reported values (\eg de~Jong \& Lacey 2000; Shen \etal 2003;
C07; SS11).  The $gri$ band dependence of the $RL$ relation is shown
in \Table{RLgri}. As in the $VL$ relation, the $RL$ relation slope
decreases from bluer to redder bands.  Bandpass effects are least
dominant for the combination of $\Rd$ and $L^c_{23.5}$ and 
largest for the pair $R^c_{23.5}$ and $L^c_{\rm ext}$. 

Use of scale lengths to construct the $RL$ relation yields weak
correlations with a Pearson coefficient $r\sim 0.7$ compared to
$r\sim 0.9$ for isophotal radii.  The larger scatter of the scaling
relations based on $\Rd$ partially stems from the larger uncertainty
in the measurement of the disk scale length itself (\se{radii}). 
SS11 found a similar result; their isophotal radii measurements
yield an $RL$ $r\sim0.86$ while C07's $RL$ relation based on disk
scale lengths has $r\sim 0.65$.  Our use of $R_{23.5}$ is thus an
improvement over C07 and a confirmation of SS11's results (see also C96).

The different scatters for our $RL$ relations are listed in \Table{RL},
with the tightest value, $\sigma_{RL}=0.068$, obtained for the combination
of $R_{23.5}$ and $L^c_{\rm ext}$.  The $RL$ scatter $\sigma_{RL}=0.05$
reported by SS11 for their isophotal $RL$ relation is tighter than our
or any other previous similar assessment (e.g., Shen \etal 2003; 
Pizagno \etal 2005; C07; AR08; see Fig. 3 of SS11).  Eliminating
SS11's outlier clipping procedure increases their reported $RL$
scatter to $\sigma_{RL}=0.06$, in closer agreement with our value.

\subsection{The R-V (Size-Velocity) Relation}\label{sec:R-V}

The third scaling relation combines the galaxy size, $R$, and 
velocity, $V$.  \Table{RV} shows our orthogonal fit parameters for
the $RV$ relation.  For instance, the $i$-band $RV$ slopes for
Sample D are:
\be
R_{\rm d}\propto V^{1.82\pm0.14} \;\;\;\;\;\;\;\;\;\;\;\;  
R_{23.5} \propto V^{1.52\pm0.07}.
\label{eq:RVrel}
\ee

The observed log scatter of the $R_{23.5}-V$ relation is 0.154 dex.
\Table{RVgri} gives the slopes and scatter for Sample D in all
three $gri$ bands. 

We can compare the $i$-band $RV$ parameters with those of S08 and SS11.  
S08's bivariate fit to a sub-sample of 699 S07 galaxies yielded
$R^{\textrm{S08,bisec}}_{23.5} \propto V^{1.24\pm0.01}$ and
a scatter $\sigma^{\textrm{S08,bisec}}_{RV}=0.07$.  In order
to compare directly with S08, we compute bivariate fits for
the $RV$ relation of our Sample D yielding:
$R^{\textrm{bisec}}_{23.5} \propto V^{1.37\pm0.01}$ and
$\sigma^{\textrm{bisec}}_{RV}=0.145$. Our and S08's results
are significantly different.
Comparison with SS11's similar, yet larger, dataset yields
somewhat different results: $R^{\textrm{SS11,ortho}}_{23.5} \propto V^{1.44\pm0.02}$
and a scatter $\sigma^{\textrm{S11,ortho}}_{RV}=0.123$.  SS11's $RV$ slope
is statistically equivalent to ours; the smaller scatter results
again from outlier clipping.

Due to the larger scatter in their $RV$ relation, C07 derived
their $RV$ slope by combining the $VR$ and $RL$ relations and
requiring that if $V\propto L^{\alpha}$ and $R \propto L^{\beta}$
then $R\propto V^{\alpha / \beta}$.  The C07 $RV$ relation had
a (log) slope of $1.10\pm0.12$ dex.  Our $RV$ relations listed
in \Table{RV} are all steeper and obey intrinsically the rule
$R\propto V^{\alpha / \beta}$ (with $R \propto V^{0.41/.27}$).
It this sense, this study supercedes C07. 

$R_{23.5}$ remains the ideal radial metric for the tightest
correlation with $V$ but, of the three scaling relation combinations
($VL$, $RL$ and $RV$), the $RV$ relation is broadest.
This weakens its appeal for cosmological (e.g., distance-measuring)
applications. 

\subsection{Distance Dependence of the $VRL$ Relation}\label{sec:distances}

We calculate the effect of a percent distance error $\Delta D / D$
in the $LV$, $RL$ and $RV$ planes as,
\beqa
\Delta \log L = \frac{5 \Delta D }{2.5 \ln (10) D}&=&0.868\frac{\Delta D}{D}\\
\Delta \log R = \frac{\Delta D}{\ln (10) D}&=& 0.434\frac{\Delta D}{D}.
\label{eq:RLerrs}
\eeqa

\begin{figure*}[htb]
\centering
\includegraphics[width=0.9\textwidth]{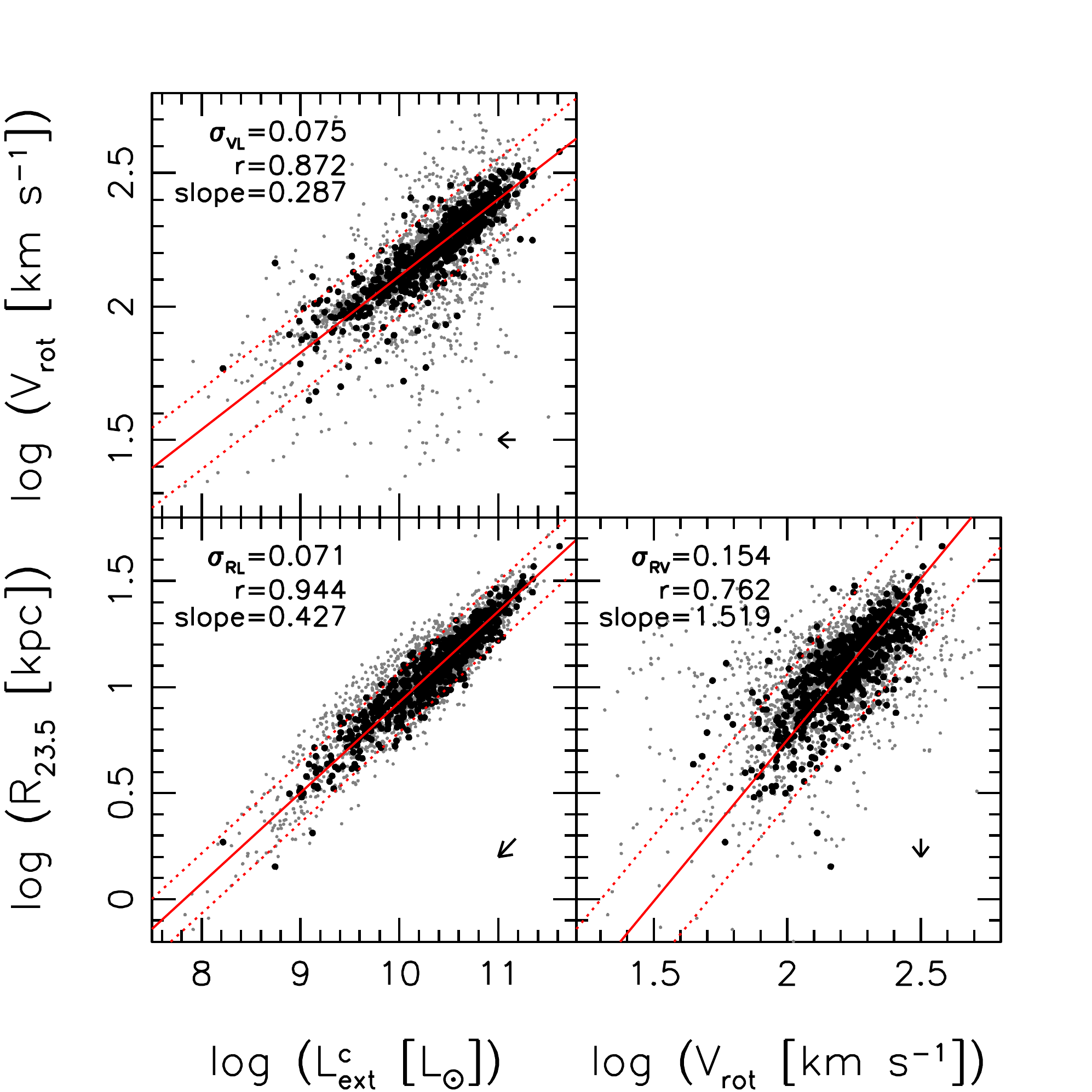}
\caption{The $VRL$ relation with $V_{\rm rot}$, $R^c_{23.5}$
  and $L^c_{\rm ext}$. The red solid line is the orthogonal
  fit to the galaxy sub-sample D shown with black points.
  The red dashed lines are 2-$\sigma$ deviations and the 
  gray points show Sample A. The 1-$\sigma$ scatter, Pearson $r$
  coefficient and slope of the $VL$ ({\it top left}), $RL$ ({\it bottom left})
  and $RV$ ({\it bottom right}) relations are shown in the top left corner
  of each panel. The magnitude and direction of the $20$\% distance
  uncertainties are shown in the lower right corners of each
  panel. The fitting parameters are listed in Tables \ref{tab:VL},
  \ref{tab:RL} and \ref{tab:RV}.}
\label{fig:VRLBest}
\end{figure*}

We show the $20$\% distance uncertainty in the bottom right corner
of each \Fig{VRLBest} plot. Note that the propagated radial uncertainty
$\Delta \log R$ is twice as large as the uncertainty in luminosity
$\Delta \log L$ and that distance uncertainties in the $RL$ relation
move points along a slope of 1/2. Since the slope of the $RL$ relation is
$\sim 0.5$ (\Table{RL}), distance uncertainties play no role
in the $RL$ relation scatter. 

Recall that the distances used to calculate luminosity and physical
radius of our galaxies are derived from the CMB redshift velocities
listed by S05 and S07 (see \Eq{distance}).  S05 and S07 also present
``peculiar velocity corrected'' distances (in \Eq{distance}, 
$V_0=V_{\rm CMB}+V_{\rm pec}$); these were computed to minimize
the scatter in the Tully-Fisher (VL) relation for each galaxy
cluster in their sample.  The peculiar velocity estimates can
thus be very large and even unrealistic as they absorb all
possible sources of scatter.  We therefore caution against the
use of such velocities in scaling relation studies. 

\begin{figure*}[htb]
\centering
\includegraphics[width=\textwidth]{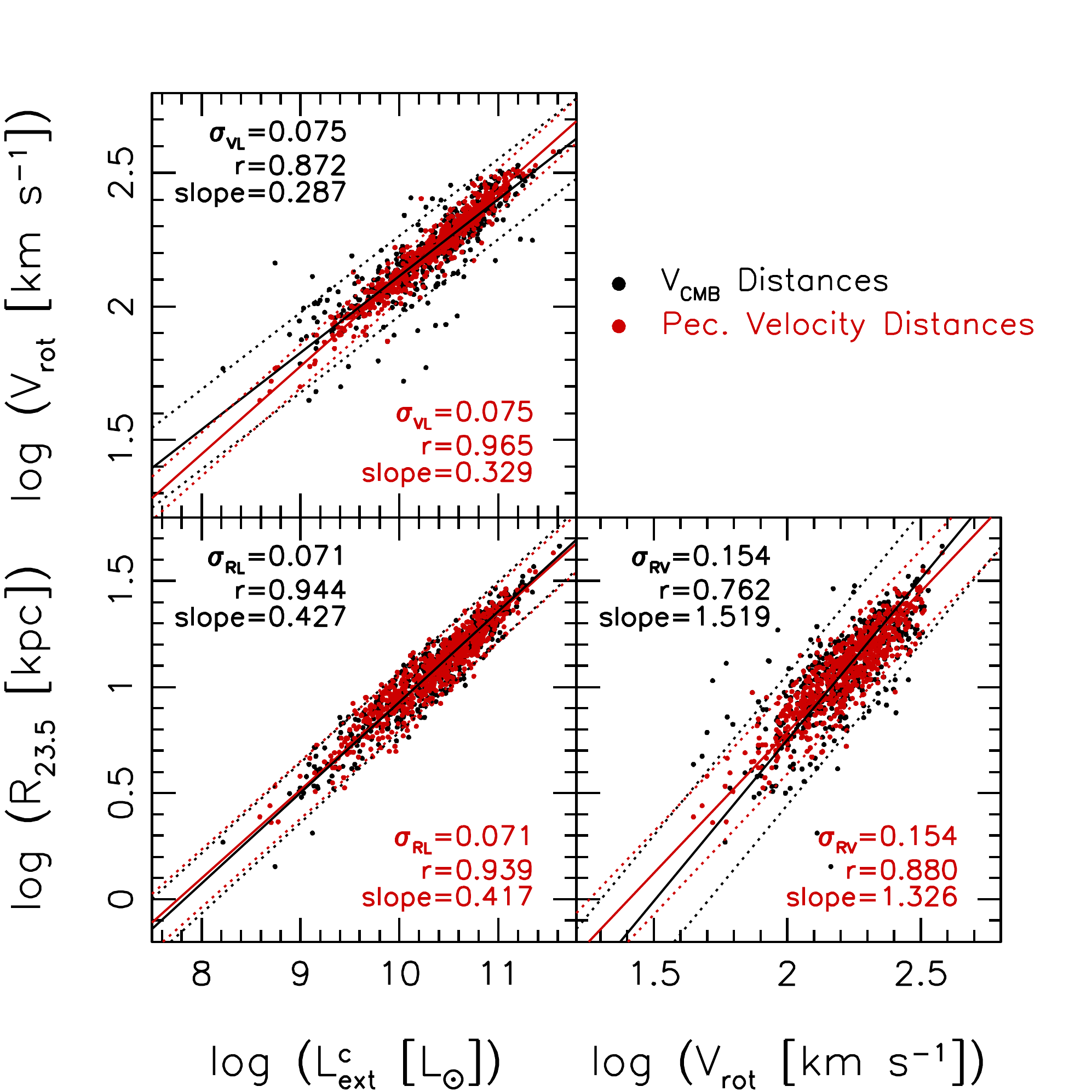}
\caption{The $VRL$ relation for the galaxy Sample D with
  distances derived from $V_{\rm CMB}$ redshifts (black points)
  and the peculiar velocity corrected distances (red points) 
  from S05/S07, with luminosities in the $i$-band. The line type
  and parameters are as in \Fig{VRLBest}. 
  The black points, fits and line statistics are as in \Fig{VRLBest}.}
\label{fig:VRLDcorr}
\end{figure*}

\Fig{VRLDcorr} shows the $VRL$ relation computed with both 
$V_{\rm CMB}$ distance data (black points and lines) and
peculiar velocity data (red points and lines).  Standard
statistics are shown.  The $V_{\rm CMB}$ data in black
are the same as in \Fig{VRLBest}.  Note the significant 
difference in slopes due to an over-interpretation of the 
scatter as being due solely by peculiar velocities. 

\subsection{Dependence on Morphology, Colour and Surface Brightness}\label{sec:Morph}

Much effort has been invested in the study of the $VRL$ relation
scatter, with particular emphasis on the Tully-Fisher $VL$ relation.
The search for the ``third'' parameter has included, among others,
morphology, colour, mass-to-light ratios and surface brightness
(Zwaan \etal 1995; CR99; Verheijen 2001; Kannappan \etal 2002;
C07; SS11 to cite a few).
In this section we highlight the $VRL$ dependence as a function
of morphology, colour and surface brightness.  The full quantitative
dependence analysis will be presented elsewhere. 

\begin{figure*}[t]
\centering
\includegraphics[width=0.53\textwidth]{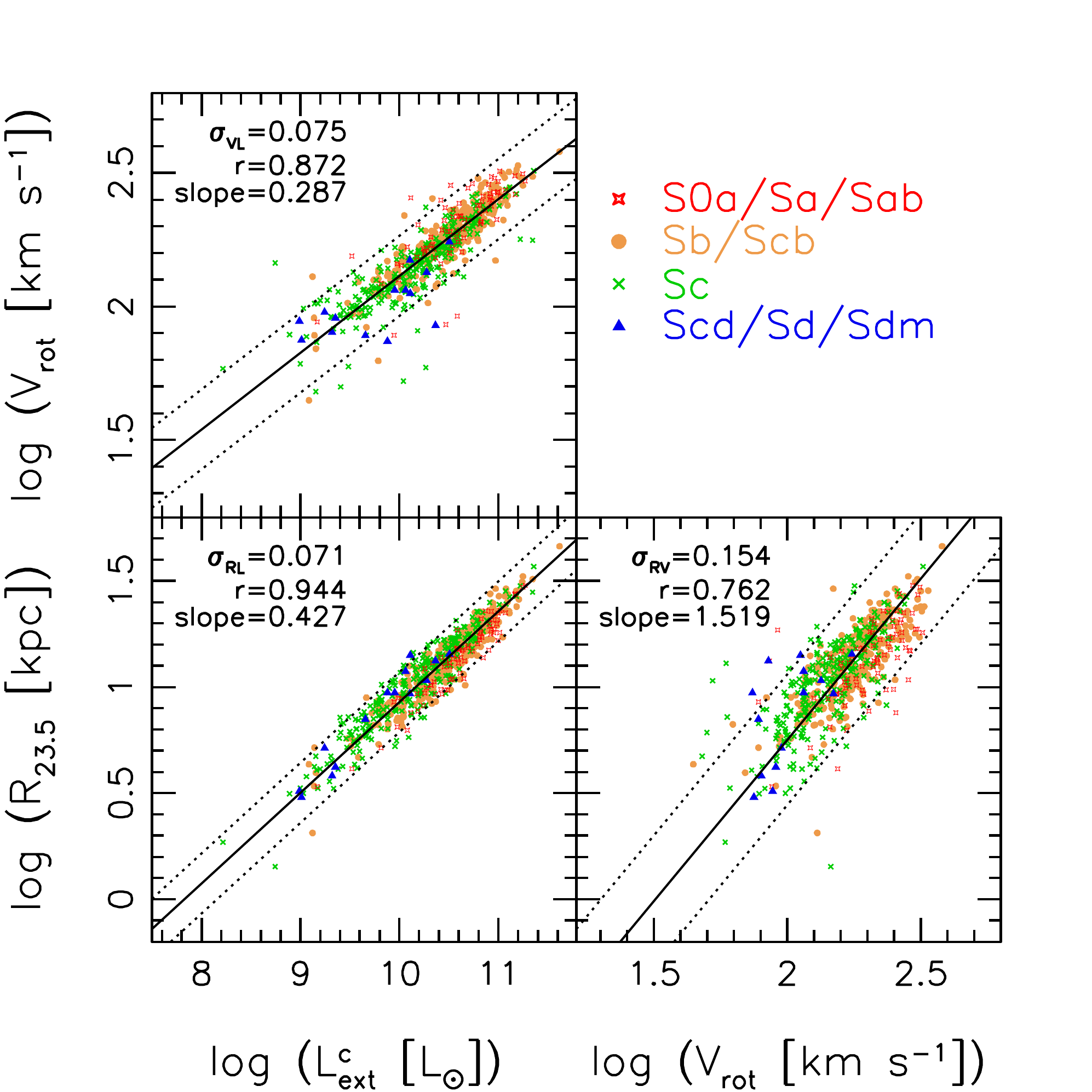}\includegraphics[width=0.53\textwidth]{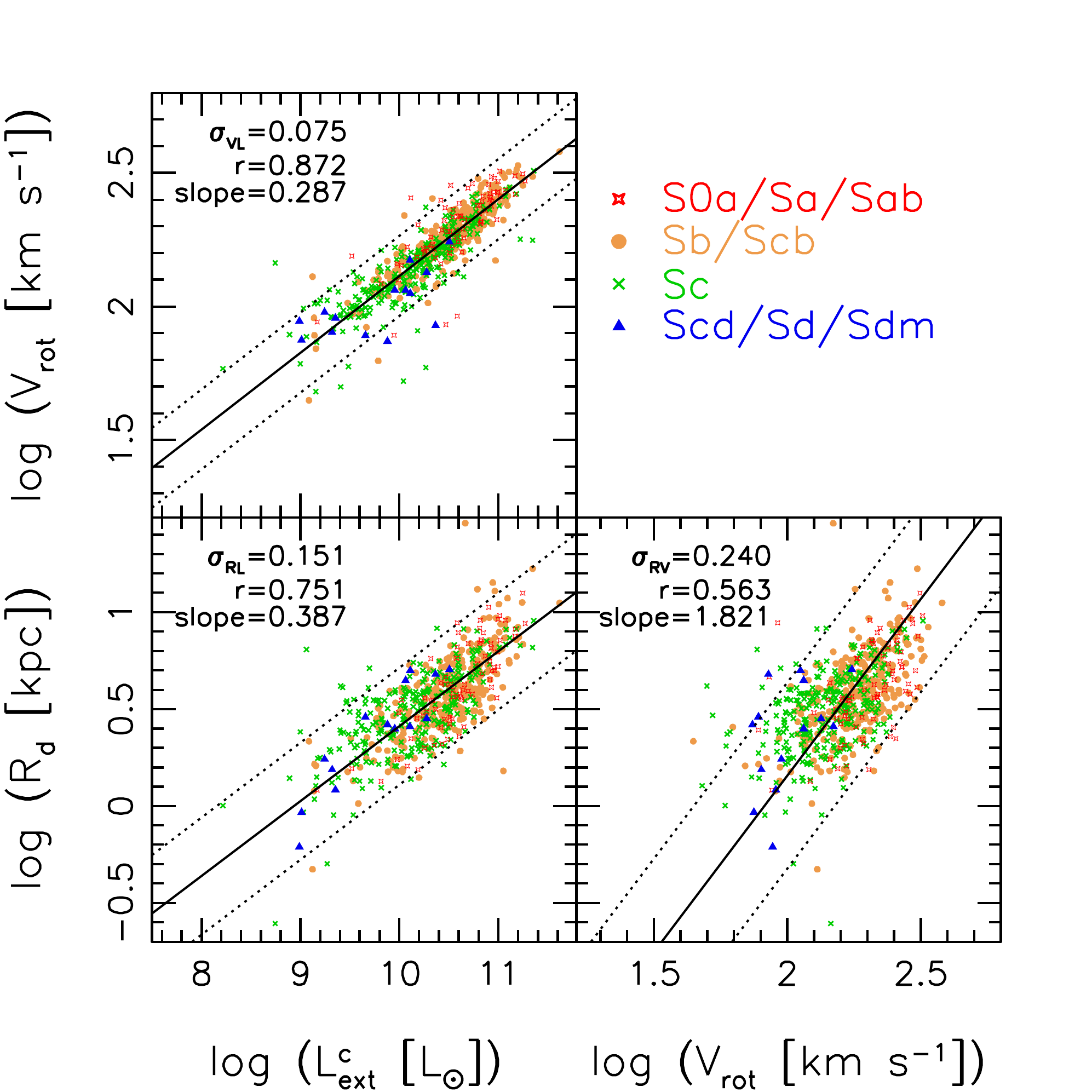}
\caption{Dependence of the $VRL$ relation on morphology displayed
  with both radial scaling parameters $R_{23.5}$ (left) and $\Rd$ (right). 
  The slope and 2-$\sigma$ deviation of the orthogonal fits to Sample D
  are shown as solid and dotted black lines, respectively.
  The data points are coloured according to Hubble type, as provided by S05 (but not S07).}
\label{fig:VRLMorph}
\end{figure*}

\Fig{VRLMorph} presents the morphological dependence of the $VRL$
relation for Sample A; the RL and RV relations on the left
and right figures use $R_{23.5}$ and $\Rd$, respectively. 
The $RL$ slope are expected to become shallower from early to late-type
galaxies (Shen \etal 2003), as we see for $\Rd$.  The isophotal 
RL relation is however free from such a trend. 
Fitting different $VRL$ relations for different morphologies
is beyond the scope of this work; however we do confirm the same
trends as measured by C07.

\begin{figure*}[t]\centering
\includegraphics[width=0.53\textwidth]{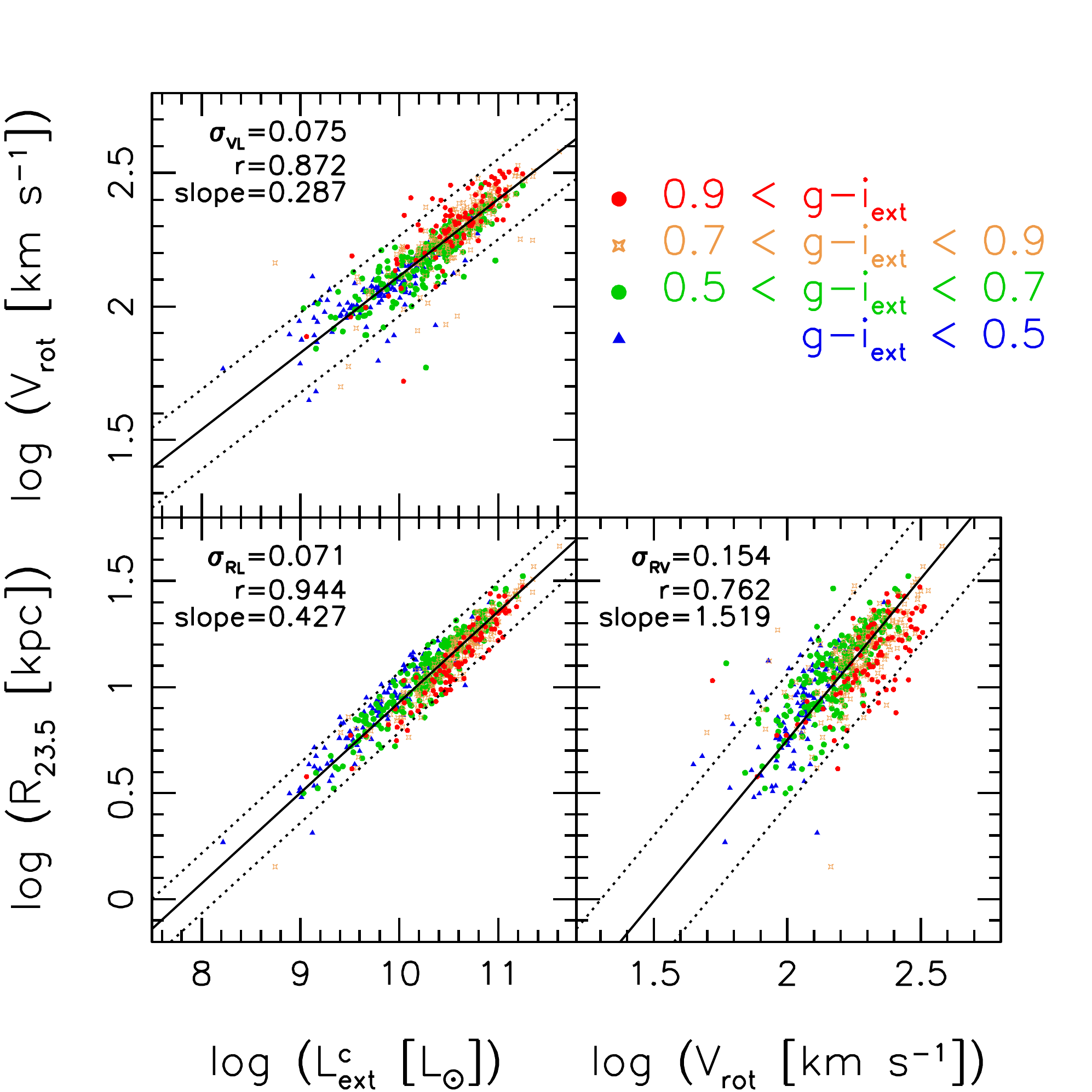}\includegraphics[width=0.53\textwidth]{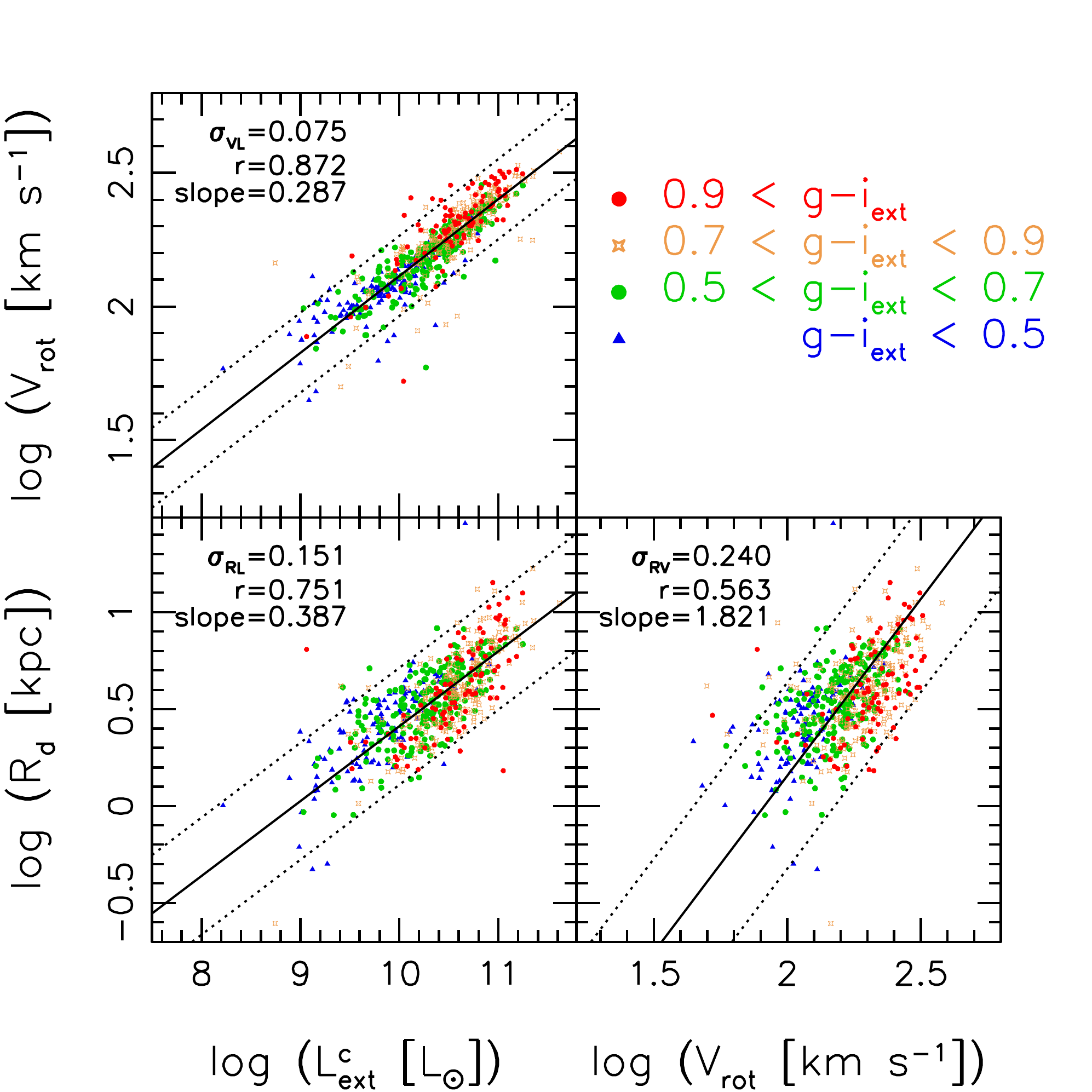}
\caption{Same as \Fig{VRLMorph} but against colour, $(g-i)_{\rm ext}$.}
\label{fig:VRLCol}
\end{figure*}

\begin{figure*}[t]\centering
\includegraphics[width=0.53\textwidth]{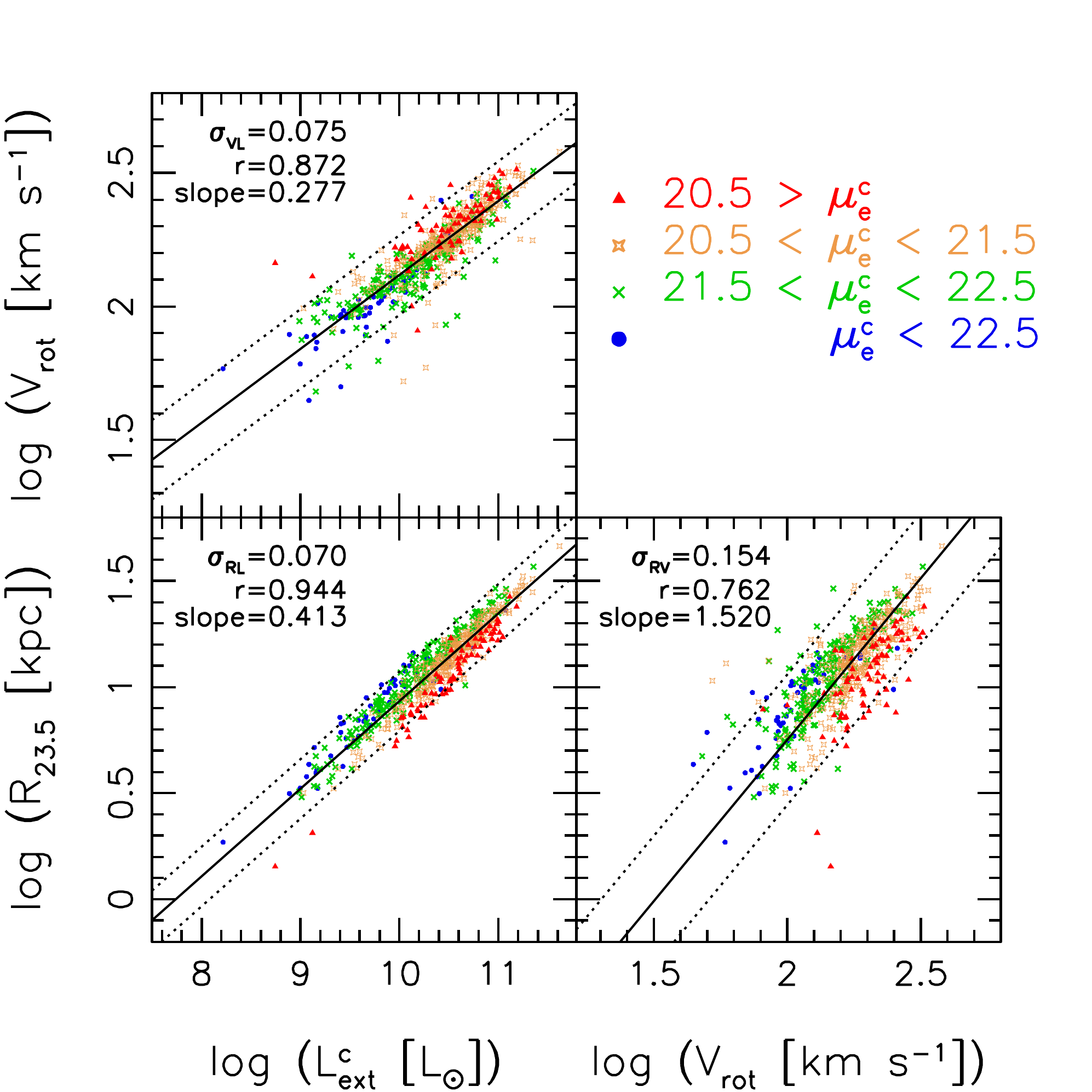}\includegraphics[width=0.53\textwidth]{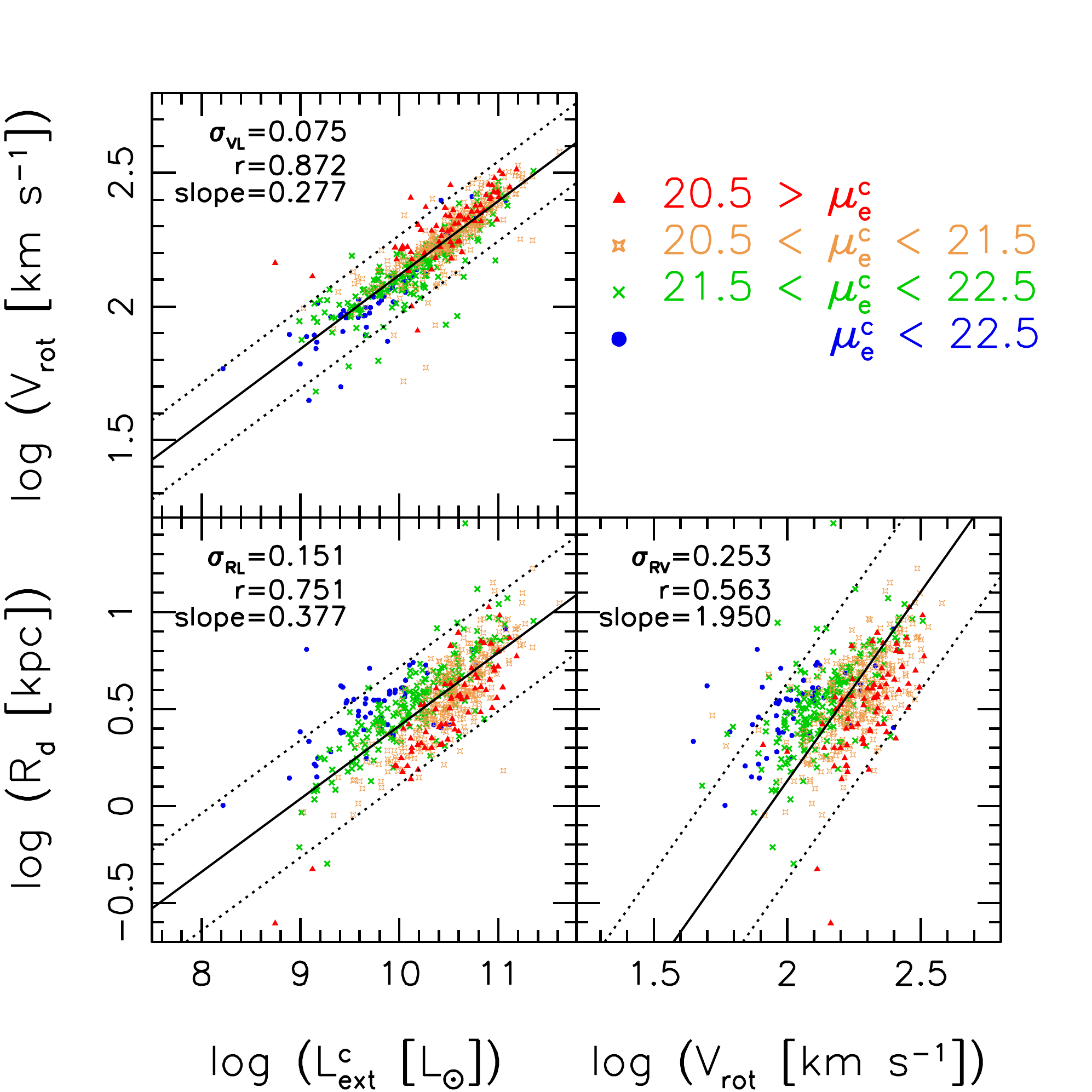}
\caption{Same as \Fig{VRLMorph} but against surface brightness, $\mu_e^c$.
}
\label{fig:VRLmuE}
\end{figure*}

Galaxy colour has been examined as a possible agent in the $VRL$
scatter, with larger variations in bluer bands due to star formation
(see Kannappan \etal 2002; C07; P07). We show the colour dependence
of the $VRL$ relation in \Fig{VRLCol} for Sample D only for clarity. 
We observe a slightly larger scatter of the $VRL$ relation for bluer
galaxies.  Clearly the full Sample A (shown in \Fig{VRLMorph}) shows
the largest scatter in the redder, early-type galaxies. 

Overall this confirms that redder, brighter galaxies are faster rotators
and that the largest galaxies are the brightest and reddest late-type
galaxies.

\Fig{VRLmuE} shows the dependence of the $VRL$ relations on surface
brightness.  The effective surface brightness, $\mu^c_e$,
is colour-coded with red for highest surface brightness (HSB) galaxies
and blue for low surface brightness (LSB) galaxies. The $RL$ relation
illuminates the conjunction of high concentration galaxies (small $R$,
large $L$) with HSBs and low concentrations with LSBs.  One might
naively expect that HSBs are faster rotators than more extended
LSB galaxies for a given luminosity, though no dependence of
the $VL$ relation on surface brightness has ever been
found (\eg Sprayberry \etal 1995; Zwaan \etal 1995; CR99; D07; SS11).
Note that if the velocity were measured at relatively small radii,
a surface brightness dependence of the $VL$ relation would be expected
(e.g., Catinella \etal 2007).
$VL$ and $RL$ residuals are only weakly correlated indicating that
for a given luminosity, the disk size does not affect the rotation
speed.  The intrinsic scatter in the $VL$ relation may then be attributed
to scatter in the dark matter fraction and stellar mass-to-light ratio
(D07). 

The non-correlation of $\mu^c_e$ in the $VL$ residuals has often
been interpreted as proof for a correlation of surface brightness
with dark matter content. The latter is however difficult to assess
given large uncertainties in stellar population and galaxy formation
models.  Various factors contributing to the $VRL$ scatter include
star formation, the initial mass function of stars, halo and disk
spin parameters, feedback, adiabatic contraction of the halo,
correlation of halo parameters and more (Navarro \& Steinmetz 2000;
Firmani \& Avila-Reese 2000; D07; AR08; S11, D11).
A detailed, quantitative model analysis of these data will be presented
elsewhere. 


\section{The Baryonic Tully-Fisher Relation}\label{sec:BTF}

The Baryonic Tully-Fisher relation (BTF) relates the total baryonic
(stellar + gaseous) mass of a spiral galaxy to its rotational
velocity. For their intermediate-size samples (a few hundred galaxies),
McGaugh \etal (2000), Trachternach \etal (2008), and Stark \etal (2009)
found that the BTF is a more ``fundamental'' relation than the luminous
(Tully-Fisher) or stellar mass Tully-Fisher (STF) relation.

We can test these postulates with S05's larger compilation since it includes
gas masses for thousands of galaxies.  We can indeed sum the stellar masses,
computed from SDSS luminosities and stellar population models
(e.g., Bell \& de~Jong 2001), with the S05 gas masses to obtain
a baryonic mass for each galaxy.

McGaugh \etal (2000) also noted a kink in the luminous TF relation
of the low-mass galaxies with $V\lsim100$ \kms; those would have
systematically lower luminosities than predicted by the TF fit at
higher mass. By considering the total baryonic mass instead, they
showed that this effect disappears and galaxies of both low and high
baryonic assume the same BTF relation. This result is not confirmed by
Geha \etal (2006) who do not see a deviation from the luminous
high mass TF relation for a sample of 101 very low-mass dwarf
galaxies all with $V\lsim 100$ \kms.
Our sample does not permit a critical assessment of the McGaugh versus
Geha dilemma as it includes too few galaxies with $V\lsim100$ \kms\
and $M_{\rm gas} > M_*$\footnote{Note that the tension with Geha
\etal (2006) is reduced if the improved data base of Stark \etal (2009)
is used instead of McGaugh \etal (2000).}.
However, it is worth computing the scatter and slope of the BTF
down to those velocities. To our knowledge, ours is the largest
BTF sample to date.

For the computation of our galaxies' STF and BTF relations, we
use as the fiducial luminosity parameter our extrapolated light
measurement $L^c_{ext}$ (\se{magnitudes}) which provides the
tightest correlation with rotational velocity (\se{VRL}).  We convert
our $(g-i)_{\rm ext}$ galaxy colours into an $i$-band stellar
mass-to-light ratios, $M_*/L_i$, using the prescription of
Bell \etal (2003) with a -0.10 offset to reproduce the Chabrier (2003) IMF.  

The neutral gas mass is calculated from the corrected \hi\ line flux
measurement, $S_c$, in S05 following Haynes \& Giovanelli (1984):
\be
M_{\hi}=2.355\times 10^{5}S_c D^2 \;\;\;\; (\text{M}_{\odot}),
\label{eq:hi}
\ee
\noindent where $D$ is the distance of the galaxy in kpc
(see \Eq{distance}). The contribution from helium and other
metals to the gas mass is approximated as in McGaugh \etal (2000);
\be
M_{\rm gas}=1.4M_{\rm \hi}  \;\;\;\; (\text{M}_{\odot}).
\ee
The baryonic mass is the sum of the stellar and gas contributions,
\be
M_{\rm bar}=M_*+M_{\rm gas} \;\;\;\; (\text{M}_{\odot})
\ee

\begin{figure*}[htb]\centering
\includegraphics[width=\textwidth]{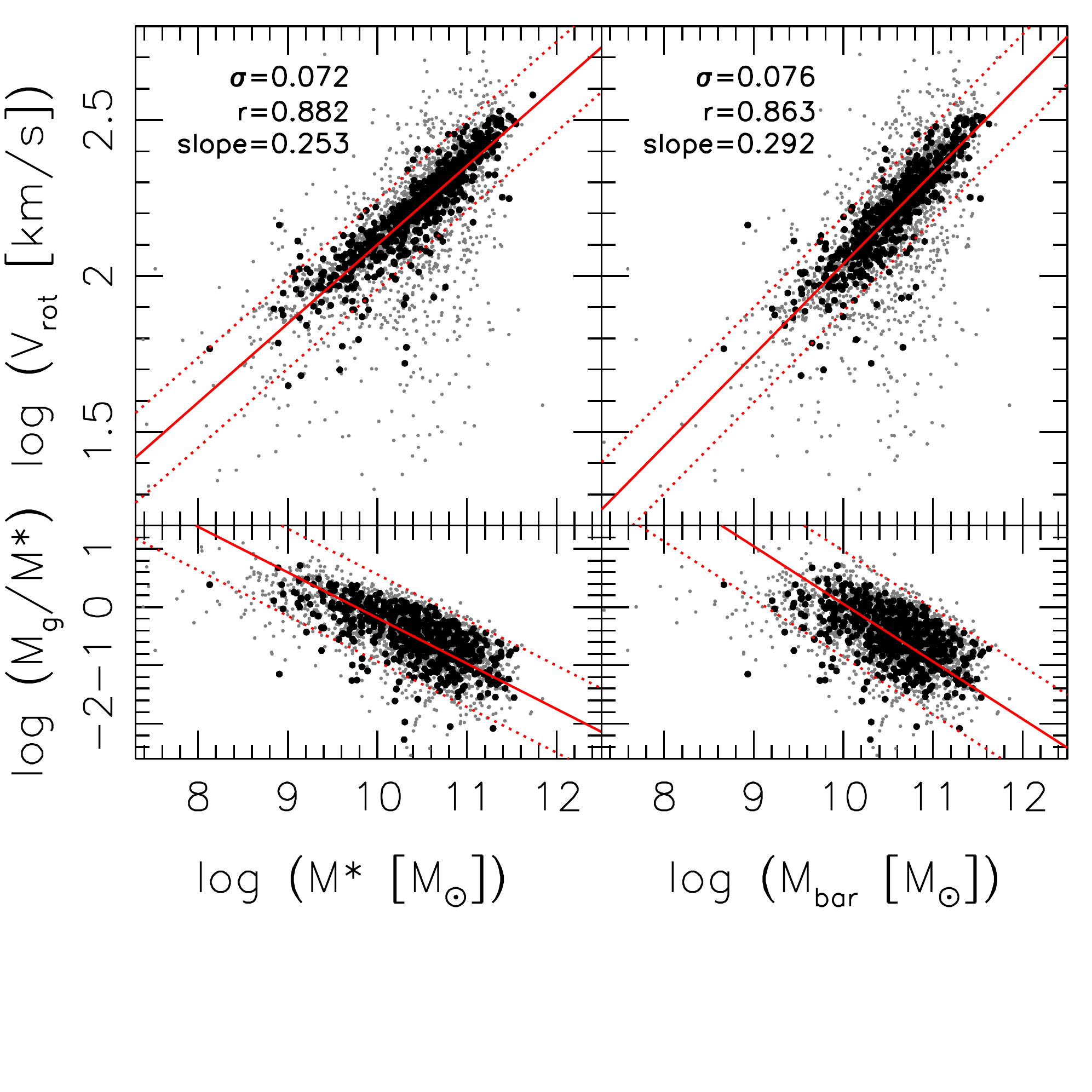}
\caption{The stellar mass {\it (top left)} and baryonic Tully-Fisher
  {\it (top right)} relations.  See text about the computation of
  masses.  Ratio of gas mass, M$_{\rm g}$, to stellar mass, M$^*$, as
  a function of M$^*$ {\it (bottom left)} and baryonic mass 
  {\it (bottom right)}.  The slope and 2-$\sigma$ deviation of orthogonal
  fits for Sample D (black points) are shown as solid and dotted red
  lines, respectively. Sample A galaxies are shown in gray.}
\label{fig:TFBTF}
\end{figure*}

The baryonic sample contains $\sim$5\% fewer galaxies than the STF and
TF sample, given the limited availability of \hi\ flux measurements.
\Fig{TFBTF} shows the STF (top left) and BTF (top right)
relations, respectively; the orthogonal fit results are tabulated
in \Table{TFBTF} for our galaxy Sample D.  We show in the left and
right bottom panels of \Fig{TFBTF} and in the last two entries of
\Table{TFBTF} the gas contributions to the total stellar and
baryonic masses expressed as $\log (M_{\rm g}/M_*)$, the logarithmic
ratio of gas to stellar mass. For our best Sample D, the STF and BTF
relations are fit as:
\beqa
V \propto M_*^{0.25\pm0.01} \;\;\;\;\;\;\;\;\;\;\;\; 
V \propto M_{\rm bar}^{0.29\pm0.01}
\eeqa

\noindent with measured scatters $\sigma_{VM*}=0.072$ and $\sigma_{VM_{bar}}=0.076$, 
respectively.  

These scatters are comparable to those of the basic $VL$ relation (\se{V-L}),
which is remarkable given the greater number structural parameters involved. 
Given the uncertainty in assessing the assumed errors on $M_*/L$, $L$, one
would be hard-pressed to identify any differences between the STF and BTF
scatters; they are both comparable.  However, while the observed scatters
of the TF, STF and BTF relations are comparable, both the STF and the BTF
relations may have intrinsically smaller scatter than the $VL$ relation
and might thus be more fundamental.

\begin{figure*}[htb]\centering
\includegraphics[width=\textwidth]{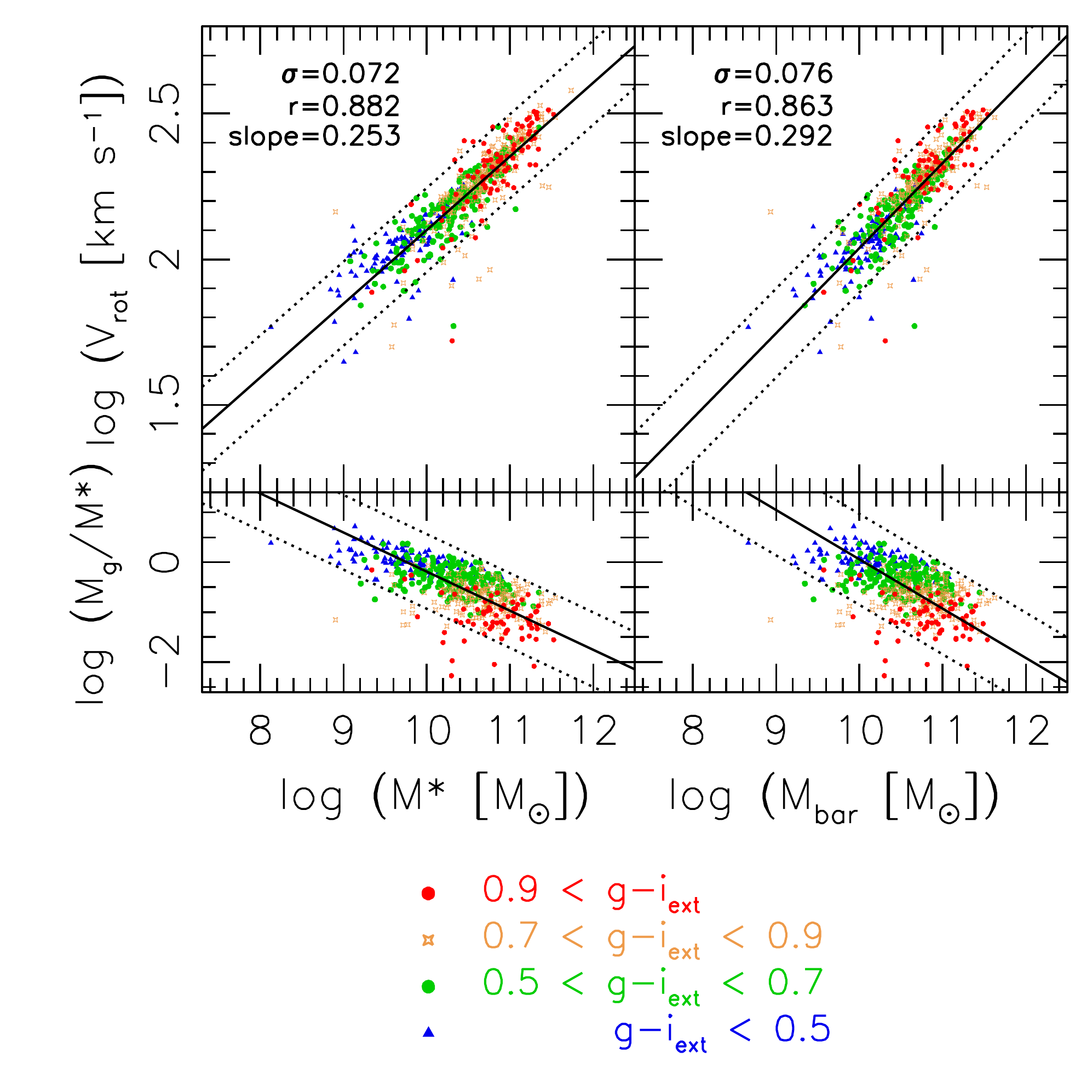}
\caption{Dependence of the stellar mass and baryonic Tully-Fisher
 relations on $g-i$ colour.}
\label{fig:TFBTF_col}
\end{figure*}

\Fig{TFBTF_col} shows the dependence of the STF and BTF relations of the
$(g-i)_{\rm ext}$ colour, as well as their gas contributions, for
sample D.  The bottom panels show that the ratio of gas to stellar
and baryonic mass is well-defined by colour: larger more massive
(luminous) galaxies have a smaller gas fraction than smaller, bluer
galaxies. The gradient in gas mass fraction, from the least to most
massive galaxies, causes most of the steepening of the BTF relation
compared to the STF.

The Pearson $r$ correlation coefficient suggests a mildly tighter
STF relation (versus the BTF).  This may be partially attributed to the
$D^2$ distance-dependence of the \hi\ gas mass calculated in \Eq{hi}
which introduces additional scatter into the BTF.  
It is also interesting to note that for their semi-analytic model, 
Dutton, van den Bosch, \& Dekel (2010) showed that galaxy star
formation rates (SFR) are more tightly correlated with stellar
mass than baryonic mass (see their Fig.~15).  However, to determine
which of the STF or BTF relation is more fundamental will require
far more accurate measurement errors for the stellar and gas masses.
For instance, our gas mass estimates include a correction for helium
and heavier elements but do not include molecular hydrogen or the warm
ionized gas which could increase the baryonic fraction and steepen
the BTF slope slightly (see Fukugita \& Peebles 2004; Bregman 2009;
McGaugh \etal 2010). 
A discussion of these effects is beyond the scope of the present study, 
but should be addressed elsewhere soon. 

The uncertainty in the stellar mass-to-light ratio, $M_*/L$,
propagates through both STF and BTF relations and different
calibrations will therefore yield different slopes and scatters.
McGaugh \etal (2005) explored three variants of $M_*/L$ and their
effect on the scatter of the resulting BTF. Their first $M_*/L$
assumed a maximum disk hypothesis.
The second method used the $B-V$ colour prescription of Bell \etal (2003),
and the third estimate was based on modified Newtonian dynamics (MOND). 
McGaugh \etal (2005) concluded that the MOND $M_*/L$'s yield the tightest
($r=0.99$) BTF with a slope of $0.25$.  However, a $M_{\rm bar}-V_{\rm flat}$
relation with slope 0.25 and zero scatter is intrinsic to MOND, 
and its recovery via MOND $M_*/L$'s should not be surprising.
Furthemore, the MOND prediction applies for McGaugh's velocity
measure $V_{flat}$ (McGaugh 2000), whereas our BTF relation relies on
\hi\ line widths, $W_{\rm F50}$.  In dwarf galaxies, $W_{\rm F50}$
is likely to under-estimate $V_{flat}$, while in massive spirals
$W_{\rm F50}$ is likely to over-estimate $V_{flat}$ (Verheijen 2001).
The BTF slope based on \hi\ line widths should thus be steeper than
that based on $V_{flat}$ measurements, as observed.  McGaugh (2005)
certainly found a range of BTF slopes from 0.25 to 0.33 depending
on the velocity estimator.

Our computation of the BTF relation used the \textit{sub-maximal}
$M_*/L$ values of Bell \etal (2003). Adopting a maximal disk decreases
the BTF slope from $0.292$ to $0.290$; the effect is thus negligible.
Bell \& de~Jong (2001) find the same slope of $\sim$0.29 for the BTF
derived with maximal $M_*/L$'s in the $B$, $R$, $I$ and $K$ bands.

AR08 also studied 76 disk galaxies with masses derived from re-scaled
$M_*/L$'s from Bell \& de~Jong (2001) and obtained a STF slope of
0.274$\pm$0.012 and a BTF slope of 0.306$\pm$0.012, in good agreement
with our own results.  A shallower slope is however found by
Gurovich \etal (2010) who fit a bivariate BTF for a small sample
of 21 galaxies with \hi\ line widths and the $V$- and $H$-band
luminosities to find a BTF slope $\sim 0.26$.  Here is a case
where the use of line widths does not yield a shallow BTF slope;
the small size and bivariate fitting method however complicate
the interpretation of this different result.  To our knowledge,
use of the velocity measure $V_{flat}$ always yields shallower
BTF relations.

A more extensive census of BTF relation slopes is presented
in Gurovich \etal  Overall, BTF studies report $V-\Mbar$ slopes
in the range 0.25-0.33.  However, care in comparing BTF relations
from different authors, using different definitions of $V_{circ}$
(e.g., whether V$_{max}$, $W_{\rm F50}$, $V_{flat}$, etc.) and 
different fitting methods (e.g., bivariate vs orthogonal fits)
must always be taken.


\section{Conclusion}\label{sec:conclusion}

We have compiled a catalog of 3041 spiral galaxies within $cz <28 000$
\kms\ with multiband photometric parameters extracted from SDSS images
and corrected \hi\ velocity widths and distances taken from Springob
\etal (2005, 2007).  We have extracted $g,r$ and $i$-band surface brightness
profiles from SDSS images and derived well-calibrated radial and
luminous measurements from each galaxy. Our main findings are as
follows:

\begin{enumerate}

\item Our radial and luminosity parameters extracted from isophotal 
  fits of SDSS images yield a scatter improvement of the $VL$ and $RL$
  relations of $\sim 8\%$ and $\sim 30\%$ compared to similar relations
  constructed with SDSS DR7 Petrosian parameters.

\item We find that the $i$-band isophotal radius $R_{23.5}$ correlates
  best with the galaxian rotational velocity,$V_{\rm rot}$. While
  $R_{23.5}$ is a more robust measurement than the scale length $R_{\rm d}$,
  for instance in terms of the $RL$ relation scatter, $R_{\rm d}$ holds
  greater significance for comparisons with galaxy formation models.

\item Both the extrapolated and isophotal $i$-band luminosities
  provide the tightest correlations with galaxian rotational
  velocity. The luminosity within 2.2 stellar disk scale lengths (peak
  of the baryonic rotation curve) is a poorer tracer of $V_{\rm rot}$.

\item We construct the $VRL$ relation in the $gri$ bands finding
  general agreement with the slopes and scatters reported elsewhere.
  Our main scaling relations in the $i$-band are summarized as:
\beqa
V \propto L_{\rm 23.5}^{0.27\pm0.01} \;\;\;\;\;\;\;\;\;\;\;\; \phantom{V} \\
R_{\rm 23.5}\propto L_{\rm 23.5}^{0.40\pm0.01} \;\;\;\;\;\;\;\;\;\;\;\;\
R_{\rm d} \propto L_{\rm 23.5}^{0.35\pm0.02} \\
R_{23.5} \propto V^{1.52\pm0.07} \;\;\;\;\;\;\;\;\;\;\;\;
R_{\rm d}\propto V^{1.82\pm0.14} \\
\eeqa

We find little correlation of the $VRL$ scatters with morphology,
colour or surface brightness.  Scaling relations for the SDSS $g$
and $r$ bands are provided in the text. 

\item We have transformed our galaxy colours into stellar masses and
  added these to the gas mass from the available \hi\ fluxes. These
  masses yield the largest stellar Tully-Fisher (STF) and baryonic TF
  relations (BTF) to date. For the STF and BTF, we find
\beqa
V \propto M_*^{0.25\pm0.01} \;\;\;\;\;\;\;\;\;\;\;\; 
V \propto M_{\rm bar}^{0.29\pm0.01}
\eeqa
both with high Pearson $r$ correlation coefficients of
$\sim$0.9.  

The observed scatters of the TF, STF and BTF relations are all comparable. 
To decide which of the STF or the BTF relation is more fundamental
requires a detailed error investigation which is beyond the scope
of the present paper.  Our reported BTF slope (0.29) matches that
of previous similar investigations based on \hi\ or \ha\ line widths. 
Use of the velocity measure $V_{flat}$ would yield a shallower BTF
relation slope of 0.25.
\end{enumerate} 

\bigskip 


We would like to thank Kristine Spekkens and Am\'elie Saintonge for
valuable discussions about the comparison of our respective results
and Karen Masters for useful advice about the Galaxy Zoo pipeline.
We also acknowledge enlightening discussions with Stacy McGaugh
about BTF relations.

SC acknowledges the support of a Discovery grant from the Natural
Sciences and Engineering Research Council of Canada and AAD was
supported by a CITA National Fellowship.
MM was supported by NASA through SAO Award Number 2834-MIT-SAO-4018,
issued by the Chandra X-ray Observatory Center on behalf of NASA (\#NAS8-03060)

Funding for the SDSS has been provided by the Alfred P. Sloan
Foundation, the Participating Institutions, the National Science
Foundation, the US Department of Energy, the National Aeronautics and
Space Administration, the Japanese Monbukagakusho, the Max Planck
Society, and the Higher Education Funding Council for England. The
SDSS Web Site is http://www.sdss.org/.  The SDSS is managed by the
Astrophysical Research Consortium for the Participating
Institutions. The Participating Institutions are the American Museum
of Natural History, Astrophysical Institute Potsdam, University of
Basel, Cambridge University, Case Western Reserve University,
University of Chicago, Drexel University, Fermilab, the Institute for
Advanced Study, the Japan Participation Group, The Johns Hopkins
University, the Joint Institute for Nuclear Astrophysics, the Kavli
Institute for Particle Astrophysics and Cosmology, the Korean
Scientist Group, the Chinese Academy of Sciences (LAMOST), Los Alamos
National Laboratory, the Max-Planck-Institute for Astronomy, the
Max-Planck-Institute for Astrophysics, New Mexico State University,
The Ohio State University, University of Pittsburgh, University of
Portsmouth, Princeton University, the United States Naval Observatory,
and the University of Washington.


\clearpage 
 
\begin{landscape}
\begin{deluxetable}{cccccccccccccccccccc}
\label{tab:datatable}
\tablenum{1}
\tablecolumns{20}
\tabletypesize{\scriptsize}
\tablewidth{0pc}
\tablecaption{Table of $i$-band Galaxy Parameters$^a$}
\tablehead{
\colhead{A/UGC} & 
\colhead{$i$} & 
\colhead{$V_{\rm CMB}$} & 
\colhead{$V_{\rm rot}$} & 
\colhead{T} & 
\colhead{$R_{\rm e}$} & 
\colhead{$R_{\rm 23.5}$} & 
\colhead{$R_{\rm d}$} & 
\colhead{m$^c_{\rm ext}$} & 
\colhead{m$^c_{\rm 23.5}$} & 
\colhead{m$^c_{\rm 2.2}$} & 
\colhead{$\mu_0^c$} & 
\colhead{$\mu_{\rm e}^c$} & 
\colhead{$g-r$} & 
\colhead{$g-i$} & 
\colhead{$C_{28,i}$} & 
\colhead{$\log M_{\hi}$} & 
\colhead{$\log M_*$} & 
\colhead{$\log M_{\rm bar}$} & 
\colhead{$\log M_{\rm T}$} \\
\colhead{(1)} & 
\colhead{(2)} &
\colhead{(3)} &
\colhead{(4)} & 
\colhead{(5)} & 
\colhead{(6)} & 
\colhead{(7)} & 
\colhead{(8)} & 
\colhead{(9)} & 
\colhead{(10)} & 
\colhead{(11)} & 
\colhead{(12)} & 
\colhead{(13)} & 
\colhead{(14)} & 
\colhead{(15)} & 
\colhead{(16)} & 
\colhead{(17)} & 
\colhead{(18)} & 
\colhead{(19)} & 
\colhead{(20)}
}
\startdata
       100002 &       71.6 &      5178. &      148.9 &         3. &      12.57 &      31.81 &       6.41 &      14.04 &      14.13 &      14.69 &      18.48 &      20.59 &       0.71 &       1.02 &       3.86 &       9.36 &      10.46 &      10.50 &      11.21\\
       100006 &       33.3 &      4621. &       91.2 &         5. &      18.06 &      23.37 &       4.76 &      13.97 &      14.11 &      15.19 &      17.86 &      21.73 &       0.47 &       0.80 &       2.64 &       9.19 &      10.01 &      10.09 &      10.65\\
       100018 &       69.4 &     11680. &      216.8 &         3. &       6.27 &      24.32 &       6.62 &      14.26 &      14.35 &      14.60 &      19.41 &      20.84 &       0.50 &       0.76 &       4.49 &       9.98 &      10.93 &      11.00 &      11.42\\
       100020 &       55.0 &      6058. &      157.4 &         3. &       5.65 &      14.52 &       5.56 &      14.59 &      14.78 &      14.85 &      20.02 &      20.29 &       0.38 &       0.64 &       3.51 &       9.71 &       9.98 &      10.22 &      10.92\\
       100025 &       18.7 &      7245. &      318.1 &         3. &       8.88 &      18.00 &       4.08 &      13.87 &      13.95 &      14.64 &      18.41 &      20.90 &       0.52 &       0.84 &       2.51 &       9.47 &      10.47 &      10.53 &      11.62\\

\hline
\enddata
\tablenotetext{a}{The full table will be presented in the refereed journal paper.}
\end{deluxetable}
\end{landscape}
\clearpage

\begin{deluxetable}{lrrrr|lrrrr}
\label{tab:skySubErr}
\tablecolumns{10}
\tablenum{2}
\tablewidth{0pc}
\tablecaption{Photometric Uncertainties for UGC 5651 with $\pm$0.2\% and $\pm$1.0\% Sky Error}
\tablehead{
\colhead{ Radius (\%) } & 
\colhead{ +0.2\% } & 
\colhead{ -0.2\% } & 
\colhead{ +1.0\% } & 
\colhead{ -1.0\% } & 
\colhead{ m (mag) } & 
\colhead{ +0.2\% } & 
\colhead{ -0.2\% } &
\colhead{ +1.0\% } & 
\colhead{ -1.0\% }  
}
\startdata
    $\Delta R_{\rm e}$  &      -1.70 &       1.88 &      -5.62 &      21.00 &  $\Delta m_{\rm ext}$  &       0.03 &      -0.03 &       0.15 &      -0.16\\
     $\Delta R_{23.5}$  &      -0.69 &       1.00 &      -3.93 &       5.14 &  $\Delta$ m$_{23.5}$   &       0.01 &      -0.01 &       0.07 &      -0.08\\
    $\Delta R_{\rm d}$  &      -9.65 &       6.19 &     -24.26 &      27.16 &  $\Delta$ m$_{\rm d}$  &      -0.03 &       0.00 &      -0.04 &      -0.09\\
                        &            &            &            &            &  $\Delta$ m$_{2.2}$    &       0.08 &      -0.05 &       0.21 &      -0.20\\

\enddata
\end{deluxetable}
\clearpage

\begin{deluxetable}{rc|ccccc}
\label{tab:VL}
\tablecolumns{7}
\tablenum{3}
\tablewidth{0pc}
\tablecaption{Orthogonal Fits for the $i$ band $VL$ Relation \break $\log V=a+b \log L$}
\tablehead{
\colhead{ Sample } & 
\colhead{  } & 
\colhead{  $a \pm$ $\Delta a$ } & 
\colhead{  $b \pm$ $\Delta b$ } & 
\colhead{ $\sigma_{VL}$} & 
\colhead{ $\sigma_{LV}$} & 
\colhead{ Pearson $r$ } 
}
\startdata
 $V_{\rm rot}|L^c_{\rm ext}$ & & & & & & \\
  A  &     3041 &  -0.688 $\pm$   0.060 &   0.279 $\pm$   0.006 &   0.141 &   0.504 &   0.711\\
  B  &     1725 &  -0.829 $\pm$   0.079 &   0.292 $\pm$   0.008 &   0.136 &   0.467 &   0.729\\
  C  &     1076 &  -0.670 $\pm$   0.054 &   0.279 $\pm$   0.005 &   0.075 &   0.269 &   0.885\\
  D  &      652 &  -0.761 $\pm$   0.086 &   0.287 $\pm$   0.008 &   0.075 &   0.261 &   0.872\\
 $V_{\rm rot}|L^c_{23.5}$ & & & & & & \\
  A  &     3041 &  -0.503 $\pm$   0.056 &   0.263 $\pm$   0.005 &   0.139 &   0.529 &   0.718\\
  B  &     1725 &  -0.619 $\pm$   0.074 &   0.274 $\pm$   0.007 &   0.134 &   0.491 &   0.737\\
  C  &     1076 &  -0.525 $\pm$   0.050 &   0.266 $\pm$   0.005 &   0.074 &   0.278 &   0.888\\
  D  &      652 &  -0.595 $\pm$   0.079 &   0.273 $\pm$   0.008 &   0.074 &   0.269 &   0.877\\

\hline
\enddata
\end{deluxetable}
\clearpage

\begin{deluxetable}{rc|ccccc}
\label{tab:VLgri}
\tablecolumns{7}
\tablenum{4}
\tablewidth{0pc}
\tablecaption{Orthogonal Fits for the $gri$ band $VL$ Relations
              of Sample D \break $\log V=a+b \log L$}
\tablehead{
\colhead{ Band } & 
\colhead{ N } & 
\colhead{  $a \pm$ $\Delta a$ } & 
\colhead{  $b \pm$ $\Delta b$ } & 
\colhead{ $\sigma_{VL}$} & 
\colhead{ $\sigma_{LV}$} & 
\colhead{ Pearson $r$ } 
}
\startdata
 $V_{\rm rot}|L^c_{\rm ext}$ & & & & & & \\
  g  &      652 &  -0.995 $\pm$   0.108 &   0.312 $\pm$   0.010 &   0.083 &   0.266 &   0.844\\
  r  &      652 &  -0.808 $\pm$   0.092 &   0.294 $\pm$   0.009 &   0.077 &   0.264 &   0.864\\
  i  &      652 &  -0.761 $\pm$   0.086 &   0.287 $\pm$   0.008 &   0.075 &   0.261 &   0.872\\
 $V_{\rm rot}|L^c_{23.5}$ & & & & & & \\
  g  &      652 &  -0.691 $\pm$   0.095 &   0.286 $\pm$   0.009 &   0.082 &   0.287 &   0.848\\
  r  &      652 &  -0.624 $\pm$   0.083 &   0.278 $\pm$   0.008 &   0.075 &   0.271 &   0.871\\
  i  &      652 &  -0.595 $\pm$   0.079 &   0.273 $\pm$   0.008 &   0.074 &   0.269 &   0.877\\

\hline
\enddata
\end{deluxetable}
\clearpage

{\tiny
\begin{deluxetable}{rc|ccccc}
\label{tab:RL}
\tablecolumns{7}
\tablenum{5}
\tablewidth{0pc}
\tablecaption{Orthogonal Fits for the $i$ band $RL$ Relation with Uncorrected
 Radii \break $\log R=a+b \log L$}
\tablehead{
\colhead{ Sample } & 
\colhead{ N } & 
\colhead{  $a \pm$ $\Delta a$ } & 
\colhead{  $b \pm$ $\Delta b$ } & 
\colhead{ $\sigma_{RL}$} & 
\colhead{ $\sigma_{LR}$} & 
\colhead{ Pearson $r$ } 
}
\startdata
 $R_{\rm d}|L^c_{\rm ext}$ & & & & & & \\
  A  &     3041 &  -3.312 $\pm$   0.075 &   0.377 $\pm$   0.007 &   0.171 &   0.454 &   0.743\\
  B  &     1725 &  -3.531 $\pm$   0.104 &   0.396 $\pm$   0.010 &   0.165 &   0.417 &   0.763\\
  C  &     1076 &  -3.028 $\pm$   0.118 &   0.349 $\pm$   0.011 &   0.153 &   0.440 &   0.744\\
  D  &      652 &  -3.454 $\pm$   0.179 &   0.387 $\pm$   0.017 &   0.151 &   0.392 &   0.751\\
 $R_{\rm d}|L^c_{23.5}$ & & & & & & \\
  A  &     3041 &  -2.821 $\pm$   0.083 &   0.332 $\pm$   0.008 &   0.185 &   0.557 &   0.689\\
  B  &     1725 &  -2.974 $\pm$   0.120 &   0.345 $\pm$   0.012 &   0.182 &   0.527 &   0.702\\
  C  &     1076 &  -2.677 $\pm$   0.121 &   0.317 $\pm$   0.012 &   0.162 &   0.513 &   0.704\\
  D  &      652 &  -2.996 $\pm$   0.193 &   0.345 $\pm$   0.019 &   0.163 &   0.475 &   0.700\\
 $R_{23.5}|L^c_{\rm ext}$ & & & & & & \\
  A  &     3041 &  -3.251 $\pm$   0.041 &   0.419 $\pm$   0.004 &   0.096 &   0.230 &   0.918\\
  B  &     1725 &  -3.500 $\pm$   0.041 &   0.441 $\pm$   0.004 &   0.073 &   0.166 &   0.953\\
  C  &     1076 &  -2.925 $\pm$   0.060 &   0.390 $\pm$   0.006 &   0.089 &   0.228 &   0.914\\
  D  &      652 &  -3.345 $\pm$   0.067 &   0.427 $\pm$   0.006 &   0.071 &   0.165 &   0.944\\
 $R_{23.5}|L^c_{23.5}$ & & & & & & \\
  A  &     3041 &  -2.948 $\pm$   0.039 &   0.393 $\pm$   0.004 &   0.098 &   0.249 &   0.915\\
  B  &     1725 &  -3.147 $\pm$   0.044 &   0.410 $\pm$   0.004 &   0.079 &   0.191 &   0.946\\
  C  &     1076 &  -2.692 $\pm$   0.058 &   0.370 $\pm$   0.006 &   0.091 &   0.247 &   0.908\\
  D  &      652 &  -3.052 $\pm$   0.075 &   0.402 $\pm$   0.007 &   0.076 &   0.189 &   0.935\\

\hline
\enddata
\end{deluxetable}
}

{\tiny
\begin{deluxetable}{rc|ccccc}
\label{tab:RLgri}
\tablecolumns{7}
\tablenum{6}
\tablewidth{0pc}
\tablecaption{Orthogonal Fits for the $gri$ band $RL$ Relations of Sample D \break
$\log R=a+b \log L$}
\tablehead{
\colhead{ Band } & 
\colhead{ N } & 
\colhead{  $a \pm$ $\Delta a$ } & 
\colhead{  $b \pm$ $\Delta b$ } & 
\colhead{ $\sigma_{RL}$} & 
\colhead{ $\sigma_{LR}$} & 
\colhead{ Pearson $r$ } 
}
\startdata
 $R_{\rm d}|L^c_{\rm ext}$ & & & & & & \\
  g  &      652 &  -3.831 $\pm$   0.183 &   0.424 $\pm$   0.018 &   0.145 &   0.343 &   0.763\\
  r  &      652 &  -3.476 $\pm$   0.169 &   0.389 $\pm$   0.016 &   0.144 &   0.371 &   0.761\\
  i  &      652 &  -3.454 $\pm$   0.179 &   0.387 $\pm$   0.017 &   0.151 &   0.392 &   0.751\\
 $R_{\rm d}|L^c_{23.5}$ & & & & & & \\
  g  &      652 &  -3.037 $\pm$   0.191 &   0.351 $\pm$   0.019 &   0.165 &   0.470 &   0.674\\
  r  &      652 &  -2.963 $\pm$   0.177 &   0.342 $\pm$   0.017 &   0.158 &   0.464 &   0.701\\
  i  &      652 &  -2.996 $\pm$   0.193 &   0.345 $\pm$   0.019 &   0.163 &   0.475 &   0.700\\
 $R_{23.5}|L^c_{\rm ext}$ & & & & & & \\
  g  &      652 &  -3.917 $\pm$   0.072 &   0.474 $\pm$   0.007 &   0.071 &   0.150 &   0.945\\
  r  &      652 &  -3.471 $\pm$   0.066 &   0.438 $\pm$   0.006 &   0.069 &   0.157 &   0.947\\
  i  &      652 &  -3.345 $\pm$   0.067 &   0.427 $\pm$   0.006 &   0.071 &   0.165 &   0.944\\
 $R_{23.5}|L^c_{23.5}$ & & & & & & \\
  g  &      652 &  -3.446 $\pm$   0.078 &   0.434 $\pm$   0.008 &   0.073 &   0.167 &   0.942\\
  r  &      652 &  -3.143 $\pm$   0.073 &   0.410 $\pm$   0.007 &   0.074 &   0.181 &   0.937\\
  i  &      652 &  -3.052 $\pm$   0.075 &   0.402 $\pm$   0.007 &   0.076 &   0.189 &   0.935\\

\hline
\enddata
\end{deluxetable}
}

{\tiny
\begin{deluxetable}{rc|ccccc}
\label{tab:RV} 
\tablecolumns{7}
\tablenum{7}
\tablewidth{0pc}
\tablecaption{Orthogonal Fits for the $i$ band $RV$ Relation \break
$\log R=a+b \log V$}
\tablehead{
\colhead{ Sample } & 
\colhead{ N } & 
\colhead{  $a \pm$ $\Delta a$ } & 
\colhead{  $b \pm$ $\Delta b$ } & 
\colhead{ $\sigma_{RV}$} & 
\colhead{ $\sigma_{VR}$} & 
\colhead{ Pearson $r$ } 
}
\startdata
 $R_{\rm d}|V_{\rm rot}$ & & & & & & \\
  A  &     3041 &  -2.944 $\pm$   0.161 &   1.606 $\pm$   0.073 &   0.312 &   0.194 &   0.428\\
  B  &     1725 &  -2.972 $\pm$   0.212 &   1.614 $\pm$   0.096 &   0.310 &   0.192 &   0.437\\
  C  &     1076 &  -3.087 $\pm$   0.196 &   1.661 $\pm$   0.089 &   0.226 &   0.136 &   0.592\\
  D  &      652 &  -3.484 $\pm$   0.315 &   1.821 $\pm$   0.141 &   0.240 &   0.132 &   0.563\\
 $R_{23.5}|V_{\rm rot}$ & & & & & & \\
  A  &     3041 &  -1.901 $\pm$   0.094 &   1.356 $\pm$   0.042 &   0.227 &   0.168 &   0.610\\
  B  &     1725 &  -1.871 $\pm$   0.109 &   1.339 $\pm$   0.049 &   0.212 &   0.158 &   0.655\\
  C  &     1076 &  -2.151 $\pm$   0.106 &   1.475 $\pm$   0.048 &   0.158 &   0.107 &   0.761\\
  D  &      652 &  -2.287 $\pm$   0.149 &   1.519 $\pm$   0.066 &   0.154 &   0.101 &   0.762\\

\hline
\enddata
\end{deluxetable}
}

\begin{deluxetable}{rc|ccccc}
\label{tab:RVgri} 
\tablecolumns{7}
\tablenum{8}
\tablewidth{0pc}
\tablecaption{Orthogonal Fits for the $gri$ band $RV$ Relations of Sample D \break
$\log R=a+b \log V$}
\tablehead{
\colhead{ Band } & 
\colhead{ N } & 
\colhead{  $a \pm$ $\Delta a$ } & 
\colhead{  $b \pm$ $\Delta b$ } & 
\colhead{ $\sigma_{RV}$} & 
\colhead{ $\sigma_{VR}$} & 
\colhead{ Pearson $r$ } 
}
\startdata
 $R_{\rm d}|V_{\rm rot}$ & & & & & & \\
  g  &      652 &  -3.303 $\pm$   0.310 &   1.733 $\pm$   0.139 &   0.229 &   0.133 &   0.565\\
  r  &      652 &  -3.303 $\pm$   0.293 &   1.732 $\pm$   0.132 &   0.230 &   0.133 &   0.563\\
  i  &      652 &  -3.484 $\pm$   0.315 &   1.821 $\pm$   0.141 &   0.240 &   0.132 &   0.563\\
 $R_{23.5}|V_{\rm rot}$ & & & & & & \\
  g  &      652 &  -2.439 $\pm$   0.155 &   1.539 $\pm$   0.069 &   0.162 &   0.105 &   0.742\\
  r  &      652 &  -2.313 $\pm$   0.148 &   1.514 $\pm$   0.066 &   0.155 &   0.102 &   0.758\\
  i  &      652 &  -2.287 $\pm$   0.149 &   1.519 $\pm$   0.066 &   0.154 &   0.101 &   0.762\\

\hline
\enddata
\end{deluxetable}

\begin{deluxetable}{rc|cc|ccc}
\label{tab:TFBTF}
\tablecolumns{7}
\tablenum{9}
\tablewidth{0pc}
\tablecaption{Orthogonal Fits for the Stellar Mass and Baryonic Tully-Fisher Relations \break
 $Y|X$, $\log Y = a + b \log X$}
\tablehead{
\colhead{ Sample D } & 
\colhead{ N } & 
\colhead{ $a \pm \Delta a$} & 
\colhead{ $b \pm \Delta b$ } & 
\colhead{ $\sigma_{YX}$} & 
\colhead{ $\sigma_{XY}$} & 
\colhead{  Pearson $r$  }
}
\startdata
$V|M_*$                                  &      652 & $ -0.429 \pm   0.068 $ & $  0.253 \pm   0.006 $ &   0.072 &   0.292 &   0.882\\
$V|M_{\rm bar}$                          &      562 & $ -0.879 \pm   0.102 $ & $  0.292 \pm   0.010 $ &   0.076 &   0.267 &   0.863\\
$(M_{\rm g}/M_*)|M_*$                      &      562 & $  7.635 \pm   0.430 $ & $ -0.782 \pm   0.042 $ &   0.374 &   0.474 &  -0.652\\
$(M_{\rm g}/M_*)|M_{\rm bar}$              &      562 & $  9.943 \pm   0.816 $ & $ -0.988 \pm   0.077 $ &   0.455 &   0.450 &  -0.515\\

\enddata
\end{deluxetable}
\clearpage


\clearpage

\clearpage

\clearpage

\clearpage

\clearpage

\clearpage

\clearpage

\clearpage

\clearpage

\clearpage

\clearpage

\clearpage

\clearpage

\clearpage

\clearpage

\clearpage

\clearpage

\clearpage

\clearpage

\clearpage

\clearpage

\clearpage

\clearpage

\end{document}